\documentclass[preprint,12pt]{elsarticle}

\usepackage{amssymb}
\usepackage{amsmath}
\usepackage{amsthm}
\theoremstyle{remark}
\newtheorem*{remark*}{\textbf{Remark}}
\usepackage{float}
\usepackage{booktabs}
\usepackage{verbatim}
\usepackage{graphicx}
\usepackage{pgfplots} 
\usepackage{comment}
\usepackage{adjustbox}
\usepackage{subcaption} 
\usepackage{xcolor}
\usepackage{microtype}

\usepackage{nomencl}
\usepackage{stmaryrd}
\makenomenclature

\usepackage{acronym}
\usepackage[acronym]{glossaries} 
\usepackage{hyperref}



\pgfplotsset{height=2cm,compat=1.15}
\usepgfplotslibrary{fillbetween}

\vspace{-10cm}

\journal{\fontsize{9}{11}\selectfont Computer Methods in Applied Mechanics and Engineering \hspace{0.3cm}}

\begin{document}

\begin{frontmatter}



\title{A shell-to-shell cohesive line element for efficient modeling of interfacial cracking in overmolded stiffened panels}

\author[1,2]{Sérgio G. F. Cordeiro \corref{cor1}}
\ead{s.ferreiracordeiro@tudelft.nl}
\author[2]{Boyang Chen}
\author[1]{Frans P. van der Meer}

\address[1]{Faculty of Civil Engineering and Geosciences, Delft University of Technology, The Netherlands}
\address[2]{Faculty of Aerospace Engineering, Delft University of Technology, The Netherlands}

\cortext[cor1]{Corresponding author}

\begin{abstract}

{\color{blue}The growing use of thermoplastics in lightweight structures requires efficient numerical methods to predict debonding in overmolded parts. In this work, a novel structural cohesive element is proposed as an efficient alternative to conventional cohesive elements for modeling debonding in thermoplastic composite panels with overmolded stiffeners.} Three-node, {\color{blue}higher-order hybrid/mixed} shell elements {\color{blue}based on the Kirchhoff hypothesis} are used to model thin panels and stiffeners. {\color{blue}The novel kinematics allows to obtain the jump vector at any point over the cohesive surface from the shell displacement approximations evaluated at the element edges. The weakly enforced higher-order continuity in skin and stiffener displacements enables debonding analysis on coarse meshes. The framework is suitable for analyzing debonding in skin-stiffener structures with non-constant damage through the stiffener thickness.} The model is verified for mode I, mode II and mixed-mode benchmark problems. A debonding problem is analyzed with both standard 3D cohesive elements and the proposed element. The results show that the element size in the proposed models can be much larger than that in the standard model, with more than 95\% reduction in CPU time. The debonding analysis of a complex stiffened panel is also presented to demonstrate the intended use of the proposed element for simulating debonding in structural~components.

\end{abstract}

\begin{graphicalabstract}

\vspace{20pt}

\includegraphics[width=\textwidth]{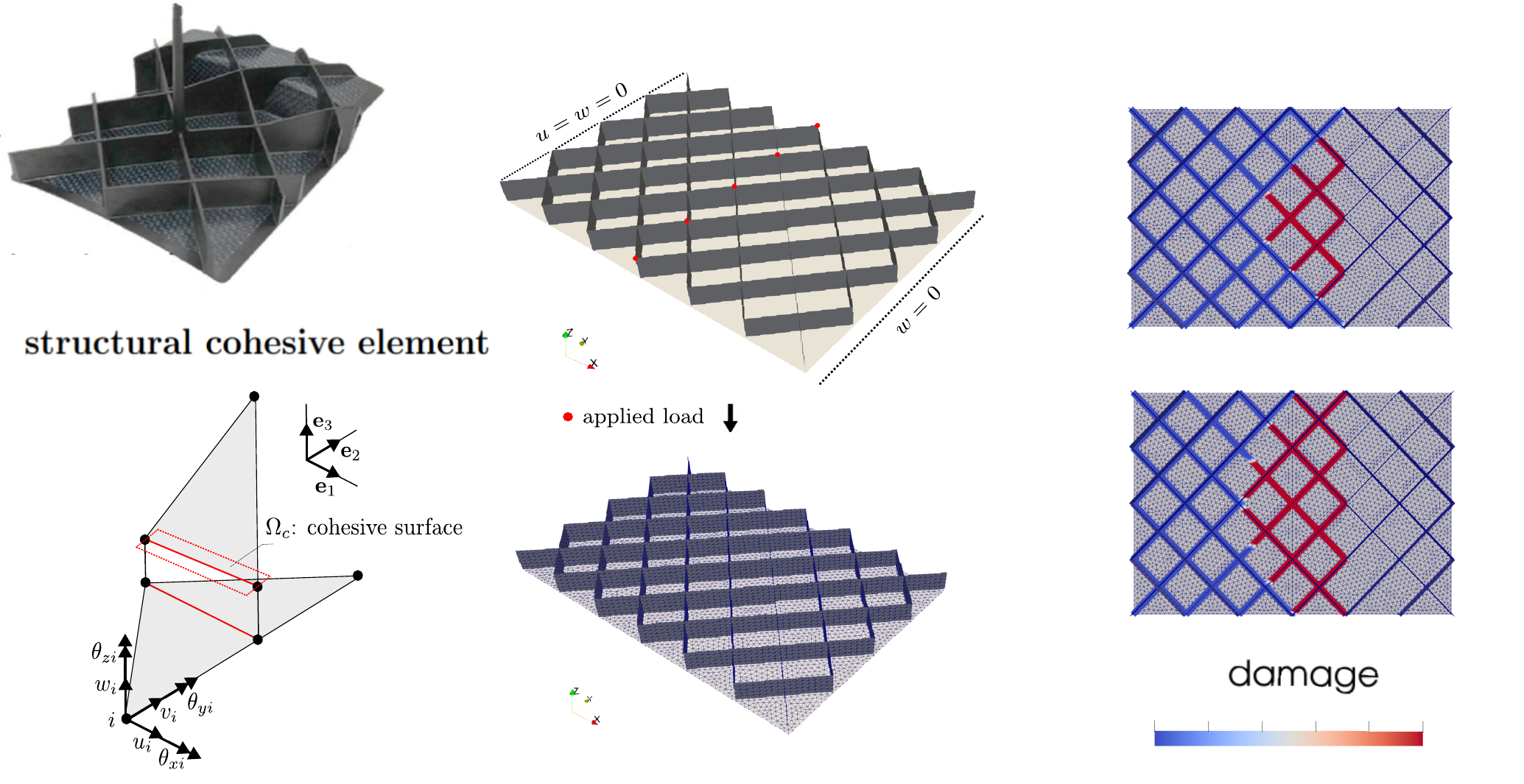}
\end{graphicalabstract}

\begin{highlights}
\item A {\color{blue}higher-order} shell-to-shell cohesive line element is developed for modeling skin-stiffener debonding between overmolded parts;

\item The element yielded stable cohesive crack propagation on coarse meshes;

\item The results show that the proposed element can reduce the CPU time by more than 95\% compared to standard cohesive elements.

\end{highlights}


\end{frontmatter}


\section{Introduction}
\label{sec:introduction}

Fiber reinforced polymer composites are increasingly adopted in the aerospace industry to meet lightweighting requirements. In particular, laminated stiffened panels made of thermoplastic composites can be fusion bonded, without the need of additional materials such as adhesives or bolts, resulting in more weight-savings, faster processing cycles and the possibility to manufacture composite parts with more complex geometries. The overmolding of a dissimilar material in a thermoplastic laminate is also an alternative that allows for high performance stiffened structural parts \cite{Akkermanetal2020}. However, the failure behaviour of these fusion bonded/overmolded thermoplastic composites strongly depends on the processing conditions \cite{Valverdeetal2018, Valverdeetal2020, Akkermanetal2020, Neveuetal2022}. At present, the lack of {\color{blue}efficient} and reliable performance prediction tools forms an obstacle to the widespread use of thermoplastic composites.

Describing the mechanical behavior of composites through finite element simulations has been traditionally a challenge because of the  complex anisotropic, inhomogeneous and multiscale nature of these materials. The most common approach in finite element modelling of composite materials is based on macroscale modelling, where layered shells, stacked solid or stacked shell elements are used to represent the laminate. However, such modelling strategies alone cannot explicitly capture interlaminar damage. To simulate delamination between laminate plies, cohesive elements are commonly introduced between adjacent layers \cite{Turonetal2006, Turonetal2010}. The cohesive elements are developed based on the cohesive zone model, proposed by Dugdale and Barenblatt \cite{Dugdale1960, Barenblatt1962}, in which a fracture process zone exists along the interface, ahead of the stress-free crack tip. A traction-separation relationship describes how the interfacial stresses and damage evolve with respect to the interfacial openings. Standard cohesive elements are usually developed for use between two solid elements to model their debonding \cite{Qiuetal2001, CamanhoandDavila2002, YangandCox2005}.

Another common failure mode of laminated composite panels is skin-stiffener debonding due to relatively low interface strengths \cite{Akterskaiaetal2018}. In thermoplastic composites, even if good welding occurs, the weaker zone moves to the laminate side at the interface between the matrix and the reinforcement, resulting in a weaker zone at the welding area \cite{GiustiandLucchetta2020}. Cohesive element models are also a versatile approach for the modeling of skin-stiffener separation in composite stiffened panels \cite{BalzaniandWagner2010, Akterskaiaetal2018, GiustiandLucchetta2023}. For instance, Balzani and Wagner \cite{BalzaniandWagner2010} examined debonding between skin and stiffener with cohesive elements, implemented as surface interface elements between shells. Akterskaia et al. \cite{Akterskaiaetal2018} developed a global-local methodology to evaluate skin–stiffener debonding with two accuracy levels: a fast global analysis to identify critical areas using a coarse mesh, followed by refined local submodels where cohesive elements simulate detailed damage propagation. The fracture strength of thermoplastic composite T-joints was investigated through numerical simulations in \cite{GiustiandLucchetta2023, Hofmanetal2026}. While \cite{GiustiandLucchetta2023} evaluated mode I fracture strength using a mode-I cohesive model, \cite{Hofmanetal2026} employed a framework combining a mixed-mode cohesive model with an anisotropic viscoplasticity model for the laminate.

Although widely used for delamination and  skin–stiffener debonding modeling, standard cohesive elements require very fine meshes because the element size must be much smaller than the cohesive-zone length. High stress gradients develop within the cohesive zone during failure in composites \cite{Qiuetal2001, YangandCox2005}. Therefore, a sufficiently refined mesh is required to accurately capture the solution. The fine mesh requirement of cohesive-element modeling has drawn the attention of many researchers in the past \cite{Turonetal2007, Yangetal2010, vanderMeeretal2012, Doetal2013, Luetal2018, BalducciandChen2024, Aietal2025}.  The so-called structural cohesive elements, which conform to {\color{blue}$C^1$ continuous} structural elements, i.e., beams, plates and shells, are a recent approach that has demonstrated potential for delamination analysis with coarse meshes \cite{RussoandChen2020,BalducciandChen2024,Aietal2025}. {\color{blue}Shell models based on the Kirchhoff hypothesis have previously been adopted in the formulation of the structural cohesive elements \cite{Bazilevsetal2018, BalducciandChen2024, Aietal2025}. The advantages of such models stem from the fact that the higher-order smooth representation of the shell mid-surface displacement fields allows to adopt relatively coarse discretizations without sacrificing solution accuracy. In multi-layer delamination modelling, compliant cohesive interfaces can still provide the transverse shear deformation expected in a laminate that isolated Kirchhoff models cannot capture.} For instance, Balducci and Chen \cite{BalducciandChen2024} developed a structural cohesive element for delamination problems which is compatible with TUBA3 plate elements \cite{Bell1969}. Their results showed that the formulation allowed for the use of larger elements in delamination analysis. However, the curvature degrees of freedom (DoFs) of the element make it more complicated to set boundary conditions. Recently, Ai et al. \cite{Aietal2025} extended the Allman \cite{Allman1976} triangular Kirchhoff plate element, which has only DoFs commonly used by engineers, for the modeling of composite plies. They also developed the corresponding structural {\color{blue}cohesive element} designed for delamination modeling. {\color{blue}Even though the hybrid shell formulation employed in \cite{Aietal2025} does not actually give a smooth representation of the shell mid-surface transversal displacement in a strong sense, it does enforce higher-order continuity in a weak sense, which appears to be enough to give more accurate results than the standard models \cite{Aietal2025}.}


This article introduces a new structural cohesive element formulation specifically designed for the skin-stiffener debonding problem. First, the plate element from \cite{Aietal2025} is extended to 3D shell models by using the triangular membrane element with drilling DoFs developed by Bergan and Felippa \cite{BerganandFelippa1985}. This {\color{blue}ensures higher-order approximations for the in-plane displacements and avoids membrane locking issues in case of in-plane bending of the stiffeners, while preserving a simple shell formulation based on a 3-node triangular element with DoFs commonly used to impose boundary conditions in structural analysis} \cite{Felippa2003, Boutagouga2021}. Then, the novel structural {\color{blue}cohesive element}, namely the shell-to-shell cohesive line element, is proposed for the efficient modeling of skin-stiffener debonding in overmolded panels. {\color{blue} The main novelty is the proposed kinematics that allows to obtain the jump vector over the skin-stiffener interface from the higher-order shell mid-surface displacement fields evaluated at the element edges. The model is suitable for analyzing debonding in skin-stiffener structures with non-constant damage
through the stiffener thickness. It is designed to capture the overall skin–stiffener debonding response when the panel failure is governed solely by interface damage. Further developments are required to account for intralaminar damage or plasticity.}

\vspace{5pt}

The paper is outlined as follows. The governing equations and finite element approximations for the shell element are briefly reviewed in Section \ref{sec:K-L shell element}. The proposed structural cohesive element is presented in Section \ref{sec:shell-to-shell CE}. Numerical results are presented in Section \ref{sec:results}. The model is first verified using mode I, mode II, and mixed-mode classical benchmarks. The performance of the proposed cohesive element is then demonstrated through comparisons with standard 3D cohesive elements in debonding problems. The debonding analysis of a complex stiffened panel is also presented for illustrative purposes. In Section \ref{sec:conclusions}, we draw the conclusions of this work.

\section{{\color{blue}Hybrid/mixed higher-order shell element}}
\label{sec:K-L shell element}

\begin{figure}[H]
\centering
\scalebox{0.40}{\includegraphics{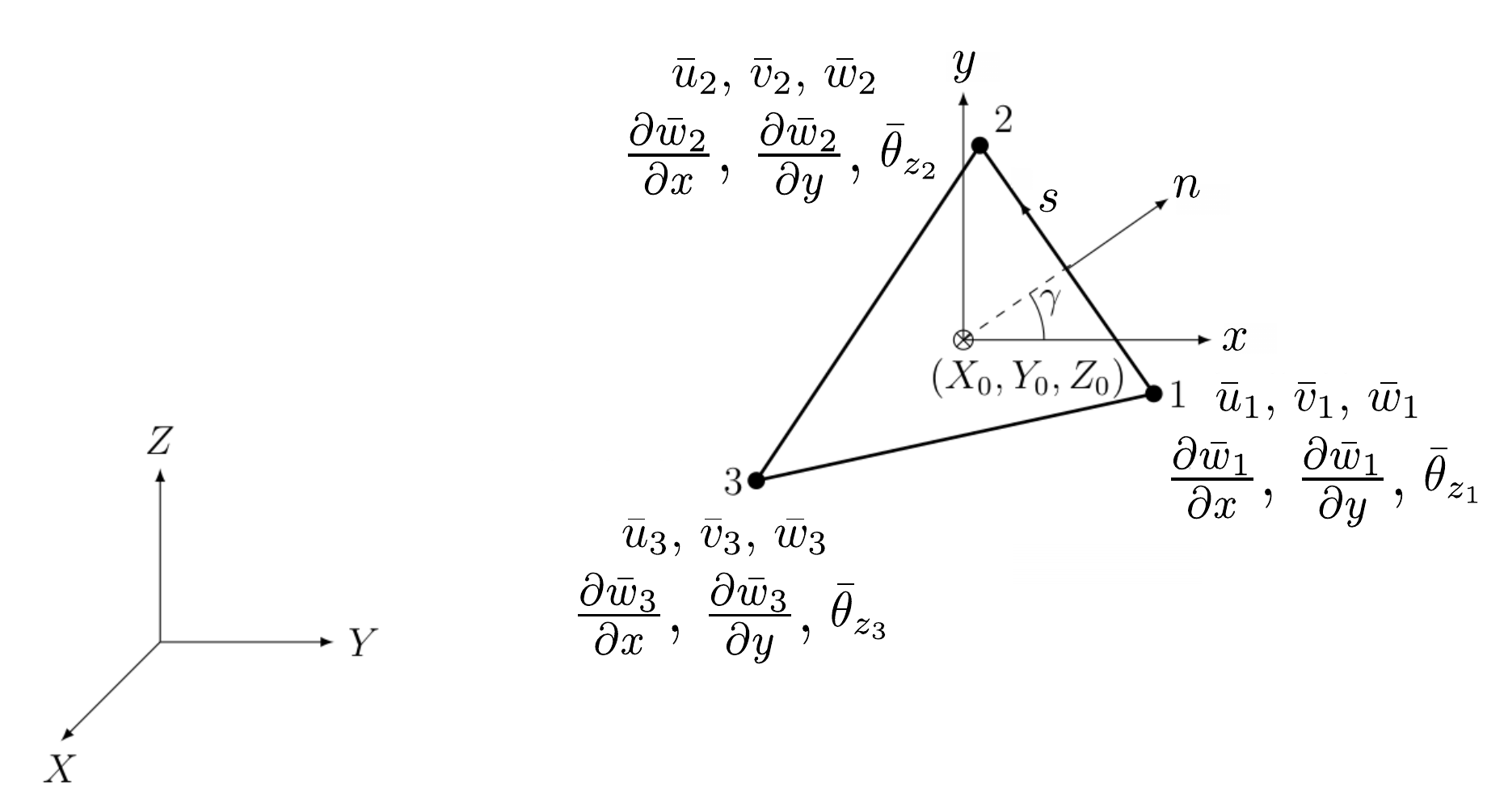}}
	\caption{\color{blue}Three–node higher-order shell element.}
	\label{fig:ShellElement}
\end{figure}

The {\color{blue}three-node higher-order} shell element is based on the membrane element with drilling DoFs developed in {\color{blue}\cite{BerganandNygard1984,BerganandFelippa1985,Felippa1989}}, and the Kirchhoff plate element developed in \cite{Allman1976, Aietal2025}. Figure \ref{fig:ShellElement} shows the element and the nodal DoFs. A local coordinate system with an origin at the centroid is defined. The local axes are represented by $x$, $y$ and $z$, to distinguish them from the global ones, $X$, $Y$ and $Z$. The area of the element is $A^{e}$. The anti-clockwise coordinate along the element boundary is denoted by $s$, and $n$ is the exterior normal. The angle between the normal $n$ and
the local axis $x$ is $\gamma$. 
\vspace{-5pt}
\vspace{-5pt}

The in-plane displacements in the $x$ and $y$ directions defined over $A^{e}$ are $u(x, y)$ and $v(x, y)$, respectively. Independent in-plane displacements $\overline{u}_n(s)$ and $\overline{u}_s(s)$ are assumed along the boundary $\partial A^{e}$. The membrane DoFs are the nodal displacements $\overline{u}$, $\overline{v}$ and the nodal rotation $\overline{\theta}_z$. The out-of-plane plate displacement defined over the domain is $w(x, y)$. An independent out-of-plane boundary displacement, $\overline{w}(s)$, and its compatible normal derivative, $\partial \overline{w}/\partial n(s)$, are assumed along the boundary $\partial A^{e}$. The plate DoFs at each node include the displacement $\overline{w}$, and the two rotations $\partial \overline{w}/\partial x =-\overline{\theta}_y$ and $\partial \overline{w}/\partial y =\overline{\theta}_x$, defined according to Kirchhoff kinematics.

\subsection{Membrane element with drilling DoFs}

Consider a membrane surface domain $A$ with boundary $\partial A: \partial A_{\color{blue}N} \cup \partial A_{\mathrm{d}}$ in a plane stress condition. Membrane forces  $\mathbf{N}^{*} =
\bigl\lfloor N_{nn}^{*},\ N_{ns}^{*} \bigr\rfloor^{\mathrm{T}}$ are  prescribed on $\partial A_{\color{blue}N}$, whereas in-plane displacements $\mathbf{d}^{*}=\bigl\lfloor u_{n}^{*},\ u_{s}^{*} \bigr\rfloor^{\mathrm{T}}$ are prescribed on $\partial A_{\mathrm{d}}$. The internal fields are in-plane displacements $\mathbf{u}=\bigl\lfloor u,\ v \bigr\rfloor^{\mathrm{T}}$, membrane forces $\mathbf{N}=\bigl\lfloor N_{\text{xx}},\ N_{\text{yy}},\ N_{\text{xy}} \bigr\rfloor^{\mathrm{T}}$, membrane strains $\boldsymbol{\varepsilon}_{m}=\bigl\lfloor \varepsilon_{\text{xx}},\ \varepsilon_{\text{yy}},\ \gamma_{\text{xy}} \bigr\rfloor^{\mathrm{T}}$, and given forces per unit area $\mathbf{q}^{*}= \bigl\lfloor q_{\text{x}}^{*},\ q_{\text{y}}^{*}\bigr\rfloor^{\mathrm{T}}$. The internal field equations are: 
\begin{align} 
        {\color{blue}\boldsymbol{\varepsilon}_{\mathrm{m}}=\boldsymbol{\nabla}_{\mathrm{m}}\mathbf{u}, \hspace{20pt} \boldsymbol{N}=\mathbf{A} \boldsymbol{\varepsilon}_{\mathrm{m}}, \hspace{20pt} \boldsymbol{\nabla}_{\mathrm{m}}^{\mathrm{T}} \mathbf{N}+\mathbf{q}^{*}=\mathbf{0} \hspace{10pt} \text{in $A$}}
\end{align}
where
\begin{align} \nonumber
        \boldsymbol{\nabla}^{\mathrm{T}}_{\mathrm{m}}=\left[\begin{array}{ccc}
             \displaystyle \frac{\partial(\cdot)}{\partial x} & 0 &  \displaystyle \frac{\partial(\cdot)}{\partial y} \\[1.0em] 
            0 &  \displaystyle \frac{\partial(\cdot)}{\partial y} &  \displaystyle \frac{\partial(\cdot)}{\partial x} 
            \end{array}\right], \hspace{5pt} \text{and} \hspace{5pt} \mathbf{A}=\left[\begin{array}{ccc}
            A_{11} & A_{12} &  A_{16} \\
            A_{12} & A_{22} &  A_{26} \\
            A_{16} & A_{26}  & A_{66}
            \end{array}\right]
\end{align}

\noindent are the membrane strain operator and the membrane stiffness matrix, respectively. The hybrid/mixed parametrized energy functional of $n$ membrane elements connected through their boundary displacements, as given by \cite{Felippa1989}, can be written as 
\begin{align} \nonumber
        \Psi_{\mathrm{m}}(\tilde{\mathbf{u}}, \tilde{\mathbf{N}}, \bar{\mathbf{d}}) = \sum_{e=1}^{n} \left\{\frac{1}{2}(1-\gamma)\int_{A^{e}} \mathbf{N}^{u\mathrm{T}} \boldsymbol{\varepsilon}_{\mathrm{m}}^u \mathrm{~d} A
        -\frac{1}{2} \gamma \int_{A^{e}} \tilde{\mathbf{N}}^{T} \boldsymbol{\varepsilon}_{\mathrm{m}}^N \mathrm{~d} A \right.\\
        \left.+\gamma \int_{A^{e}}\tilde{\mathbf{N}}^{T} \boldsymbol{\varepsilon}_{\mathrm{m}}^u \mathrm{~d} A-\mathbb{P}^{e}_{\mathrm{m}}\right\}, 
\end{align}
\noindent where $\gamma$ is a scalar, {\color{blue}and $\mathbb{P}^{e}_{\mathrm{m}}(\tilde{\mathbf{u}}, \tilde{\mathbf{N}}, \bar{\mathbf{d}})$ is the membrane forcing potential, analogous to the forcing potential of the d-generalized variational principle given by \cite{Felippa1989}, which enforces in a weak sense $\tilde{\mathbf{u}}$ and $\bar{\mathbf{d}}$ to be equal at $\partial A$.}

For the finite element discretization, it is assumed that
\begin{align}
        \tilde{\mathbf{u}}=\boldsymbol{\phi}_{u}\mathbf{q}_{u}, \hspace{7pt} \tilde{\mathbf{N}}=\boldsymbol{\phi}_{N}\mathbf{q}_{N} \hspace{7pt} \text{in} \hspace{7pt} A^{e}, \quad \text{and} \quad \bar{\mathbf{d}}=\boldsymbol{\phi}_{d} \mathbf{q}_{d} \hspace{7pt} \text{in} \hspace{7pt} \partial A^{e}.  
    \label{Membrane_Approximations}
\end{align}
\noindent {\color{blue}The matrices $\boldsymbol{\phi}_{u}$, $\boldsymbol{\phi}_{N}$ and $\boldsymbol{\phi}_{d}$ collect generalized shape functions for internal displacements, internal stresses and boundary displacements, respectively. The column vectors $\mathbf{q}_{u}$, $\mathbf{q}_{N}$ and $\mathbf{q}_{d}$ collect the corresponding generalized degrees of freedom. 
The membrane element from \cite{BerganandFelippa1985} approximates membrane forces as constant $\tilde{\mathbf{N}}= \overline{\mathbf{N}}$. Besides, the $\tilde{\mathbf{u}}$ approximation is decomposed into rigid body ($r$), constant strain ($c$), and higher-order ($h$) displacements}
\begin{align}
       {\color{blue}\tilde{\mathbf{u}}=\left\{\begin{array}{l}
 \tilde{u} \\
 \tilde{v}
\end{array}\right\}=\sum_{i=1}^9 \boldsymbol{\phi}_{u_i} q_{u_i}=\boldsymbol{\phi}_{u}\mathbf{q}_{u}=\boldsymbol{\phi}_{r} \mathbf{q}_{\mathrm{r}}+\boldsymbol{\phi}_{c} \mathbf{q}_{\mathrm{c}}+\boldsymbol{\phi}_{h} \mathbf{q}_{\mathrm{h}},}  
    \label{Displacement_decomposition}
\end{align}
{\color{blue}where the rigid body, constant strain and higher-order modes $\boldsymbol{\phi}_{r}$, $\boldsymbol{\phi}_{c}$ and $\boldsymbol{\phi}_{h}$ are given in \cite{BerganandFelippa1985}. The internal fields $\tilde{\mathbf{u}}$ and $\tilde{\mathbf{N}}$ may be discontinuous across elements. On the other hand, the boundary displacement field $\bar{\mathbf{d}}$ have the same value on adjacent elements, as the approximation adopted for $\bar{\mathbf{d}}$ on an edge separating two elements is uniquely interpolated by nodal values $\mathbf{q}_{d}=\overline{\mathbf{u}}$:}
\vspace{-5pt}
\begin{align}
    {\color{blue}\overline{\mathbf{u}}=\left\{\overline{u}_1, \overline{v}_1, \overline{\theta}_{z_1},\overline{u}_2, \overline{v}_2, \overline{\theta}_{z_2}, \overline{u}_3, \overline{v}_3, \overline{\theta}_{z_3} \right\}^{\mathrm{T}},}  
    \label{Membrane_Dofs}
\end{align}
{\color{blue}where $\theta_{z}$ is given in geometrically linear fashion as $\theta_{z}=\displaystyle \frac{1}{2}\left(\frac{\partial v}{\partial x}-\frac{\partial u}{\partial y}\right)$.}

\noindent {\color{blue}The boundary displacement $\bar{\mathbf{d}}_{ij}=\left\{\bar{u}_{n}, \bar{u}_{s}\right\}^{\mathrm{T}}$ along side $i-j$ in the side-coordinate system $(n, s)$ is interpolated by beam-type shape functions using the nodal in-plane displacements $u,v$ and in-plane rotations $\theta_{z}$. The relation between $\mathbf{q}_u$ and the membrane DoFs $\overline{\mathbf{u}}$ is given by}
\begin{align}
    {\color{blue}\overline{\mathbf{u}}=\mathbf{G} \mathbf{q}_{u}=
    \left[\begin{array}{c}
\mathbf{G}_{r} \quad \mathbf{G}_{c} \quad \mathbf{G}_{h}
\end{array}\right] \mathbf{q}_{u}=
\mathbf{G}_{r} \mathbf{q}_{r}+\mathbf{G}_{c} \mathbf{q}_{c}+\mathbf{G}_{h} \mathbf{q}_{h},}
    \label{au_to_qm_transformation}
\end{align}
{\color{blue}where the ($9\times3$) matrices $\mathbf{G}_{r}$, $\mathbf{G}_{c}$ and $\mathbf{G}_{h}$ are given in \cite{BerganandFelippa1985}. The resulting $9\times9$ matrix $\mathbf{G}$ is non-singular and may be inverted to give}
\begin{align}
    {\color{blue}\mathbf{q}_{u}=
    \left\{\begin{array}{ccc}
        \mathbf{q}_{r} &
        \mathbf{q}_{c} &
        \mathbf{q}_{h}
    \end{array}\right\}^{\mathrm{T}}=
    \mathbf{G}^{-1} \bar{\mathbf{u}}=
    \left[\begin{array}{ccc}
        \mathbf{H}_{r} &
        \mathbf{H}_{c} &
        \mathbf{H}_{h}
    \end{array}\right]^{\mathrm{T}} \bar{\mathbf{u}}.}
    \label{qm_to_au_transformation}
\end{align}

{\color{blue}Inserting the approximations into the functional $\Psi_{\mathrm{m}}^{e}$ of an element, and making it stationary, yields the element equations. The internal displacement decomposition in Eq. (\ref{qm_to_au_transformation}) induces a partitioned version of the element equations, which can be reduced by using Eq. (\ref{Displacement_decomposition}) and static condensation, as discussed in detail in \cite{Felippa1989}. The reduced element equations are given by}
\begin{align}
    {\color{blue}\left[\mathbf{K}_{\mathrm{b}}+(1-\gamma) \mathbf{K}_{\mathrm{h}}\right] \bar{\mathbf{u}}=\mathbf{f}_{\mathrm{m}}  \quad \Rightarrow \quad \mathbf{K}_{\mathrm{m}} \bar{\mathbf{u}}=\mathbf{f}_{\mathrm{m}}}
    \label{Final_Membrane_System}
\end{align}

\noindent {\color{blue}in which $\mathbf{K}_{\mathrm{m}}$ is the membrane stiffness matrix and $\gamma=1/2$ is recommended in \cite{BerganandFelippa1985} as an optimal value for bending behavior of regular element patches. $\mathbf{K}_{\mathrm{b}}$ and $\mathbf{K}_{\mathrm{h}}$ are, respectively, the basic and higher-order membrane stiffness matrices, given by}
\begin{align}
    {\color{blue}\mathbf{K}_{\mathrm{b}}=\frac{1}{A^{e}} \overline{\mathbf{L}} \mathbf{A} \overline{\mathbf{L}}^{\mathrm{T}} \hspace{5pt} \text{and} \hspace{5pt}  \mathbf{K}_{\mathrm{h}}=\mathbf{H}_{\mathrm{h}}^{\mathrm{T}} \mathbf{K}_{qh} \mathbf{H}_{\mathrm{h}}, \hspace{3pt} \text{with} \hspace{10pt} \mathbf{K}_{qh}=\int_{A^{e}} \mathbf{B}_{\mathrm{h}}^{\mathrm{T}} \mathbf{A} \mathbf{B}_{\mathrm{h}} \mathrm{~d} A.}
    \label{K_b_and_K_qh}
\end{align}
{\color{blue}Expressions for the lumping matrix $\overline{\mathbf{L}}$ and the generalized higher-order stiffness matrix $\mathbf{K}_{qh}$ were explicitly derived in \cite{BerganandFelippa1985}. The external force vector of the membrane element is} 
\begin{align}
     {\color{blue}\mathbf{f}_{\mathrm{m}} = \mathbf{f}_{N^{*}}+\mathbf{H}_{\mathrm{r}}^{\mathrm{T}} \mathbf{f}_{qr^{*}}+\frac{1}{A^{e}} \overline{\mathbf{L}} \mathbf{f}_{qc^{*}}+\mathbf{H}_{\mathrm{h}}^{\mathrm{T}} \mathbf{f}_{qh^{*}}.}
    \label{Membrane_Stiffnes_Matrices}
\end{align} 

\noindent {\color{blue}in which $\mathbf{f}_{N^*}=\int_{\partial A^{e}_t} \boldsymbol{\phi}_{u}^{\mathrm{T}} \mathbf{N^*} \mathrm{~d} s$ and $\mathbf{f}_{qk^{*}}=\int_{A^{e}} \boldsymbol{\phi}_{k}^{\mathrm{T}} \mathbf{q}^{*} \mathrm{~d} A$, where $k=r,c,h$.}


\subsection{Kirchhoff-plate element}

Consider a plate surface domain $A$ with boundary {\color{blue}$\partial A: \partial A_{V_n} \cup \partial A_{w}$ or $\partial A:\partial A_{M_{n}} \cup \partial A_{w_{n}}$.} Kirchhoff shear forces $V_n^{*}$ and bending moments $M_{nn}^{*}$ are prescribed on boundaries of the type {\color{blue}$\partial A_{V_n}$ and $\partial A_{M_{n}}$, respectively}, whereas out-of-plane displacements $w^{*}$ and normal rotations $\partial w^{*}/\partial n$ are prescribed on boundaries of the type {\color{blue}$\partial A_{w}$ and $\partial A_{w_{n}}$, respectively.} In addition, $R_{N}^{*}$ are prescribed values of possible concentrated forces. The internal fields are the out-of-plane displacement $w$, bending moments $\mathbf{M}=\bigl\lfloor M_{\text{xx}},\ M_{\text{yy}},\ M_{\text{xy}} \bigr\rfloor^{\mathrm{T}}$, curvatures $\boldsymbol{\kappa}=\bigl\lfloor \kappa_{\text{xx}},\ \kappa_{\text{yy}},\ \kappa_{\text{xy}} \bigr\rfloor^{\mathrm{T}}$, and given out-of-plane forces per unit area $q_z^{*}$. The internal field equations are:
\begin{align} 
        {\color{blue}\boldsymbol{\kappa}=\boldsymbol{\nabla}_{\mathrm{p}}w, \hspace{20pt} \boldsymbol{M}=\mathbf{D} \boldsymbol{\kappa}, \hspace{20pt} \boldsymbol{\nabla}_{\mathrm{p}}^{\mathrm{T}} \mathbf{M}+q_{z}^{*}=0\hspace{10pt} \text{in $A$}}
\end{align}
where
\begin{align} \nonumber
        \boldsymbol{\nabla}^{\mathrm{T}}_{\mathrm{p}}=\left[
\begin{array}{ccc}
\displaystyle \frac{\partial^{2}(\cdot)}{\partial x^{2}} &
\displaystyle \frac{\partial^{2}(\cdot)}{\partial y^{2}} &
2\displaystyle\frac{\partial^{2}(\cdot)}{\partial x\,\partial y}
\end{array}
\right] \quad \text{and} \quad \mathbf{D}=\left[\begin{array}{ccc}
           D_{11} & D_{12} & D_{16} \\
           D_{12} & D_{22} & D_{26} \\
           D_{16} & D_{26} & D_{66}
            \end{array}\right]
\end{align}

\noindent are the Kirchhoff-plate differential operator and the plate stiffness matrix.  The hybrid energy functional of $n$ plate elements connected through their boundary displacements can be written as \cite{Allman1976}
\begin{align} \nonumber
        \Psi_{\mathrm{p}}\left(\tilde{w}, \bar{w} \right) =\sum_{e=1}^{n} \left\{\int_{A^{e}} U^{w}_0 \mathrm{~d} A
        +\sum_{N=1}^{3}R^w_{N}\left(\bar{w}_N-\tilde{w}_{N}\right)\right. \\
        \left.+\int_{\partial A^{e}} V^w_{n}\left(\bar{w}-\tilde{w}\right) \mathrm{~d} s - \int_{\partial A^{e}} M^w_{nn}\left(\frac{\partial \bar{w}}{\partial n}-\frac{\partial \tilde{w}}{\partial n}\right) \mathrm{~d} A 
        -\mathbb{P}^{e}_{\mathrm{p}}\right\}, 
        \label{Plate_Energy_Functional}
\end{align}
\noindent where $\tilde{w}$ and $\bar{w}$ are the domain and boundary out-of-plane
plate displacements, respectively. {\color{blue}The strain energy density $U^{w}_0$, computed from $\tilde{w}$ for symmetric laminates, and the forcing potential $\mathbb{P}^{e}_{\mathrm{p}}$ for the plate are presented in \cite{Aietal2025}. The bending moments $\mathbf{M}^w=\bigl\lfloor M^w_{\text{xx}},\ M^w_{\text{yy}},\ M^w_{\text{xy}} \bigr\rfloor^{\mathrm{T}}$ give rise to a normal bending moment $M^w_{nn}$, and resultant Kirchhoff shear force $V^w_{n}$, on the element boundary, together with concentrated forces $R^{w}_{N\,(N=1,2,3)}$ at the vertices \cite{Allman1976}.

For the finite element discretization, it is assumed that}
\begin{align} \nonumber
        {\color{blue}\tilde{w}=A_1+A_2 x+A_3 y+\alpha_1 x^2+\alpha_2 x y+\alpha_3 y^2+\alpha_4 x^3}\\
        {\color{blue}+\alpha_5 x^2 y+\alpha_6 x y^2+\alpha_7 y^3}  
    \label{Plate_Approximations}
\end{align}

\noindent {\color{blue}where the three coefficients $A_1, A_2$ and $A_3$ represent arbitrary rigid body movement. With this choice of $\tilde{w}$, and applying the Green’s theorem to transform the strain energy integral, it is possible to rewrite the plate functional given in Eq. (\ref{Plate_Energy_Functional}) as}
\begin{align} \nonumber
        {\color{blue}\Psi_{\mathrm{p}}\left(\tilde{w}, \bar{w} \right) =\sum_{m} \left\{-\int_{A^{e}} U^{w}_0 \mathrm{~d} A
        +\sum_{N=1}^{3}R^w_{N}\bar{w}_N
        +\int_{\partial A^{e}} V^w_{n}\bar{w} \mathrm{~d} s\right.} \\
         {\color{blue}\left.-\int_{\partial A^{e}} M^w_{nn}\frac{\partial \bar{w}}{\partial n} \mathrm{~d} s 
       -\sum_{N=1}^{3}R_{N}^{*}\bar{w}_N 
       -\int_{\partial A^{e}}V_n^{*} \bar{w} \mathrm{~d} s
       +\int_{\partial A^{e}} M_{nn}^{*} \frac{\partial \bar{w}}{\partial n} \mathrm{~d} s\right\},} 
        \label{Modified_Plate_Energy_Functional}
\end{align}
\noindent {\color{blue}Substituting Eq. (\ref{Plate_Approximations}) into strain energy density $U^{w}_0$, the plate strain energy can be rewritten as}
\begin{align} 
        {\color{blue}\int_{A^{e}} U^{w}_0 \mathrm{~d} x \mathrm{~d} y=\frac{1}{2} \boldsymbol{\alpha}^{\mathrm{T}} \mathbf{H} \boldsymbol{\alpha},} 
    \label{Discrete_Strain_Energy}
\end{align}
\noindent {\color{blue}where $
\boldsymbol{\alpha}=\left\{\alpha_1, \alpha_2, \alpha_3, \alpha_4, \alpha_5, \alpha_6, \alpha_7\right\}^{\mathrm{T}}
$, and the matrix $\mathbf{H}$ was derived for symmetric laminates in \cite{Aietal2025}. Linear normal bending moment $\bar{M}_{nn}$ and constant Kirchhoff shear force $\bar{V}_{n}$ distributions are
assumed along the element edges \cite{Allman1976}. Folowing \cite{Allman1976,Aietal2025}, the twelve generalized forces required to completely specify the linear distribution of boundary tractions are expressed in terms of the displacement field in equation (\ref{Plate_Approximations}), resulting in}
\begin{align} 
    {\color{blue}\mathbf{Q}^{w}=\mathbf{B}^{\mathrm{T}} \boldsymbol{\alpha},}
    \label{Generalized_Forces_alpha_relation}
\end{align}

\noindent {\color{blue}where $\mathbf{Q}^{w}$ is the vector of the twelve generalized forces}
\begin{align} 
    {\color{blue}\mathbf{Q}^{w}=\left\{R_1, R_2, R_3, V_n^{12}, V_n^{23}, V_n^{31}, M_{nn}^{12}, M_{nn}^{21}, M_{nn}^{23}, M_{nn}^{32}, M_{nn}^{31}, M_{nn}^{13}\right\}^{\mathrm{T}}}
    \label{Generalized_Forces}
\end{align}

\noindent {\color{blue}in which the superscript $w$ was suppressed from its components for brevity, and the matrix $\mathbf{B}$ was carefully derived for symmetric laminates in \cite{Aietal2025}.}

\noindent {\color{blue}The 2nd, 3rd and 4th terms in Eq. (\ref{Modified_Plate_Energy_Functional}) denote the generalized forces work}
\begin{align} 
    {\color{blue}\sum_{N=1}^3 R^{w}_N \bar{w}_N+\int_{\partial A^{e}} V^{w}_n \bar{w} \mathrm{~d} s-\int_{\partial A^{e}} M^{w}_{nn} \frac{\partial \bar{w}}{\partial n} \mathrm{~d} s=\mathbf{q}^{\mathrm{T}}\mathbf{Q}^{w}.}
    \label{Generalized_Forces_Work}
\end{align}

\noindent {\color{blue}The generalized displacements $\mathbf{q}$, corresponding to the generalized forces in Eq. (\ref{Generalized_Forces}) are given in \cite{Allman1976}. A matrix $\mathbf{T}$ was derived in \cite{Allman1976} to relate the plate nodal DoFs} 
\begin{align} 
    {\color{blue}\overline{\mathbf{w}}=\left\{\bar{w}_1, \frac{\partial \bar{w}_1}{\partial x}, \frac{\partial \bar{w}_1}{\partial y}, \bar{w}_2, \frac{\partial \bar{w}_2}{\partial x}, \frac{\partial \bar{w}_3}{\partial y}, \bar{w}_3, \frac{\partial \bar{w}_3}{\partial x}, \frac{\partial \bar{w}_3}{\partial y}\right\}^{\mathrm{T}}}
    \label{plate_Dofs}
\end{align}
{\color{blue}to the vector $\mathbf{q}$ by means of: $\mathbf{q}=\mathbf{T} \overline{\mathbf{w}}$. The derivation of $\mathbf{T}$ assumes cubic approximation for $\bar{w}$ and linear variation for $\partial \bar{w} / \partial n$ along each side $i$-$j$. Substituting $\mathbf{q}=\mathbf{T} \overline{\mathbf{w}}$ and  Eqs. (\ref{Discrete_Strain_Energy})-(\ref{Generalized_Forces_Work}) into Eq. (\ref{Modified_Plate_Energy_Functional}), the plate energy functional for a finite element under prescribed boundary loads is}
\begin{align} 
    {\color{blue}\Psi^{e}_{\mathrm{p}}=-\frac{1}{2} \boldsymbol{\alpha}^{\mathrm{T}} \mathbf{H} \boldsymbol{\alpha}+\boldsymbol{\alpha}^{\mathrm{T}}(\mathbf{B T}) \overline{\mathbf{w}}-\mathbf{Q}^{* \mathrm{~T}} \mathbf{T} \overline{\mathbf{w}},}
    \label{Discrete_Plate_Energy_Functional}
\end{align}

\noindent {\color{blue}where the vector $\mathbf{Q}^*$ represents external forces obtained by replacing generalized forces in $\mathbf{Q}$ with the prescribed quantities. The first and second terms in Eq. (\ref{Discrete_Plate_Energy_Functional}) denote the internal work $ U^{e}$ of the plate element, whereas the third term corresponds to minus the external work $W^{e}$. The principle of stationary potential energy, applied to Eq. (\ref{Discrete_Plate_Energy_Functional}) by considering variations for $\delta \boldsymbol{\alpha}$ and $\delta \overline{\mathbf{w}}$, allows us to obtain the relation}
\begin{align} 
    {\color{blue}\boldsymbol{\alpha}=\mathbf{H}^{-1}(\mathbf{B}\mathbf{T}) \overline{\mathbf{w}}}
    \label{alpha_Dofs_Relation}
\end{align}
{\color{blue}by setting the coefficient of the arbitrary variation to zero \cite{Aietal2025}. Performing the variation of $U^{e}$, with $\boldsymbol{\alpha}$ substituted by Eq. (\ref{alpha_Dofs_Relation}), gives}
\begin{align} 
    {\color{blue}\delta U^{e}=\delta \overline{\mathbf{w}}^{\mathrm{T}} \mathbf{K}_{\mathrm{p}} \overline{\mathbf{w}}, \quad \Rightarrow  \quad \mathbf{K}_{\mathrm{p}}=(\mathbf{B T})^{\mathrm{T}} \mathbf{H}^{-1}(\mathbf{B}\mathbf{T}),}
    \label{Discrete_Internal_Work_Variation}
\end{align}

\noindent {\color{blue}where $\mathbf{K}_{\mathrm{p}}$ is the stiffness matrix of the plate element. The variation of the external work  $W^{e}$  gives the external force vector of the plate element in the absence of distributed pressure load}
\begin{align} 
    {\color{blue}\delta W^{e}=\delta \overline{\mathbf{w}}^{\mathrm{T}} \mathbf{T}^{\mathrm{T}} \mathbf{Q}^*=\delta \overline{\mathbf{w}}^{\mathrm{T}} \mathbf{f}_{\mathrm{p}}, \quad \Rightarrow \quad \mathbf{f}_{\mathrm{p}}=\mathbf{T}^{\mathrm{T}} \mathbf{Q}^*.}
    \label{Discrete_External_Work_Variation}
\end{align}

\subsection{{\color{blue}Shell element}}
\label{sec:shell_element_equation}
{\color{blue}For symmetric laminates, membrane–bending coupling is absent. Hence, the membrane and plate stiffness matrices can be assembled directly into the stiffness matrix of the flat shell element}
\begin{align}
    {\color{blue}\mathbf{K}_{\mathrm{s}}=
    \left[\begin{array}{cc}
        \mathbf{K}_{\mathrm{m}} & 0 \\
        0 & \mathbf{K}_{\mathrm{p}}
    \end{array}\right].}  
    \label{Shell_Stiffness_Matrix}
\end{align}

\noindent {\color{blue}The DoF vector of the shell element is simply $ \mathbf{q}_{\mathrm{s}}=\{\overline{\mathbf{u}}, \overline{\mathbf{w}}\}^{\mathrm{T}}$ and the force vector $\mathbf{f}_{\mathrm{s}}=\{\mathbf{f}_{\mathrm{m}},\mathbf{f}_{\mathrm{p}}\}^{\mathrm{T}}$. The element equations are first written in the local orthogonal coordinate system ($x,y,z$), defined from the triangular element geometry.  The local basis vectors $\mathbf{e}_{1}$, $\mathbf{e}_{2}$ and $\mathbf{e}_{3}$ are constructed from global nodal coordinates $\mathbf{x}_i$, $i=1,2,3$, and an orientation vector $\mathbf{v}$ lying on the triangle plane} 
\begin{align} \nonumber
    {\color{blue}\mathbf{e}_{3}=\frac{\mathbf{x}_{12} \times \mathbf{x}_{13}}{ \left| \left| \mathbf{x}_{12} \times \mathbf{x}_{13} \right| \right|}, \quad
    \mathbf{e}_{1} = \frac{\mathbf{v}-(\mathbf{v} \cdot \mathbf{e}_{3})\mathbf{e}_{3}}{\left| \left|\mathbf{v}-(\mathbf{v} \cdot \mathbf{e}_{3})\mathbf{e}_{3}\right| \right|}, \quad
    \mathbf{e}_{2} = \mathbf{e}_{3} \times \mathbf{e}_{1},}
\end{align}
\noindent {\color{blue}where $\mathbf{x}_{ij}=\mathbf{x}_{j}-\mathbf{x}_{i}$. For the DoFs of a node $i$: $\mathbf{q}^{l}_{i} =\left[\bar{u}_i, \bar{v}_i, \bar{w}_i, \frac{\partial \bar{w}_i}{\partial x}, \frac{\partial \bar{w}_i}{\partial y}, \bar{\theta}_{z_i}\right]^{\mathrm{T}}$, written in the local system, the following transformations hold}
\begin{align} \nonumber
     {\color{blue}\mathbf{q}^{l}_{i}=\mathbf{L}_{q} \mathbf{q}^{g}_{i},  \quad \quad \quad \mathbf{q}^{g}_{i}=\mathbf{L}^{\mathrm{T}}_{q} \mathbf{q}^{l}_{i}, \quad \quad \quad \mathbf{L}_{q}=\operatorname{diag}
     \left[\mathbf{L}_{l}, \quad \mathbf{L}_{w\theta} \right]}
\end{align}
\noindent {\color{blue}where $\mathbf{q}^{g}_{i}=\left[\bar{u}_{X_i}, \bar{u}_{Y_i}, \bar{u}_{Z_i}, \bar{\theta}_{X_i}, \bar{\theta}_{Y_i}, \bar{\theta}_{Z_i}\right]^{\mathrm{T}}$ are the nodal DoFs written with respect to the global system and $\mathbf{L}_{l}=\left[\mathbf{e}^{\mathrm{T}}_{1}, \mathbf{e}^{\mathrm{T}}_{2}, \mathbf{e}^{\mathrm{T}}_{3}\right]^{\mathrm{T}}$ is a orthogonal transformation. The transformation $\mathbf{L}_{w\theta}$ takes care of writing the rotational DoFs in the local system, and accounts for $\partial \bar{w}_i/ \partial x=-\bar{\theta}_{y_i}$ and $\partial \bar{w}_i/ \partial y = \bar{\theta}_{x_i}$. A new ordering $ \mathbf{q}^{l}_{\text {shell}} =\left[\mathbf{q}^{l}_{1}, \mathbf{q}^{l}_{2},\mathbf{q}^{l}_{3}\right]^{\mathrm{T}}$ is defined for the DoFs of the local shell element. For the new DoF vector, the following transformations hold}
\begin{align} \nonumber
     {\color{blue}\mathbf{q}^{l}_{\mathrm{s}}=\mathbf{R} \mathbf{q}^{g}_{\mathrm{s}},  \quad \quad \quad \mathbf{q}^{g}_{\mathrm{s}}=\mathbf{R}^{\mathrm{T}} \mathbf{q}^{l}_{\mathrm{s}}, \quad \quad \quad \mathbf{R}=\operatorname{diag}
     \left[\mathbf{L}_{q}, \quad \mathbf{L}_{q}, \quad \mathbf{L}_{q} \right]}
\end{align}
\noindent {\color{blue}where $\mathbf{q}^{g}_{\mathrm{s}}$ is the global DoF vector. Element stiffness matrix and element force vector are then transformed from local to global frame as $\mathbf{K}^{g}_{\mathrm{s }}=\mathbf{R}^{\mathrm{T}}\mathbf{K}^{l}_{\mathrm{s}} \mathbf{R}$ and $\mathbf{f}^{g}_{\mathrm{s}}=\mathbf{R}^{\mathrm{T}} \mathbf{f}^{l}_{\mathrm{s}}$.}


\section{The shell-to-shell cohesive line interface model}
\label{sec:shell-to-shell CE}

\subsection{Kinematics: Displacement jump and DoFs}

\begin{figure}[H]
\centering
\scalebox{0.40}{\includegraphics{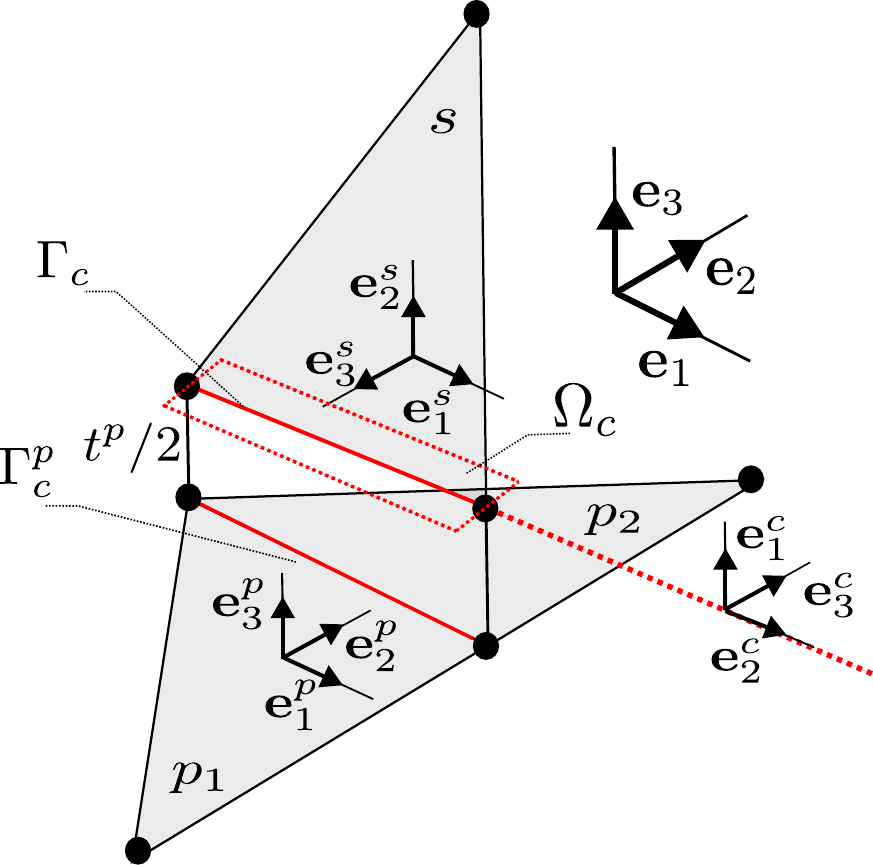}}
	\caption{The shell-to-shell cohesive line element.}
	\label{fig:ShellShellCE}
\end{figure}

The shell-to-shell cohesive line element must be kinematically compatible with two shell elements in the panel ($p_1$ and $p_2$) and the stiffener element ($s$), as illustrated in Figure \ref{fig:ShellShellCE}.  The panel and the stiffener have constant thickness $t^{p}$ and $t^{s}$, respectively. The displacement jump over the cohesive surface $\Omega_c=\left[-t^{s}/2,t^{s}/2\right] \times \Gamma_c$ is computed from the displacement jump over its central line $\Gamma_c$, in addition to a contribution from the jump in rotations over $\Gamma_c$ 

\begin{equation}
    \boldsymbol{\Delta}\left(\mathbf{x}\right)= \llbracket \mathbf{u}_{\Gamma_c} \rrbracket + z^{c} \llbracket \boldsymbol{\theta}_{\Gamma_c} \rrbracket, \hspace{40pt} \mathbf{x} \in \Omega_c
    \label{JumpDecomposition}
\end{equation}

\noindent where $\boldsymbol{\Delta}$ is the displacement jump vector, written with respect to the cohesive local system $\mathbf{e}_{i}^{c}$, and $\llbracket \mathbf{u}_{\Gamma_c} \rrbracket$, $\llbracket \boldsymbol{\theta}_{\Gamma_c} \rrbracket$ are, respectively, displacement and rotation jumps over the central line $\Gamma_c$, which are defined as
\begin{equation}
    \llbracket \mathbf{u}_{\Gamma_c} \rrbracket=\mathbf{u}^{c+}_{\Gamma_c}-\mathbf{u}^{c-}_{\Gamma_c}, \hspace{20pt} \llbracket \boldsymbol{\theta}_{\Gamma_c} \rrbracket=\boldsymbol{\theta}^{c+}_{\Gamma_c}-\boldsymbol{\theta}^{c-}_{\Gamma_c}.
    \label{DispRotJumps}
\end{equation}

 \noindent The central line displacements $\mathbf{u}^{c+}_{\Gamma_c}$ and $\mathbf{u}^{c-}_{\Gamma_c}$ are given by
\begin{align}    
    \mathbf{u}^{c+}_{\Gamma_c} = \mathbf{L}_{cs}\mathbf{u}^{s}_{\Gamma_c}, \quad \mathbf{u}^{c-}_{\Gamma_c} = \mathbf{L}_{cp}\mathbf{u}^{p}_{\Gamma_c}
    \label{CentralDisplacementDefinition}
\end{align}
where the transformations
\begin{equation}
    \mathbf{L}_{c p} = \left[\begin{array}{lll}
    \mathbf{e}_{1}^{c} \cdot \mathbf{e}_{1}^{p} & \mathbf{e}_{1}^{c} \cdot \mathbf{e}_{2}^{p} & \mathbf{e}_{1}^{c} \cdot \mathbf{e}_{3}^{p}\\
    \mathbf{e}_{2}^{c} \cdot \mathbf{e}_{1}^{p} & \mathbf{e}_{2}^{c} \cdot \mathbf{e}_{2}^{p} & \mathbf{e}_{2}^{c} \cdot \mathbf{e}_{3}^{p}\\
    \mathbf{e}_{3}^{c} \cdot \mathbf{e}_{1}^{p} & \mathbf{e}_{3}^{c} \cdot \mathbf{e}_{2}^{p} & \mathbf{e}_{3}^{c} \cdot \mathbf{e}_{3}^{p}
    \end{array}\right], \hspace{5pt}  \mathbf{L}_{c s} = \left[\begin{array}{lll}
    \mathbf{e}_{1}^{c} \cdot \mathbf{e}_{1}^{s} & \mathbf{e}_{1}^{c} \cdot \mathbf{e}_{2}^{s} & \mathbf{e}_{1}^{c} \cdot \mathbf{e}_{3}^{s}\\
    \mathbf{e}_{2}^{c} \cdot \mathbf{e}_{1}^{s} & \mathbf{e}_{2}^{c} \cdot \mathbf{e}_{2}^{s} & \mathbf{e}_{2}^{c} \cdot \mathbf{e}_{3}^{s}\\
    \mathbf{e}_{3}^{c} \cdot \mathbf{e}_{1}^{s} & \mathbf{e}_{3}^{c} \cdot \mathbf{e}_{2}^{s} & \mathbf{e}_{3}^{c} \cdot \mathbf{e}_{3}^{s}
    \end{array}\right] 
    \label{TransformationMatrices}
\end{equation}
are computed from the basis vectors: $\mathbf{e}_{i}^{p}$, $\mathbf{e}_{i}^{s}$ and $\mathbf{e}_{i}^{c}$ in Fig. \ref{fig:ShellShellCE}. The stiffener and panel displacement fields along $\Gamma_c$ are defined based on the classical Kirchhoff plate kinematics
\begin{align}
        & \mathbf{u}^{s}_{\Gamma_c}=\left\{\begin{array}{c}
        u^{s}|_{\Gamma_c} \\
        v^{s}|_{\Gamma_c} \\
        w^{s}|_{\Gamma_c}
        \end{array}\right\}, \hspace{20pt} \mathbf{u}^{p}_{\Gamma_c}=\left\{\begin{array}{c}
        u^{p}|_{\Gamma_{c}^{p}} \\
        v^{p}|_{\Gamma_{c}^{p}} \\
        w^{p}|_{\Gamma_{c}^{p}}
        \end{array}\right\}+\frac{t^{p}}{2}\left\{\begin{array}{c}
        -w_{\color{blue},x}^{p}|_{\Gamma_{c}^{p}} \\
        -w_{\color{blue},y}^{p}|_{\Gamma_{c}^{p}} \\
        0
        \end{array}\right\}, 
\end{align}
in which $(\cdot)|_{\Gamma}$ stands for the internal fields evaluated at the edge $\Gamma$, and the notation $f_{\color{blue},x} =  \partial f/\partial x$, $f_{\color{blue},y} = \partial f/\partial y$ is adopted herein for derivatives. 

The central line rotations $\boldsymbol{\theta}^{c+}_{\Gamma_c}$ and $\boldsymbol{\theta}^{c-}_{\Gamma_c}$ which cause motion out-of the stiffener plane are 
\begin{align} \label{RotationDefinition_s}
        \boldsymbol{\theta}^{c+}_{\Gamma_c} &={\color{blue}\boldsymbol{\Theta}^{c+}_{\Gamma_c} \mathbf{e}_{3}} =\left\{\begin{array}{c}
        \phantom{-}\theta_x^{s}|_{\Gamma_c}  \\
        -\theta_y^{s}|_{\Gamma_c}  \\
        0
        \end{array}\right\}=\left\{\begin{array}{c}
        w_{\color{blue},y}^{s}|_{\Gamma_c}  \\
        w_{\color{blue},x}^{s}|_{\Gamma_c}  \\
        0
        \end{array}\right\}, \\ \label{RotationDefinition_p}
         \boldsymbol{\theta}^{c-}_{\Gamma_c}&= {\color{blue}\boldsymbol{\Theta}^{c-}_{\Gamma_c} \mathbf{e}_{3}} =\left\{\begin{array}{c}
        \phantom{-}\theta_x^{p}|_{\Gamma_{c}} \\
        -\theta_z^{p}|_{\Gamma_{c}} \\
        0
        \end{array}\right\}=\left\{\begin{array}{c}
        w_{\color{blue},y}^{p}|_{\Gamma_{c}^{p}} \\
        -\frac{1}{2}\left(v_{\color{blue},x}^{p}-u_{\color{blue},y}^{p}\right)|_{\Gamma_{c}^{p}} \\
        0
        \end{array}\right\}.
\end{align}
where the skew-symmetric small rotation tensors (or spin tensors) ${\color{blue}\boldsymbol{\Theta}^{c+}_{\Gamma_c}}=\mathbf{L}_{cs}{\color{blue}\boldsymbol{\Theta}^{s}_{\Gamma_c}}\mathbf{L}^{T}_{cs}$ and ${\color{blue}\boldsymbol{\Theta}^{c-}_{\Gamma_c}}=\mathbf{L}_{cp}{\color{blue}\boldsymbol{\Theta}^{p}_{\Gamma_c}}\mathbf{L}^{T}_{cp}$ are computed with the transformation matrices (\ref{TransformationMatrices}), defined for the basis vectors illustrated in Fig. \ref{fig:ShellShellCE}. {\color{blue}Notice that $\theta_x=w_{,y}$ and $\theta_z=-1/2\left[v_{,x}-zw_{,yx}-\left(u_{,y}-zw_{,xy}\right)\right]$ in Eq.(\ref{RotationDefinition_p}) do not depend on the local coordinate $z$.}

Considering Eqs. (\ref{DispRotJumps})-(\ref{CentralDisplacementDefinition}) and (\ref{RotationDefinition_s})-(\ref{RotationDefinition_p}), the central line  displacement and rotation jumps are given by
\begin{align}  \label{CentralLineJump}
    &\llbracket \mathbf{u}_{\Gamma_c} \rrbracket =\left\{\begin{array}{c}
        v^{s}|_{\Gamma_c} - w^{p}|_{\Gamma_c^{p}} \\[4pt]
        u^{s}|_{\Gamma_c} - u^{p}|_{\Gamma_c^{p}}+\frac{t^{p}}{2}w_{\color{blue},x}^{p}|_{\Gamma_c^{p}} \\[4pt]
        -w^{s}|_{\Gamma_c}-v^{p}|_{\Gamma_c^{p}}+\frac{t^{p}}{2}w_{\color{blue},y}^{p}|_{\Gamma_c^{p}}
        \end{array}\right\},\\
        &\llbracket \boldsymbol{\theta}_{\Gamma_c} \rrbracket =\left\{\begin{array}{c}
        w_{\color{blue},y}^{s}|_{\Gamma_c} - w_{\color{blue},y}^{p}|_{\Gamma_c} \\[4pt]
        w_{\color{blue},x}^{s}|_{\Gamma_c} + \frac{1}{2}\left(v_{\color{blue},x}^{p}-u_{\color{blue},y}^{p}\right)|_{\Gamma_{c}^{p}} \\[4pt]
        0
        \end{array}\right\}.
        \label{CentralLineRotJump}
\end{align}

\noindent Due to the incompatible internal fields of the adopted shell element, the panel fields $u^{p}$, $v^{p}$, $v_{\color{blue},x}^{p}$, $u_{\color{blue},y}^{p}$, $w^{p}$, $w_{\color{blue},x}^{p}$, $w_{\color{blue},y}^{p}$ are herein defined as the average of the internal fields from panels 1 and 2, evaluated at their common edge. Notice that the local frame of reference of panel $p_2$ corresponds to the local frame of $p_1=p$, with its origin translated to the centroid of $p_2$. 

Evaluating the internal fields of the shell element along its edges requires the contribution of the DOFs associated with all its nodes. Therefore, the DoF vector of the cohesive element is defined as: $\mathbf{q}_{\mathrm{c}}=\left[\overline{\mathbf{u}}_{p_1}^{\mathrm{T}}, \overline{\mathbf{w}}_{p_1}^{\mathrm{T}},
\overline{\mathbf{u}}_{p_2}^{\mathrm{T}},
\overline{\mathbf{w}}_{p_2}^{\mathrm{T}},
\overline{\mathbf{u}}_{s}^{\mathrm{T}}, \overline{\mathbf{w}}_{s}^{\mathrm{T}}\right]^{\mathrm{T}}$, where $\overline{\mathbf{u}}_{\alpha}$ and $\overline{\mathbf{w}}_{\alpha}$, $\alpha = s, \hspace{2pt} p_1, \hspace{2pt} p_2$, are the membrane and plate DoF vectors for the first panel, second panel and stiffener elements, respectively.

 \begin{remark*}
    The vector $\mathbf{q}_{\mathrm{c}}$ contains repeated Dofs, whose contributions can be properly handled during the assembly of the global system of equations.
\end{remark*}

The central task of cohesive element formulations is to find the {\color{blue}kinematic operator} $\mathbf{B}_{\mathrm{c}}$ that relates the opening vector to the nodal DoFs
\begin{equation}
    \boldsymbol{\Delta}\left(\mathbf{x}\right)=\left[\Delta_{\mathrm{I}}, \Delta_{\mathrm{II}}, \Delta_{\mathrm{III}}\right]^{\mathrm{T}}=\mathbf{B}_{\mathrm{c}} \mathbf{q}_{\mathrm{c}}, \quad \mathbf{x} \in \Omega_c . 
    \label{Jump_Dofs_relation_I}
\end{equation}

 \noindent Examining the expression of $\boldsymbol{\Delta}$ in Eq. (\ref{JumpDecomposition}), it is possible to write

 \begin{equation}
    \boldsymbol{\Delta}=\left[\mathbf{B}_{\mathrm{c_d}} + z^{c} \mathbf{B}_{\mathrm{c_\theta}}\right]\mathbf{q}_\mathrm{c} = \mathbf{B}_{\mathrm{c}}\mathbf{q}_{\mathrm{c}}
    \label{Jump_Dofs_relation_II}
\end{equation}

 \noindent in which the operators $\mathbf{B}_{\mathrm{c_d}}$ and $\mathbf{B}_{\mathrm{c_\theta}}$ are defined from the shell internal fields, evaluated at $\Gamma_c$ and $\Gamma_c^{p}$. Therefore, they are functions solely of the parametric coordinates $s^{s} \in \Gamma_c$ and $s^{p} \in \Gamma_c^{p}$. From the expressions of $\llbracket \mathbf{u}_{\Gamma_c} \rrbracket$ and $\llbracket \boldsymbol{\theta}_{\Gamma_c} \rrbracket$ in Eqs. (\ref{CentralLineJump})-(\ref{CentralLineRotJump}), we can see that the $\mathbf{B}_{\mathrm{c_d}}$ and $\mathbf{B}_{\mathrm{c_\theta}}$ matrices shall be composed of sub-matrices that relate the approximation terms
\begin{align}  \nonumber
    &\tilde{u}^{s}, \tilde{v}^{s}, \tilde{w}^{s}, \tilde{u}^{p_1}, \tilde{v}^{p_1}, \tilde{w}^{p_1}, \tilde{w}_{\color{blue},x}^{p_1}, \tilde{w}_{\color{blue},y}^{p_1}, \tilde{u}^{p_2}, \tilde{v}^{p_2}, \tilde{w}^{p_2}, \tilde{w}_{\color{blue},x}^{p_2}, \tilde{w}_{\color{blue},y}^{p_2}, \\ \nonumber
    & \tilde{w}_{\color{blue},x}^{s}, \tilde{w}_{\color{blue},y}^{s}, \tilde{w}_{\color{blue},y}^{p_1}, \tilde{w}_{\color{blue},y}^{p_2}, \tilde{v}_{\color{blue},x}^{p_1}, \tilde{v}_{\color{blue},x}^{p_2}, \tilde{u}_{\color{blue},y}^{p_1}, \tilde{u}_{\color{blue},y}^{p_2}. 
\end{align}
to the nodal DoFs in  $\mathbf{q}_{\mathrm{c}}$.

The in-plane displacements $\tilde{u}^{\alpha}$, $\tilde{v}^{\alpha}$  are those from Eq (\ref{Displacement_decomposition}). These approximations, and their derivatives, can be expressed in terms of the membrane nodal Dofs $\overline{\mathbf{u}}^{\alpha}$. The out-of-plane displacements $\tilde{w}^{\alpha}$ and the rotations $\tilde{w}_{\color{blue},x}^{\alpha}$ and $\tilde{w}_{\color{blue},y}^{\alpha}$ are those derived from the cubic $\tilde{w}^{\alpha}$ approximation in Eq. (\ref{Plate_Approximations}), which were explicitly expressed in terms of the plate nodal Dofs $\overline{\mathbf{w}}^{\alpha}$ in \cite{Aietal2025}. All these approximations and the corresponding shape function matrices $\mathbf{N}_{u}^{\alpha}$, $\mathbf{N}_{v}^{\alpha}$, $\mathbf{N}_{v_{\color{blue},x}}^{\alpha}$, $\mathbf{N}_{u_{\color{blue},y}}^{\alpha}$, $\mathbf{N}_{w}^{\alpha}$, $\mathbf{N}_{w_{\color{blue},x}}^{\alpha}$, $\mathbf{N}_{w_{\color{blue},y}}^{\alpha}$  are summarized in \ref{app:shapefunctions}.

\subsection{\texorpdfstring{The $\mathbf{B}_{\mathrm{c}}$ matrix}{The B-CE matrix}}

Once the shape functions have been explicitly defined in \ref{app:shapefunctions}, the next step is to assemble the $\mathbf{B}_{\mathrm{c}}$ matrix that relates the DoF vector $\mathbf{q}_{\mathrm{c}}$ to the opening vector $\boldsymbol{\Delta}$
 \begin{equation}
    \mathbf{B}_{\mathrm{c}}=\mathbf{B}_{\mathrm{c_d}}\left(s,s^{p}\right) + z^{c} \mathbf{B}_{\mathrm{c_\theta}}\left(s,s^{p}\right),
    \label{BCE_Matrix}
\end{equation}
with $s \in \Gamma_c$, $s^{p} \in \Gamma_c^{p}$ and $z^{c} \in \left[-t^{s}/2,t^{s}/2\right]$. 

Substituting the approximations in Eqs. (\ref{Membrane_displacements_Dofs_II})-(\ref{Plate_fields_Dofs_III}) to express the $u$, $v$, $v_{\color{blue},x}$, $u_{\color{blue},y}$, $w$, $w_{\color{blue},x}$ and $w_{\color{blue},y}$ terms in Eqs. (\ref{CentralLineJump})-(\ref{CentralLineRotJump}), the expressions for $\mathbf{B}_{\mathrm{c}_{d}}$ and $\mathbf{B}_{\mathrm{c}_{\theta}}$ results
\begin{align} \nonumber
     \mathbf{B}_{\mathrm{c}_{d}}=
     \left[\begin{array}{cccccc}
    \mathbf{0} & -\frac{1}{2}\mathbf{N}_{w}^{p_1}|_{\Gamma_c^{p}} & \mathbf{0} & -\frac{1}{2}\mathbf{N}_{w}^{p_2}|_{\Gamma_c^{p}} & \mathbf{N}_{v}^{s}|_{\Gamma_c} & \mathbf{0}\\[4pt]
    -\frac{1}{2}\mathbf{N}_{u}^{p_1}|_{\Gamma_c^{p}} & \frac{t^{p}}{4}\mathbf{N}_{w_{\color{blue},x}}^{p_1}|_{\Gamma_c^{p}} &-\frac{1}{2}\mathbf{N}_{u}^{p_2}|_{\Gamma_c^{p}} & \frac{t^{p}}{4}\mathbf{N}_{w_{\color{blue},x}}^{p_2}|_{\Gamma_c^{p}} & \mathbf{N}_{u}^{s}|_{\Gamma_c} & \mathbf{0}\\[4pt]
    -\frac{1}{2}\mathbf{N}_{v}^{p_1}|_{\Gamma_c^{p}} & \frac{t^{p}}{4}\mathbf{N}_{w_{\color{blue},y}}^{p_1}|_{\Gamma_c^{p}} &-\frac{1}{2}\mathbf{N}_{v}^{p_2}|_{\Gamma_c^{p}} & \frac{t^{p}}{4}\mathbf{N}_{w_{\color{blue},y}}^{p_2}|_{\Gamma_c^{p}} & \mathbf{0} & -\mathbf{N}_{w}^{s}|_{\Gamma_c}
    \end{array}\right]
\end{align}
and 
\begin{align} \nonumber
  \mathbf{B}_{\mathrm{c}_{\theta}}=\left[\begin{array}{cccccc}
    \mathbf{0} & -\frac{1}{2}\mathbf{N}_{w_{\color{blue},y}}^{p_1}|_{\Gamma_c^{p}} & \mathbf{0} & -\frac{1}{2}\mathbf{N}_{w_{\color{blue},y}}^{p_2}|_{\Gamma_c^{p}} & \mathbf{0} & \mathbf{N}_{w_{\color{blue},y}}^{s}|_{\Gamma_c}\\
     \frac{1}{2}\mathbf{N}_{\theta_{z}}^{p_1}|_{\Gamma_c^{p}} & \mathbf{0} & \frac{1}{2}\mathbf{N}_{\theta_{z}}^{p_2}|_{\Gamma_c^{p}} & \mathbf{0} & \mathbf{0} & \mathbf{N}_{w_{\color{blue},x}}^{s}|_{\Gamma_c}\\
    \mathbf{0} & \mathbf{0} & \mathbf{0} & \mathbf{0} & \mathbf{0} & \mathbf{0}
\end{array}\right],
\end{align}
in which $\mathbf{N}_{\theta_{z}}^{p_i}=1/2\left(\mathbf{N}_{v_{\color{blue},x}}^{p_i}-\mathbf{N}_{u_{\color{blue},y}}^{p_i}\right)$, $i=1,2$, and $\mathbf{N}|_{\Gamma}$ stands for the shape function matrix $\mathbf{N}$ evaluated at the edge $\Gamma$. 


\subsection{Turon mixed-mode damage cohesive model}
{\color{blue}The mixed-mode damage model used in the present work is the well-known Turon model \cite{Turonetal2006}. The relationship between cohesive traction $\mathbf{t}_{\mathrm{c}}=\left[\mathrm{t}_{\mathrm{I}}, \mathrm{t}_{\mathrm{II}}, \mathrm{t}_{\mathrm{III}}\right]^{\mathrm{T}}$ and the opening vector $\boldsymbol{\Delta}=\left[\Delta_{\mathrm{I}}, \Delta_{\mathrm{II}}, \Delta_{\mathrm{III}}\right]^{\mathrm{T}}$ are given by}

\vspace{-5pt}

\begin{align} 
    {\color{blue}\left\{\begin{array}{c}
        \mathrm{t}_{\mathrm{I}}  \\
        \mathrm{t}_{\mathrm{II}}  \\
        \mathrm{t}_{\mathrm{III}}
        \end{array}\right\}=
    \left[\begin{array}{ccc}
        \left(1-d_{\mathrm{I}}\right) K & 0 & 0 \\
        0 & (1-d) K & 0 \\
        0 & 0 & (1-d) K
    \end{array}\right]\left\{\begin{array}{c}
        \Delta_{\mathrm{I}}  \\
        \Delta_{\mathrm{II}}  \\
        \Delta_{\mathrm{III}}
        \end{array}\right\}}
    \label{Cohesive_Model}
\end{align}

\noindent where $K$ and $d$ are the penalty stiffness and the damage variable of the cohesive model, respectively. The damage variable $d$ in this work is updated by the bi-linear cohesive law proposed by Turon et al. \cite{Turonetal2006,Turonetal2010}. The damage variable $d_{\mathrm{I}}$ under Mode I loading is distinguished from $d$ to avoid interpenetration of the top and bottom surfaces under compression
\begin{align} 
    d_{\mathrm{I}}= \begin{cases}d, & \Delta_{\mathrm{I}} \geq 0 \\ 0, & \Delta_{\mathrm{I}}<0\end{cases}.
    \label{Interpenetration_Condition}
\end{align}
{\color{blue}The parameters required by Turon’s model are the fracture energies $G_{\mathrm{I}c}$, $G_{\mathrm{II}c}$, the interface strengths $\tau_{\mathrm{I}c}$, $\tau_{\mathrm{II}c}$, the penalty stiffness $K$ and the exponent $\eta$ from the phenomenological relation between fracture toughness and mode ratio proposed by Benzeggagh and Kenane \cite{BenzeggaghandKenane1996}, from which the critical energy release rate $G_{\mathrm{c}}$ for a mixed-mode ratio is given by}
\begin{align} 
    {\color{blue}G_{\mathrm{c}}=G_{\mathrm{Ic}}+\left(G_{\mathrm{IIc}}-G_{\mathrm{Ic}}\right)\left(\frac{G_{\text {shear }}}{G_{\mathrm{T}}}\right)^\eta}
    \label{BenzeggaghandKenaneRelation}
\end{align}

\noindent {\color{blue}where $G_{\text {shear }}=G_{\mathrm{II}}+G_{\mathrm{III}}$ and $G_{\mathrm{T}}=G_{\mathrm{I}}+G_{\mathrm{II}}+G_{\mathrm{III}}$.}

{\color{blue}The proper choice of the penalty stiffness $K$ is a key factor in cohesive models. The closed-form expression} 
\begin{align} 
    {\color{blue}K=\alpha_{p} \frac{E_3}{t}}
    \label{Penalty_Stiffness}
\end{align}
\noindent {\color{blue}was developed for delamination problems in \cite{Turonetal2007} from mechanical considerations, where $E_3$ is the out-of-plane Young's modulus of the plies, $t$ is the ply thickness and $\alpha_{p}$ is a constant. For values of $\alpha_{p}$ equal or greater than 50, the loss of stiffness due to the presence of the interface is less than 2\%, which is sufficiently accurate for most problems. In the absence of a closed-form expression for the stiffness of the skin-stiffener interface, the same expression is adopted herein by considering $E_3$ and $t$ to be the out-of-plane Young’s modulus and the thickness of the panel. The study of different values of $\alpha_{p}$ on the structural response is reported in the results section.} 

\subsection{Cohesive-interface element equation}
\label{subsec:Cohesive element equation}

The variational formulation of the stiffened panel model that includes the shell-to-shell cohesive line interface model may be obtained by adding the cohesive energy variation

\begin{align} 
    \centering
    \delta \Psi_{\mathrm{c}} &= \int_{\Omega_{c}} \delta {\boldsymbol{\Delta}}^{\mathrm{T}} \mathbf{t}_{\mathrm{c}} \mathrm{~d} \Omega= \int_{\Gamma_{c}} \int_{-\frac{t^{s}}{2}}^{\frac{t^{s}}{2}} \delta {\boldsymbol{\Delta}}^{\mathrm{T}} \mathbf{t}_{\mathrm{c}}  \mathrm{~d} z^{c} \mathrm{~d} \Gamma
\end{align}
\noindent into the weak form of the equilibrium equations. Introducing the cohesive model (\ref{Cohesive_Model}) and the approximations (\ref{Jump_Dofs_relation_II}) into the cohesive virtual work, and considering the arbitrary nature of the virtual DoFs $\delta \mathbf{q}_{\mathrm{c}}$, the element equations result
\begin{align}
    \centering
    \delta \Psi_{\mathrm{c}} = \delta \mathbf{q}_{\mathrm{c}}^{\mathrm{T}}\mathbf{f}_{\mathrm{c}} =0 \quad \Leftrightarrow \quad \mathbf{f}_{\mathrm{c}}=\mathbf{0}
\end{align}
\noindent in which
\begin{align} 
    \centering
   \mathbf{f}_{\mathrm{c}} =  \int_{\Gamma_{c}} \int_{-\frac{t^{s}}{2}}^{\frac{t^{s}}{2}} \mathbf{B}_{\mathrm{c}}^{\mathrm{T}} \mathbf{t}_{\mathrm{c}}  \mathrm{~d} z^{c}\mathrm{~d}\Gamma
       \label{f_CE}
\end{align}

\noindent is the internal force vector of the cohesive element. The consistent tangent stiffness matrix of the proposed shell-to-shell cohesive line element results
\begin{align}
    \centering
    \mathbf{K}_{\mathrm{c}} &=\frac{\partial \mathbf{f}_{\mathrm{c}}}{\partial \mathbf{q}_{\mathrm{c}}}=\int_{\Gamma_{c}} \int_{-\frac{t^{s}}{2}}^{\frac{t^{s}}{2}} \mathbf{B}_{\mathrm{c}}^{\mathrm{T}} {\color{blue}\mathbf{D}_{\mathrm{c}}}\mathbf{B}_{\mathrm{c}} \mathrm{~d} z^{c} \mathrm{~d} \Gamma,
    \label{K_CE}
\end{align}

\noindent where ${\color{blue}\mathbf{D}_{\mathrm{c}}}=\partial \mathbf{t}_{\mathrm{c}}/ \partial \boldsymbol{\Delta}$ is the consistent tangent matrix of Turon's model, as derived in \cite{vanderMeerandSluys2010}. {\color{blue}The consistency of the variational formulation is preserved by using the shell approximation fields in the definition of $\Delta$.}

The integrals in Eqs. (\ref{f_CE}) and (\ref{K_CE}) are {\color{blue}difficult to evaluate analytically}, mainly due to the complex nonlinear behavior of the {\color{blue}cohesive traction vector $\mathbf{t}_{\mathrm{c}}$ and tangent matrix $\mathbf{D}_{\mathrm{c}}$} after damage initiation. Thus, {\color{blue}numerical integration is applied with a standard Gaussian integration scheme.} Earlier works have shown that using a higher number of
quadrature points improves the solution accuracy and smoothness in delamination simulations \cite{RussoandChen2020, Álvarezetal2014, BalducciandChen2024, Aietal2025}. Accordingly, numerical integration is performed herein using a 13-point Gauss–Legendre quadrature along $\Gamma_c$ and a 5-point Gauss–Legendre quadrature along $z^{c} \in \left[-t^{s}/2,t^{s}/2\right]$.

The ordering of shell DoFs defined in Section \ref{sec:shell_element_equation} is also adopted for the DoFs of the panels and stiffener elements: \noindent $\mathbf{q}^{l}_{\text {p}_1}$, $\mathbf{q}^{l}_{\text {p}_2}$, $\mathbf{q}^{l}_{\text {s}}$. Therefore
\begin{align} \nonumber
     \mathbf{q}_{\mathrm{c}}^{l\mathrm{T}}=\left[\mathbf{q}_{\text {p}_1}^{l\mathrm{T}}, \mathbf{q}_{\text {p}_2}^{l\mathrm{T}},
        \mathbf{q}_{\text {s}}^{l\mathrm{T}}\right]^{\mathrm{T}}.
\end{align}
        
For this DoF vector, the following transformations hold
\begin{align} \nonumber
     \mathbf{q}^{l}_{\mathrm{c}}=\mathbf{R}_{\mathrm{c}} \mathbf{q}^{g}_{\mathrm{c}},  \quad \quad  \mathbf{q}^{g}_{\mathrm{c}}=\mathbf{R}_{\mathrm{c}}^{T} \mathbf{q}^{l}_{\mathrm{c}}, \quad \quad \mathbf{R}_{\mathrm{c}}=\left[\begin{array}{lll}
        \mathbf{R}_{\text{p}_1} & \mathbf{0} & \mathbf{0}\\
        \mathbf{0} & \mathbf{R}_{\text{p}_2} & \mathbf{0}\\
        \mathbf{0} & \mathbf{0} & \mathbf{R}_{\text{s}}
        \end{array}\right],
\end{align}

\noindent in which $\mathbf{q}^{g}_{\mathrm{c}}$ is the DoF vector of the cohesive element in the global coordinate system, and $\mathbf{R}_{\text{p}_1}$, $\mathbf{R}_{\text{p}_2}$, $\mathbf{R}_{\text{s}}$ corresponds to the previously defined transformation $\mathbf{R}$, defined for the first panel, second panel and stiffener frames of reference. The internal force vector and tangent stiffness matrix of the cohesive element, rewritten with respect to the new DoF ordering, are denoted as: $\mathbf{f}^{l}_{\mathrm{c}}$ and $\mathbf{K}^{l}_{\mathrm{c}}$. Applying the orthogonal transformation $\mathbf{R}_{\mathrm{c}}$, it is possible to obtain the element force vector and tangent stiffness matrix written for the global coordinate system
\begin{align}
    \mathbf{f}^{g}_{\mathrm{c}} = \mathbf{R}_{\mathrm{c}}^{T} \mathbf{f}^{l}_{\mathrm {c}} \text{,} \quad \quad \mathbf{K}^{g}_{\mathrm{c}} = \mathbf{R}_{\mathrm{c}}^{T}\mathbf{K}^{l}_{\mathrm{c}} \mathbf{R}_{\mathrm{c}}.
    \label{Global_Cohesive_System}
\end{align}


\section{Results}
\label{sec:results}

The proposed structural cohesive element (CE) is first verified using three classical unidirectional laminate benchmarks. A Mode I skin–stiffener debonding test is then used to assess the performance of the structural CEs relative to 3D solid modeling using standard CEs. Finally, a complex stiffened panel under three-point bending is analyzed to demonstrate the computational efficiency of the proposed approach. The analyses were performed in Jive, an open source C++ FEM library \cite{Nguyen-Thanhetal2020}. The CEs followed the Turon model \cite{Turonetal2006}, with the improved linearization proposed in \cite{vanderMeerandSluys2010}. All simulations were performed using a flexible path-following method that adopts both displacement control and energy-based arc-length control schemes \cite{Verhooseletal2009, vanderMeer2012}. 

\subsection{Model verification with benchmarks}

The shell-to-shell cohesive line model was first verified on three classical unidirectional laminate benchmarks, namely the double cantilever beam (DCB), the end-notched flexure (ENF), and the mixed-mode bending (MMB) tests. Although these benchmarks are commonly used for standard delamination model validation, {\color{blue}the setup of the shell-to-shell cohesive line models may also be applied} to reproduce these tests, as discussed in Section \ref{sec:BenchmarksDescription}. The structural CE analyses were conducted using the Free Formulation (FF) membrane element from Section \ref{sec:K-L shell element}, and standard Constant Strain Triangle (CST) membrane. {\color{blue}A CST membrane is considered by replacing the $\mathbf{N}_{u}, \mathbf{N}_{v}$ shape functions in the $\mathbf{B}_{\mathrm{c}_{d}}$ and $\mathbf{B}_{\mathrm{c}_{\theta}}$ definitions and suppressing the $\theta_{z}$ DoF.} 



\subsubsection{Description of the benchmark tests}

\begin{figure}[H]
    \centering
    \scalebox{0.63}{\includegraphics{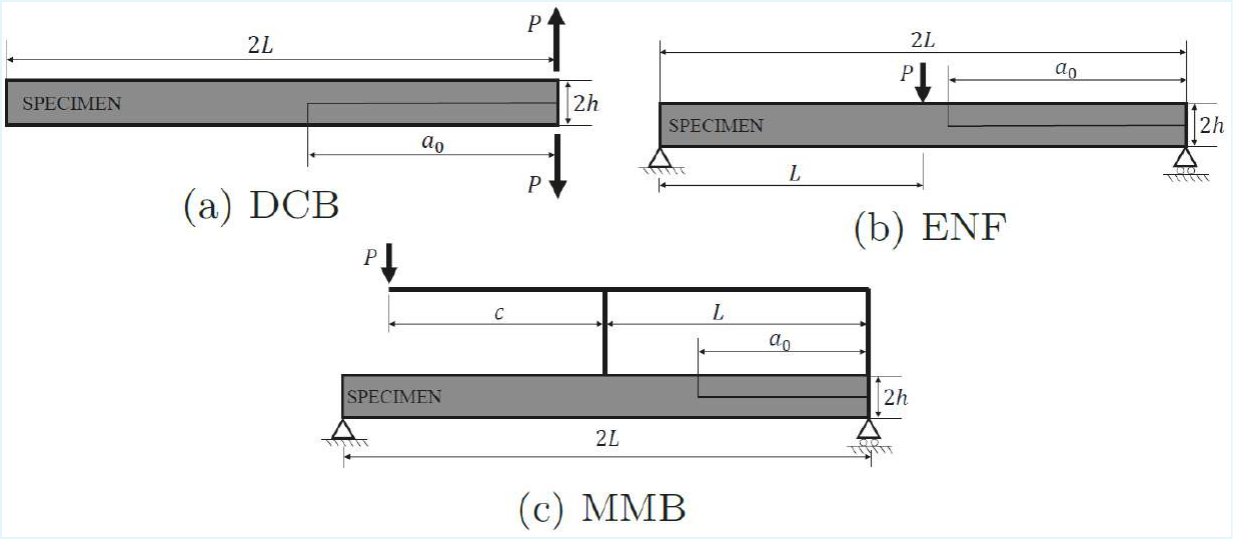}}
	\caption{DCB, ENF, and MMB test specimens \cite{Aietal2025}.}
	\label{benchmarks}
\end{figure}

The benchmarks studied herein are the same as those from \cite{Krueger2015}, with the adaptations from \cite{Aietal2025}. The geometric parameters and boundary conditions are shown in Figure \ref{benchmarks} and Table \ref{benchmarks_parameters}. 

\vspace{-10pt}
\begin{table}[H]
\centering
\footnotesize
\caption{Geometric parameters for benchmarks, parameters in $~\mathrm{mm}$.}
\begin{tabular}{@{}cccccc@{}}
    \toprule
          &   $2L$        & $a_0$  &  $h$ & $b \text{ (width)}$ & $c$\\ \midrule
    DCB & $150.0$ & $30.5$ & $1.50$  & $25.0$ & $-$ \\
    ENF & $101.6$ & $35.0$ & $2.25$  & $25.4$ & $-$  \\
    MMB & $100.8$ & $25.4$ & $2.25$  & $25.4$ & $41.3$  \\ \bottomrule
\end{tabular}
\label{benchmarks_parameters}
\end{table}

The detailed material properties are shown in Tables \ref{T300/1076_parameters} and \ref{IM7/8552_parameters}. 

\begin{table}[H]
\centering
\footnotesize
\caption{Material properties for the DCB test.}
\setlength{\tabcolsep}{10pt}
\renewcommand{\arraystretch}{1.0}
\begin{tabular}{@{}lll@{}}
    \toprule
    \textbf{Elastic constants data}~\cite{Krueger2015} \\
    \midrule
    $E_{11}=139.4~\mathrm{GPa}$ & $E_{22}=10.16~\mathrm{GPa}$ & $E_{33}=10.16~\mathrm{GPa}$ \\
    $\nu_{12}=0.30$             & $\nu_{13}=0.30$             & $\nu_{23}=0.436$ \\
    $G_{12}=4.6~\mathrm{GPa}$   & $G_{13}=4.6~\mathrm{GPa}$   & $G_{23}=3.54~\mathrm{GPa}$ \\[0.4ex]
    \midrule
    \textbf{Fracture properties data} ~\cite{Krueger2015,Luetal2019,Turonetal2010} & & \\
    \midrule
    $G_{\mathrm{I}c}=0.170~\mathrm{kJ}/\mathrm{m}^2$
      & $G_{\mathrm{II}c}=0.494~\mathrm{kJ}/\mathrm{m}^2$
      & $\eta=1.62$ \\[0.4ex]
    $\tau_{\mathrm{I}c}=30~\mathrm{MPa}$ & $\tau_{\mathrm{II}c}=60~\mathrm{MPa}$ & \\
    \bottomrule
\end{tabular}
\label{T300/1076_parameters}
\end{table}

\noindent The mixed mode ratio of the MMB specimen is $50\%$ mode II ($G_{II}/G_T = 0.5$). The methodology proposed by \cite{Turonetal2010} for accurately predicting the propagation of delamination under mixed-mode fracture with {\color{blue}CEs} was applied to the MMB models. 

\begin{table}[!ht]
\centering
\footnotesize
\caption{Material properties for the ENF and MMB tests.}
\label{tab:matprops_enf_mmb}
\setlength{\tabcolsep}{10pt}
\renewcommand{\arraystretch}{1.0}
\begin{tabular}{@{}lll@{}}
    \toprule
    \textbf{Elastic constants data}~\cite{Krueger2015} \\
    \midrule
    $E_{11}=161~\mathrm{GPa}$ & $E_{22}=11.38~\mathrm{GPa}$ & $E_{33}=11.38~\mathrm{GPa}$ \\
    $\nu_{12}=0.32$           & $\nu_{13}=0.32$            & $\nu_{23}=0.45$ \\
    $G_{12}=5.2~\mathrm{GPa}$ & $G_{13}=5.2~\mathrm{GPa}$  & $G_{23}=3.9~\mathrm{GPa}$ \\[0.4ex]
    \midrule
    \textbf{Fracture properties data} ~\cite{Krueger2015,Luetal2019,Turonetal2010} & & \\
    \midrule
    $G_{\mathrm{I}c}=0.212~\mathrm{kJ}/\mathrm{m}^2$
      & $G_{\mathrm{II}c}=0.774~\mathrm{kJ}/\mathrm{m}^2$
      & $\eta=2.1$ \\[0.4ex]
    $\tau_{\mathrm{I}c}=30~\mathrm{MPa}$ & $\tau_{\mathrm{II}c}=60~\mathrm{MPa}$ & \\
    \bottomrule
\end{tabular}
\label{IM7/8552_parameters}
\end{table}

\subsubsection{Description of the models}
\label{sec:BenchmarksDescription}

\begin{figure}[H]
    \centering
    \scalebox{0.40}{\includegraphics{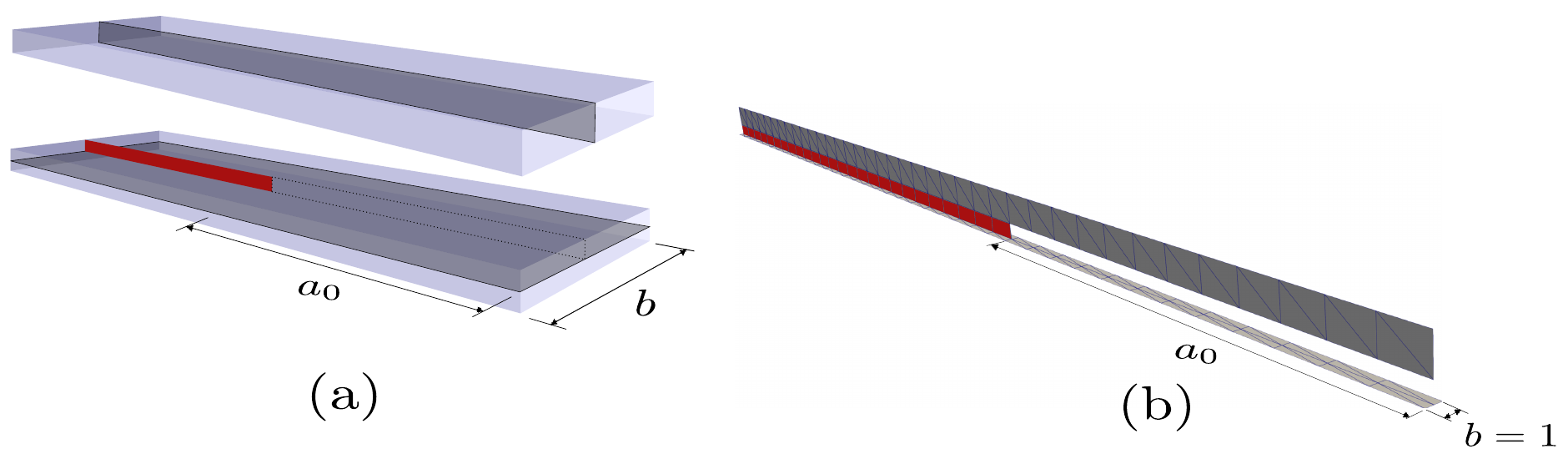}}
	\caption{{\color{blue}Shell-to-shell structural CE models to reproduce delamination tests.}}
	\label{2D_ShellShell}
\end{figure}

The setup of the shell-to-shell structural CE models used to reproduce the benchmarks is illustrated in Figure \ref{2D_ShellShell}. {\color{blue}The delamination benchmarks can be modeled with shells, adopting a horizontal mid-surface shell for the bottom}

\vspace{-10pt}
\noindent {\color{blue}ply and a vertical mid-surface shell for the top ply, as illustrated in Fig. \ref{2D_ShellShell}(a). A unit width ($b=1~\mathrm{mm}$) was considered for the shell-to-shell models, as illustrated in Fig. \ref{2D_ShellShell}(b). The load response of the model must therefore be multiplied by the real width $b$ of the specimen reported in Table \ref{benchmarks_parameters}.} On the bottom side of the delamination, two horizontal element layers were used to create a central line of nodes for the insertion of shell-to-shell CEs. On the top side, orthogonally oriented element layers with unit thickness ($t^{s}=1~\mathrm{mm}$) were used. {\color{blue}Shell-to-shell CEs - shown in red in Fig. \ref{2D_ShellShell} -} were inserted between the two sides of the delamination{\color{blue}, while the initial notch region $a_0$ remained unmeshed. Any contact interaction between the upper ply (stiffener) and the bottom ply (skin) in the ENF test is avoided by simply constraining the vertical displacement of the top-right node of the stiffener.} The analytical solutions for the benchmark problems were based on 2D beam models \cite{Williams1989}. Due to the unit width of the bottom ply and the unit thickness of the top ply, the numerical responses of the shell-to-shell models are expected to match the mechanical behavior predicted by the analytical solutions, enabling a fair comparison.

For structural CE analyses using the FF membrane, the top ply of the benchmark tests was modeled with 1, 2, and 4 element layers. In contrast, for structural CE models employing the CST membrane, a larger number of layers was used to assess potential locking issues caused by in-plane bending. Specifically, the top ply of the DCB specimen was discretized using 2, 4, and 8 element layers, whereas the top ply of the ENF and MMB specimens was discretized with 4, 8, and 16 layers. For each modeling approach — structural CEs with FF membranes and structural CEs with CST membranes — different mesh refinements along the crack propagation direction were investigated, for fixed numbers of element layers. 


\subsubsection{Load-displacement results}

The load–displacement curves obtained with the structural CE models are shown in Figures \ref{benchmarks_LvsD_results1} and \ref{benchmarks_LvsD_results2}. These curves are plotted together with the analytical solutions from \cite{Williams1989, ASTM_D5528_2007, ASTM_D6671_2022}.

The results in Fig. \ref{benchmarks_LvsD_results1} are used to compare the performance of the structural CE models with CST and FF membranes in terms of the number of element layers used for the top-ply discretization. The results show that structural CE - CST membrane models require significantly more element layers in the top-ply discretization than models based on the FF membrane. For the DCB, ENF, and MMB tests, the CST models required 4, 8, and 8 element layers, respectively, {\color{blue}with mesh refinements of 0.5 mm}, to avoid deviations from the analytical solution. In contrast, the corresponding FF models converged using {\color{blue}only 2 layers} and mesh refinements of 0.5 mm, 2.5 mm, and 0.5 mm, respectively.  

The results in Fig. \ref{benchmarks_LvsD_results2} focus on comparing the models with respect to mesh refinement in the propagation direction for fixed number of element layers for the top ply: {\color{blue}4 layers for the DCB CST-based models, 8 layers for the ENF and MMB CST-based models, and 2 layers for all the FF-based models.  Mesh refinements along propagation direction ranging from 0.5 mm to 2.0 mm, 2.5 mm to 7.5 mm, and 0.5 mm to 2.5 mm were investigated for the DCB, ENF and MMB tests, respectively. For comparison purposes, it is interesting to notice that the cohesive zone length for delamination tests can be estimated from the material properties in Tables \ref{T300/1076_parameters} and \ref{IM7/8552_parameters}, and the thickness of the plies, as discussed in \cite{HarperandHallett2008} for idealized pure mode I and II fracture conditions. Therefore, the cohesize zone length expected for the DCB and ENF tests studied herein are approximately 1.7 mm and 8.0 mm, respectively.}

{\color{blue}The results confirm that, even with a reduced number of top-ply element layers, most of the FF-based structural CE models outperform the CST-based models, providing better approximations of the peak load, even when coarser meshes are employed, as well as less unstable responses during crack propagation. The MMB test is the only exception where the improved performance of the FF-based models is not directly evident.} For this case, an additional FF-based model with eight element layers in the top-ply discretization was analyzed, as shown in Fig.  \ref{benchmarks_LvsD_results3}(b).


\begin{figure}[H]
\centering

\begin{minipage}[t]{0.48\columnwidth}
\centering
	\begin{tikzpicture}[scale=0.60]
		\begin{axis}[%
			xlabel={Displacement, mm},
			ylabel={Applied load, N},
                xlabel style={font=\large},
                ylabel style={font=\large},
                xmin=0.0, xmax=4.0,
                ymin=0.0, ymax=100.0,
                xtick distance=0.5, 
                ytick distance=10.0, 
                legend style={at={(0.0,1.00)}, anchor=north west, column sep=1ex,
                legend cell align=left, font=\small},
			table/col sep=comma,height=9cm]

            \addplot[black,mark=none, line width=1.0pt]  table [x=-v, y=F] {4.Results_data/dcb/dcb_analytical_0.txt};
            \addlegendentry{Analytical}

            \addplot[black,mark=none, line width=1.0pt, forget plot]  table [x=-v, y=F] {4.Results_data/dcb/dcb_analytical_1.txt};

            \addplot[blue,mark=x, line width=1.0pt]  table [x=-v, y=F] {4.Results_data/dcb/dcbPanelRibsCST_8elem/dcb_panelribs5DofsCST_0.5mm.txt};
		      \addlegendentry{0.5 mm - 8 layers}

            \addplot[teal,mark=x, line width=1.0pt]  table [x=-v, y=F] {4.Results_data/dcb/dcbPanelRibsCST_4elem/dcb_panelribs5DofsCST_0.5mm.txt};
		      \addlegendentry{0.5 mm - 4 layers}

            \addplot[red,mark=x, line width=1.0pt]  table [x=-v, y=F] {4.Results_data/dcb/dcbPanelRibsCST_2elem/dcb_panelribs5DofsCST_0.5mm.txt};
		      \addlegendentry{0.5 mm - 2 layers}
              
		\end{axis}
	\end{tikzpicture}
\caption*{{\color{blue}(a) DCB structural CE - CST membrane}}
\end{minipage}
\hfill
\begin{minipage}[t]{0.48\columnwidth}
\centering
	\begin{tikzpicture}[scale=0.60]
		\begin{axis}[%
			xlabel={Displacement, mm},
			ylabel={Applied load, N},
                xlabel style={font=\large},
                ylabel style={font=\large},
                xmin=0.0, xmax=4.0,
                ymin=0.0, ymax=100.0,
                xtick distance=0.5, 
                ytick distance=10.0, 
                legend style={at={(0.0,1.00)}, anchor=north west, column sep=1ex,
                legend cell align=left, font=\small},
			table/col sep=comma,height=9cm]

            \addplot[black,mark=none, line width=1.0pt]  table [x=-v, y=F] {4.Results_data/dcb/dcb_analytical_0.txt};
            \addlegendentry{Analytical}

            \addplot[black,mark=none, line width=1.0pt, forget plot]  table [x=-v, y=F] {4.Results_data/dcb/dcb_analytical_1.txt};

            \addplot[blue,mark=x, line width=1.0pt]  table [x=-v, y=F] {4.Results_data/dcb/dcbPanelRibs_4elem/dcb_panelribs6DofsBF_0.5mm.txt};
		      \addlegendentry{0.5 mm - 4 layers}

            \addplot[teal,mark=x, line width=1.0pt]  table [x=-v, y=F] {4.Results_data/dcb/dcbPanelRibs_2elem/dcb_panelribs6DofsBF_0.5mm.txt};
		      \addlegendentry{0.5 mm - 2 layers}

            \addplot[red,mark=x, line width=1.0pt]  table [x=-v, y=F] {4.Results_data/dcb/dcbPanelRibs_1elem/dcb_panelribs6DofsBF_0.5mm.txt};
		      \addlegendentry{0.5 mm - 1 layer}
              
		\end{axis}
	\end{tikzpicture}
\caption*{{\color{blue}(b) DCB structural CE - FF membrane}}
\end{minipage}

\vspace{5pt}

\begin{minipage}[t]{0.48\columnwidth}
\centering
	\begin{tikzpicture}[scale=0.60]
		\begin{axis}[%
			xlabel={Displacement, mm},
			ylabel={Applied load, N},
                xlabel style={font=\large},
                ylabel style={font=\large},
                xmin=0.0, xmax=4.0,
                ymin=0.0, ymax=2000.0,
                xtick distance=0.5, 
                ytick distance=250.0, 
                legend style={at={(0.0,1.00)}, anchor=north west, column sep=1ex,
                legend cell align=left, font=\small},
			table/col sep=comma,height=9cm]

            \addplot[black,mark=none, line width=1.0pt]  table [x=-v, y=F] {4.Results_data/enf/enf_analytical_0.txt};
            \addlegendentry{Analytical}

            \addplot[black,mark=none, line width=1.0pt, forget plot]  table [x=-v, y=F] {4.Results_data/enf/enf_analytical_1.txt};

            \addplot[black,mark=none, line width=1.0pt, forget plot]  table [x=-v, y=F] {4.Results_data/enf/enf_analytical_2.txt};

            \addplot[blue,mark=x, line width=1.0pt]  table [x=-v, y=F] {4.Results_data/enf/enfPanelRibsCST_16elem/enf_panelribs5DofsCST_0.5mm.txt};
		      \addlegendentry{0.5 mm - 16 layers}

            \addplot[teal,mark=x, line width=1.0pt]  table [x=-v, y=F] {4.Results_data/enf/enfPanelRibsCST_8elem/enf_panelribs5DofsCST_0.5mm.txt};
		      \addlegendentry{0.5 mm - 8 layers}

            \addplot[red,mark=x, line width=1.0pt]  table [x=-v, y=F] {4.Results_data/enf/enfPanelRibsCST_4elem/enf_panelribs5DofsCST_0.5mm.txt};
		      \addlegendentry{0.5 mm - 4 layers}
              
		\end{axis}
	\end{tikzpicture}
\caption*{(c) ENF structural CE - CST membrane}
\end{minipage}
\hfill
\begin{minipage}[t]{0.48\columnwidth}
\centering
	\begin{tikzpicture}[scale=0.60]
		\begin{axis}[%
			xlabel={Displacement, mm},
			ylabel={Applied load, N},
                xlabel style={font=\large},
                ylabel style={font=\large},
                xmin=0.0, xmax=4.0,
                ymin=0.0, ymax=2000.0,
                xtick distance=0.5, 
                ytick distance=250.0, 
                legend style={at={(0.0,1.00)}, anchor=north west, column sep=1ex,
                legend cell align=left, font=\small},
			table/col sep=comma,height=9cm]

            \addplot[black,mark=none, line width=1.0pt]  table [x=-v, y=F] {4.Results_data/enf/enf_analytical_0.txt};
            \addlegendentry{Analytical}

            \addplot[black,mark=none, line width=1.0pt, forget plot]  table [x=-v, y=F] {4.Results_data/enf/enf_analytical_1.txt};

            \addplot[black,mark=none, line width=1.0pt, forget plot]  table [x=-v, y=F] {4.Results_data/enf/enf_analytical_2.txt};

            \addplot[blue,mark=x, line width=1.0pt]  table [x=-v, y=F] {4.Results_data/enf/enfPanelRibs_4elem/enf_panelribs6DofsBF_2.5mm.txt};
		      \addlegendentry{2.5 mm - 4 layers}

            \addplot[teal,mark=x, line width=1.0pt]  table [x=-v, y=F] {4.Results_data/enf/enfPanelRibs_2elem/enf_panelribs6DofsBF_2.5mm.txt};
		      \addlegendentry{2.5 mm - 2 layers}

            \addplot[red,mark=x, line width=1.0pt]  table [x=-v, y=F] {4.Results_data/enf/enfPanelRibs_1elem/enf_panelribs6DofsBF_2.5mm.txt};
		      \addlegendentry{2.5 mm - 1 layer}
              
		\end{axis}
	\end{tikzpicture}
\caption*{(d) ENF structural CE - FF membrane}
\end{minipage}

\vspace{5pt}

\begin{minipage}[t]{0.48\columnwidth}
\centering
	\begin{tikzpicture}[scale=0.60]
		\begin{axis}[%
			xlabel={Displacement, mm},
			ylabel={Applied load, N},
                xlabel style={font=\large},
                ylabel style={font=\large},
                xmin=0.0, xmax=1.8,
                ymin=0.0, ymax=500,
                xtick distance=0.2, 
                ytick distance=100, 
                legend style={at={(0.0,1.00)}, anchor=north west, column sep=1ex,
                legend cell align=left, font=\small},
			table/col sep=comma,height=9cm]

            \addplot[black,mark=none, line width=1.0pt]  table [x=-v, y=F] {4.Results_data/mmb/mmb_analytical_0.txt};
            \addlegendentry{Analytical}

            \addplot[black,mark=none, line width=1.0pt, forget plot]  table [x=-v, y=F] {4.Results_data/mmb/mmb_analytical_1.txt};

            \addplot[blue,mark=x, line width=1.0pt]  table [x=-v, y=F] {4.Results_data/mmb/mmbPanelRibsCST_16elem/mmb_panelribs5DofsCST_0.5mm.txt};
		      \addlegendentry{0.5 mm - 16 layers}

            \addplot[teal,mark=x, line width=1.0pt]  table [x=-v, y=F] {4.Results_data/mmb/mmbPanelRibsCST_8elem/mmb_panelribs5DofsCST_0.5mm.txt};
		      \addlegendentry{0.5 mm - 8 layers}

            \addplot[red,mark=x, line width=1.0pt]  table [x=-v, y=F] {4.Results_data/mmb/mmbPanelRibsCST_4elem/mmb_panelribs5DofsCST_0.5mm.txt};
		      \addlegendentry{0.5 mm - 4 layers}
              
		\end{axis}
	\end{tikzpicture}
\caption*{(e) MMB structural CE - CST membrane}
\end{minipage}
\hfill
\begin{minipage}[t]{0.48\columnwidth}
\centering
	\begin{tikzpicture}[scale=0.60]
		\begin{axis}[%
			xlabel={Displacement, mm},
			ylabel={Applied load, N},
                xlabel style={font=\large},
                ylabel style={font=\large},
                xmin=0.0, xmax=1.8,
                ymin=0.0, ymax=500,
                xtick distance=0.2, 
                ytick distance=100, 
                legend style={at={(0.0,1.00)}, anchor=north west, column sep=1ex,
                legend cell align=left, font=\small},
			table/col sep=comma,height=9cm]

            \addplot[black,mark=none, line width=1.0pt]  table [x=-v, y=F] {4.Results_data/mmb/mmb_analytical_0.txt};
            \addlegendentry{Analytical}

            \addplot[black,mark=none, line width=1.0pt, forget plot]  table [x=-v, y=F] {4.Results_data/mmb/mmb_analytical_1.txt};

            \addplot[blue,mark=x, line width=1.0pt]  table [x=-v, y=F] {4.Results_data/mmb/mmbPanelRibs_4elem/mmb_panelribs6DofsBF_0.5mm.txt};
		      \addlegendentry{0.5 mm - 4 layers}

            \addplot[teal,mark=x, line width=1.0pt]  table [x=-v, y=F] {4.Results_data/mmb/mmbPanelRibs_2elem/mmb_panelribs6DofsBF_0.5mm.txt};
		      \addlegendentry{0.5 mm - 2 layers}

            \addplot[red,mark=x, line width=1.0pt]  table [x=-v, y=F] {4.Results_data/mmb/mmbPanelRibs_1elem/mmb_panelribs6DofsBF_0.5mm.txt};
		      \addlegendentry{0.5 mm - 1 layer}
              
		\end{axis}
	\end{tikzpicture}
\caption*{(f) MMB structural CE - FF membrane}
\end{minipage}
\caption{Load–displacement curves: Study on the number of top ply element layers.}
\label{benchmarks_LvsD_results1}
\end{figure}

\begin{figure}[H]
\centering

\begin{minipage}[t]{0.48\columnwidth}
\centering
	\begin{tikzpicture}[scale=0.60]
		\begin{axis}[%
			xlabel={Displacement, mm},
			ylabel={Applied load, N},
                xlabel style={font=\large},
                ylabel style={font=\large},
                xmin=0.0, xmax=4.0,
                ymin=0.0, ymax=100.0,
                xtick distance=0.5, 
                ytick distance=10.0, 
                legend style={at={(0.0,1.00)}, anchor=north west, column sep=1ex,
                legend cell align=left, font=\small},
			table/col sep=comma,height=9cm]

            \addplot[black,mark=none, line width=1.0pt]  table [x=-v, y=F] {4.Results_data/dcb/dcb_analytical_0.txt};
            \addlegendentry{Analytical}

            \addplot[black,mark=none, line width=1.0pt, forget plot]  table [x=-v, y=F] {4.Results_data/dcb/dcb_analytical_1.txt};

            \addplot[blue,mark=x, line width=1.0pt]  table [x=-v, y=F] {4.Results_data/dcb/dcbPanelRibsCST_4elem/dcb_panelribs5DofsCST_0.5mm.txt};
		      \addlegendentry{0.5 mm - 4 layers}

            \addplot[teal,mark=x, line width=1.0pt]  table [x=-v, y=F] {4.Results_data/dcb/dcbPanelRibsCST_4elem/dcb_panelribs5DofsCST_1.0mm.txt};
		      \addlegendentry{1.0 mm - 4 layers}

            \addplot[red,mark=x, line width=1.0pt]  table [x=-v, y=F] {4.Results_data/dcb/dcbPanelRibsCST_4elem/dcb_panelribs5DofsCST_2.0mm.txt};
		      \addlegendentry{2.0 mm - 4 layers}
              
		\end{axis}
	\end{tikzpicture}
\caption*{(a) DCB structural CE - CST membrane}
\end{minipage}
\hfill
\begin{minipage}[t]{0.48\columnwidth}
\centering
	\begin{tikzpicture}[scale=0.60]
		\begin{axis}[%
			xlabel={Displacement, mm},
			ylabel={Applied load, N},
                xlabel style={font=\large},
                ylabel style={font=\large},
                xmin=0.0, xmax=4.0,
                ymin=0.0, ymax=100.0,
                xtick distance=0.5, 
                ytick distance=10.0, 
                legend style={at={(0.0,1.00)}, anchor=north west, column sep=1ex,
                legend cell align=left, font=\small},
			table/col sep=comma,height=9cm]

            \addplot[black,mark=none, line width=1.0pt]  table [x=-v, y=F] {4.Results_data/dcb/dcb_analytical_0.txt};
            \addlegendentry{Analytical}

            \addplot[black,mark=none, line width=1.0pt, forget plot]  table [x=-v, y=F] {4.Results_data/dcb/dcb_analytical_1.txt};

            \addplot[blue,mark=x, line width=1.0pt]  table [x=-v, y=F] {4.Results_data/dcb/dcbPanelRibs_2elem/dcb_panelribs6DofsBF_0.5mm.txt};
		      \addlegendentry{0.5 mm - 2 layers}

            \addplot[teal,mark=x, line width=1.0pt]  table [x=-v, y=F] {4.Results_data/dcb/dcbPanelRibs_2elem/dcb_panelribs6DofsBF_1.0mm.txt};
		      \addlegendentry{1.0 mm - 2 layers}

            \addplot[red,mark=x, line width=1.0pt]  table [x=-v, y=F] {4.Results_data/dcb/dcbPanelRibs_2elem/dcb_panelribs6DofsBF_2.0mm.txt};
		      \addlegendentry{2.0 mm - 2 layers}
		\end{axis}
	\end{tikzpicture}
\caption*{(b) DCB structural CE - FF membrane}
\end{minipage}

\vspace{5pt}

\begin{minipage}[t]{0.48\columnwidth}
\centering
	\begin{tikzpicture}[scale=0.60]
		\begin{axis}[%
			xlabel={Displacement, mm},
			ylabel={Applied load, N},
                xlabel style={font=\large},
                ylabel style={font=\large},
                xmin=0.0, xmax=4.0,
                ymin=0.0, ymax=2000.0,
                xtick distance=0.5, 
                ytick distance=250.0, 
                legend style={at={(0.0,1.00)}, anchor=north west, column sep=1ex,
                legend cell align=left, font=\small},
			table/col sep=comma,height=9cm]

            \addplot[black,mark=none, line width=1.0pt]  table [x=-v, y=F] {4.Results_data/enf/enf_analytical_0.txt};
            \addlegendentry{Analytical}

            \addplot[black,mark=none, line width=1.0pt, forget plot]  table [x=-v, y=F] {4.Results_data/enf/enf_analytical_1.txt};

            \addplot[black,mark=none, line width=1.0pt, forget plot]  table [x=-v, y=F] {4.Results_data/enf/enf_analytical_2.txt};

            \addplot[blue,mark=x, line width=1.0pt]  table [x=-v, y=F] {4.Results_data/enf/enfPanelRibsCST_8elem/enf_panelribs5DofsCST_2.5mm.txt};
		      \addlegendentry{2.5 mm - 8 layers}

            \addplot[teal,mark=x, line width=1.0pt]  table [x=-v, y=F] {4.Results_data/enf/enfPanelRibsCST_8elem/enf_panelribs5DofsCST_5.0mm.txt};
		      \addlegendentry{5.0 mm - 8 layers}

            \addplot[red,mark=x, line width=1.0pt]  table [x=-v, y=F] {4.Results_data/enf/enfPanelRibsCST_8elem/enf_panelribs5DofsCST_7.5mm.txt};
		      \addlegendentry{7.5 mm - 8 layers}
              
		\end{axis}
	\end{tikzpicture}
\caption*{(c) ENF structural CE - CST membrane}
\end{minipage}
\hfill
\begin{minipage}[t]{0.48\columnwidth}
\centering
	\begin{tikzpicture}[scale=0.60]
		\begin{axis}[%
			xlabel={Displacement, mm},
			ylabel={Applied load, N},
                xlabel style={font=\large},
                ylabel style={font=\large},
                xmin=0.0, xmax=4.0,
                ymin=0.0, ymax=2000.0,
                xtick distance=0.5, 
                ytick distance=250.0, 
                legend style={at={(0.0,1.00)}, anchor=north west, column sep=1ex,
                legend cell align=left, font=\small},
			table/col sep=comma,height=9cm]

            \addplot[black,mark=none, line width=1.0pt]  table [x=-v, y=F] {4.Results_data/enf/enf_analytical_0.txt};
            \addlegendentry{Analytical}

            \addplot[black,mark=none, line width=1.0pt, forget plot]  table [x=-v, y=F] {4.Results_data/enf/enf_analytical_1.txt};

            \addplot[black,mark=none, line width=1.0pt, forget plot]  table [x=-v, y=F] {4.Results_data/enf/enf_analytical_2.txt};

            \addplot[blue,mark=x, line width=1.0pt]  table [x=-v, y=F] {4.Results_data/enf/enfPanelRibs_2elem/enf_panelribs6DofsBF_2.5mm.txt};
		      \addlegendentry{2.5 mm - 2 layers}

            \addplot[teal,mark=x, line width=1.0pt]  table [x=-v, y=F] {4.Results_data/enf/enfPanelRibs_2elem/enf_panelribs6DofsBF_5.0mm.txt};
		      \addlegendentry{5.0 mm - 2 layers}

            \addplot[red,mark=x, line width=1.0pt]  table [x=-v, y=F] {4.Results_data/enf/enfPanelRibs_2elem/enf_panelribs6DofsBF_7.5mm.txt};
		      \addlegendentry{7.5 mm - 2 layers}
              
		\end{axis}
	\end{tikzpicture}
\caption*{(d) ENF structural CE - FF membrane}
\end{minipage}

\vspace{5pt}

\begin{minipage}[t]{0.48\columnwidth}
\centering
	\begin{tikzpicture}[scale=0.60]
		\begin{axis}[%
			xlabel={Displacement, mm},
			ylabel={Applied load, N},
                xlabel style={font=\large},
                ylabel style={font=\large},
                xmin=0.0, xmax=1.8,
                ymin=0.0, ymax=500.0,
                xtick distance=0.2, 
                ytick distance=100.0, 
                legend style={at={(0.0,1.00)}, anchor=north west, column sep=1ex,
                legend cell align=left, font=\small},
			table/col sep=comma,height=9cm]

            \addplot[black,mark=none, line width=1.0pt]  table [x=-v, y=F] {4.Results_data/mmb/mmb_analytical_0.txt};
            \addlegendentry{Analytical}

            \addplot[black,mark=none, line width=1.0pt, forget plot]  table [x=-v, y=F] {4.Results_data/mmb/mmb_analytical_1.txt};

            \addplot[blue,mark=x, line width=1.0pt]  table [x=-v, y=F] {4.Results_data/mmb/mmbPanelRibsCST_8elem/mmb_panelribs5DofsCST_0.5mm.txt};
		      \addlegendentry{0.5 mm - 8 layers}

            \addplot[teal,mark=x, line width=1.0pt]  table [x=-v, y=F] {4.Results_data/mmb/mmbPanelRibsCST_8elem/mmb_panelribs5DofsCST_1.0mm.txt};
		      \addlegendentry{1.0 mm - 8 layers}

            \addplot[red,mark=x, line width=1.0pt]  table [x=-v, y=F] {4.Results_data/mmb/mmbPanelRibsCST_8elem/mmb_panelribs5DofsCST_2.5mm.txt};
		      \addlegendentry{2.5 mm - 8 layers}
              
		\end{axis}
	\end{tikzpicture}
\caption*{(e) MMB structural CE - CST membrane}
\end{minipage}
\hfill
\begin{minipage}[t]{0.48\columnwidth}
\centering
	\begin{tikzpicture}[scale=0.60]
		\begin{axis}[%
			xlabel={Displacement, mm},
			ylabel={Applied load, N},
                xlabel style={font=\large},
                ylabel style={font=\large},
                xmin=0.0, xmax=1.8,
                ymin=0.0, ymax=500,
                xtick distance=0.2, 
                ytick distance=100, 
                legend style={at={(0.0,1.00)}, anchor=north west, column sep=1ex,
                legend cell align=left, font=\small},
			table/col sep=comma,height=9cm]

            \addplot[black,mark=none, line width=1.0pt]  table [x=-v, y=F] {4.Results_data/mmb/mmb_analytical_0.txt};
            \addlegendentry{Analytical}

            \addplot[black,mark=none, line width=1.0pt, forget plot]  table [x=-v, y=F] {4.Results_data/mmb/mmb_analytical_1.txt};

            \addplot[blue,mark=x, line width=1.0pt]  table [x=-v, y=F] {4.Results_data/mmb/mmbPanelRibs_2elem/mmb_panelribs6DofsBF_0.5mm.txt};
		      \addlegendentry{0.5 mm - 2 layers}

            \addplot[teal,mark=x, line width=1.0pt]  table [x=-v, y=F] {4.Results_data/mmb/mmbPanelRibs_2elem/mmb_panelribs6DofsBF_1.0mm.txt};
		      \addlegendentry{1.0 mm - 2 layers}

            \addplot[red,mark=x, line width=1.0pt]  table [x=-v, y=F] {4.Results_data/mmb/mmbPanelRibs_2elem/mmb_panelribs6DofsBF_2.5mm.txt};
		      \addlegendentry{2.5 mm - 2 layers}            
		\end{axis}
	\end{tikzpicture}
\caption*{(f) MMB structural CE - FF membrane}
\end{minipage}
\caption{Load–displacement curves: Study on mesh refinements.}
\label{benchmarks_LvsD_results2}
\end{figure}

The comparison between CST- and FF-based models with 8 element layers shows a slight improvement when the FF membrane is employed. Very similar results to those shown in Fig. \ref{benchmarks_LvsD_results2}(d) were obtained for FF-based structural CE models with a single top-ply element layer.

\begin{figure}[H]
\centering

\begin{minipage}[t]{0.48\columnwidth}
\centering
	\begin{tikzpicture}[scale=0.60]
		\begin{axis}[%
			xlabel={Displacement, mm},
			ylabel={Applied load, N},
                xlabel style={font=\large},
                ylabel style={font=\large},
                xmin=0.0, xmax=1.8,
                ymin=0.0, ymax=500,
                xtick distance=0.2, 
                ytick distance=100, 
                legend style={at={(0.0,1.00)}, anchor=north west, column sep=1ex,
                legend cell align=left, font=\small},
			table/col sep=comma,height=9cm]

            \addplot[black,mark=none, line width=1.0pt]  table [x=-v, y=F] {4.Results_data/mmb/mmb_analytical_0.txt};
            \addlegendentry{Analytical}

            \addplot[black,mark=none, line width=1.0pt, forget plot]  table [x=-v, y=F] {4.Results_data/mmb/mmb_analytical_1.txt};

            \addplot[blue,mark=x, line width=1.0pt]  table [x=-v, y=F] {4.Results_data/mmb/mmbPanelRibsCST_8elem/mmb_panelribs5DofsCST_0.5mm.txt};
		      \addlegendentry{0.5 mm - 8 layers}

            \addplot[teal,mark=x, line width=1.0pt]  table [x=-v, y=F] {4.Results_data/mmb/mmbPanelRibsCST_8elem/mmb_panelribs5DofsCST_1.0mm.txt};
		      \addlegendentry{1.0 mm - 8 layers}

            \addplot[red,mark=x, line width=1.0pt]  table [x=-v, y=F] {4.Results_data/mmb/mmbPanelRibsCST_8elem/mmb_panelribs5DofsCST_2.5mm.txt};
		      \addlegendentry{2.5 mm - 8 layers}
              
		\end{axis}
	\end{tikzpicture}
    \caption*{(a) MMB structural CE - CST membrane}
\end{minipage}
\hfill
\begin{minipage}[t]{0.48\columnwidth}
\centering
	\begin{tikzpicture}[scale=0.60]
		\begin{axis}[%
			xlabel={Displacement, mm},
			ylabel={Applied load, N},
                xlabel style={font=\large},
                ylabel style={font=\large},
                xmin=0.0, xmax=1.8,
                ymin=0.0, ymax=500,
                xtick distance=0.2, 
                ytick distance=100, 
                legend style={at={(0.0,1.00)}, anchor=north west, column sep=1ex,
                legend cell align=left, font=\small},
			table/col sep=comma,height=9cm]

            \addplot[black,mark=none, line width=1.0pt]  table [x=-v, y=F] {4.Results_data/mmb/mmb_analytical_0.txt};
            \addlegendentry{Analytical}

            \addplot[black,mark=none, line width=1.0pt, forget plot]  table [x=-v, y=F] {4.Results_data/mmb/mmb_analytical_1.txt};

            \addplot[blue,mark=x, line width=1.0pt]  table [x=-v, y=F] {4.Results_data/mmb/mmbPanelRibs_8elem/mmb_panelribs6DofsBF_0.5mm.txt};
		      \addlegendentry{0.5 mm - 8 layers}

            \addplot[teal,mark=x, line width=1.0pt]  table [x=-v, y=F] {4.Results_data/mmb/mmbPanelRibs_8elem/mmb_panelribs6DofsBF_1.0mm.txt};
		      \addlegendentry{1.0 mm - 8 layers}

            \addplot[red,mark=x, line width=1.0pt]  table [x=-v, y=F] {4.Results_data/mmb/mmbPanelRibs_8elem/mmb_panelribs6DofsBF_2.5mm.txt};
		      \addlegendentry{2.5 mm - 8 layers}
              
		\end{axis}
	\end{tikzpicture}
    \caption*{(b) MMB structural CE - FF membrane}
\end{minipage}
\caption{Load–displacement curves: MMB structural CE models comparison.}
\label{benchmarks_LvsD_results3}
\end{figure}

\subsection{Mode I skin-stiffener debonding}

\begin{figure}[H]
\centering
\scalebox{0.20}{\includegraphics{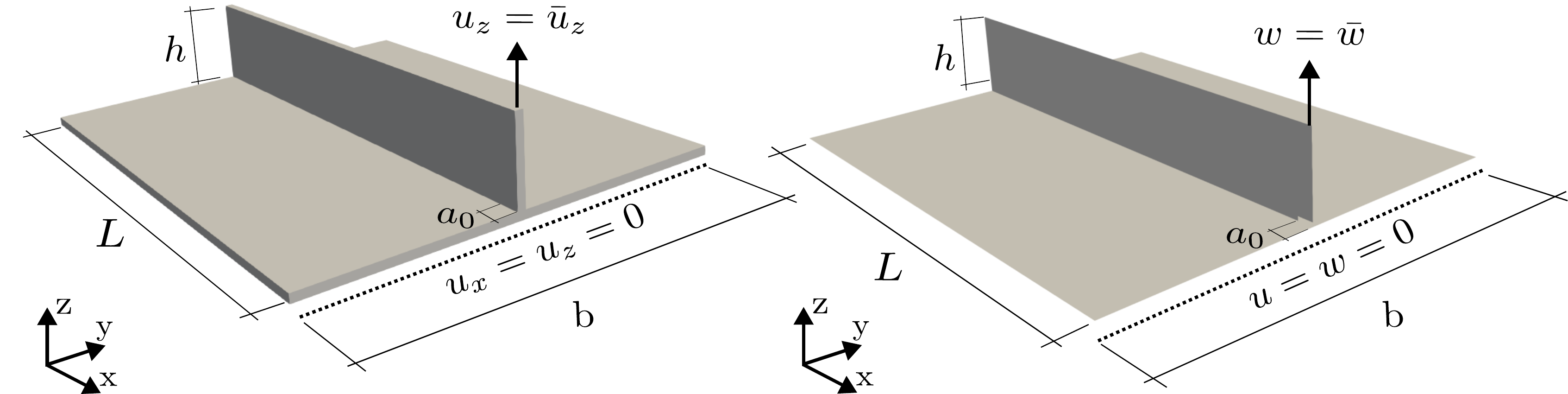}}
	\caption{Models and boundary conditions: (a) 3D model, (b) Structural model.}
	\label{fig:Mode I propagation test}
\end{figure}

The Mode I skin–stiffener debonding test was designed to ensure stable Mode I crack propagation along the interface. The geometry and boundary conditions of the 3D solid and shell models are given in Figure \ref{fig:Mode I propagation test} and Table \ref{modeIpropagation_parameters}.

\noindent A pre-crack of length $a_{0}$ was introduced at the loaded end to avoid instability, and a unit thickness $t$ was assumed for both components. Two types of models with varying skin geometries are considered for the debonding evaluation: skin models with dimensions of 60 mm $\times$ 50 mm (short debonding interface) and 120 mm $\times$ 50 mm (long debonding interface).
\begin{table}[H]
\centering
\footnotesize
\caption{Geometric parameters for the skin-stiffener debonding tests, parameters in $~\mathrm{mm}$.}
\begin{tabular}{@{}cccccc@{}}
    \toprule
    &   $L$        & $a_{0} \text{(notch)}$  &  $h$ & $b \text{ (width)}$ & $t \text{(thickness)}$\\ \midrule
    $60\times50$ models & $60.0$ & $2.0$ & $10.0$  & $50.0$ & $1.0$ \\
    $120\times50$ models & $120.0$ & $2.0$ & $10.0$  & $50.0$ & $1.0$  \\ \bottomrule
\end{tabular}
\label{modeIpropagation_parameters}
\end{table}
 
The elastic properties of the skin are the same as those presented in Table \ref{T300/1076_parameters}, whereas the elastic properties of the stiffener are presented in Table \ref{stiffener_parameters}. The interfacial fracture properties are given in Table \ref{interfacial_parameters_ModeI}. The interfacial properties follow those adopted in \cite{Aietal2025}, except that the Mode I interfacial strength $\tau_{\mathrm{I}c}$ was increased from  $30$~MPa to $60$~MPa to reduce the fracture process zone (FPZ) size.

\vspace{-10pt}

\begin{table}[H]
\centering
\footnotesize
\caption{Material properties for the stiffener.}
\setlength{\tabcolsep}{10pt}
\renewcommand{\arraystretch}{1.0}
\begin{tabular}{@{}lll@{}}
    \toprule
    \multicolumn{2}{c}{\textbf{Elastic constants data}~\cite{Neveuetal2022}}\\
    \midrule
    \hspace{1.5 cm} $E=15.5~\mathrm{GPa}$ & $\nu=0.3~\mathrm{GPa}$\\
    \bottomrule
\end{tabular}
\label{stiffener_parameters}
\end{table}
\vspace{-10pt}
\begin{table}[H]
\centering
\footnotesize
\caption{Interfacial fracture properties.}
\setlength{\tabcolsep}{10pt}
\renewcommand{\arraystretch}{1.0}
\begin{tabular}{@{}lll@{}}
    \toprule
    \multicolumn{2}{c}{\textbf{Fracture properties data}~\cite{Aietal2025}}\\
    \midrule
    $G_{\mathrm{I}c}=0.170~\mathrm{kJ}/\mathrm{m}^2$
      & $G_{\mathrm{II}c}=0.494~\mathrm{kJ}/\mathrm{m}^2$
      & $\eta=1.62$ \\[0.4ex]
    $\tau_{\mathrm{I}c}=60~\mathrm{MPa}$ & $\tau_{\mathrm{II}c}=60~\mathrm{MPa}$ & \\
    \bottomrule
\end{tabular}
\label{interfacial_parameters_ModeI}
\end{table}

\subsubsection{Description of the models}

\begin{figure}[H]
\centering
\scalebox{0.22}{\includegraphics{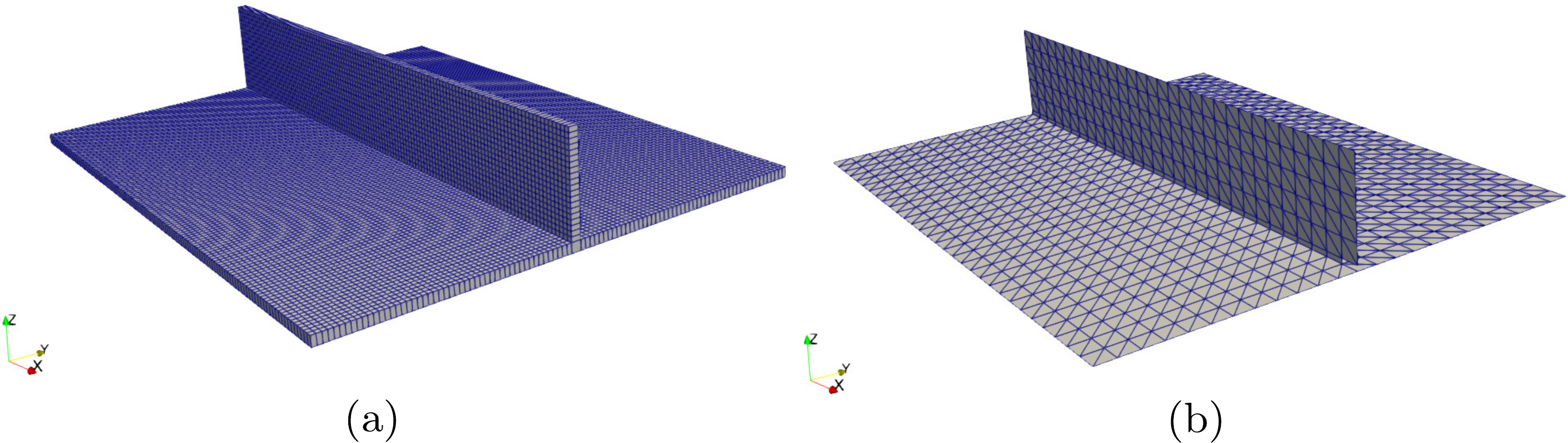}}
	\caption{60 mm $\times$ 50 mm models: (a) 3D solid CE - 0.5 mm, (b) structural CE - 2.0 mm.}
	\label{fig:Mode I propagation test meshes}
\end{figure}

The 3D solid models employed Hex8 elements for both skin and stiffener, with a single layer of elements through the thickness, and linear 8-node cohesive elements at the interface. The shell models adopted an analogous approach using single-layer shell elements and the shell-to-shell CEs.  Different mesh refinements were investigated for the 60 mm $\times$ 50 mm models, with element sizes of 2.0 mm, 1.0 mm and 0.5 mm. The 3D solid models required a fine mesh of 0.5 mm to achieve accurate and stable cohesive crack propagation, whereas the structural shell models provided accurate peak loads and stable crack growth with a coarse mesh of 2.0 mm. Figure \ref{fig:Mode I propagation test meshes}(a) illustrates the most refined 3D solid model, whereas Figure \ref{fig:Mode I propagation test meshes}(b) illustrates the least refined structural shell model. 
The 120 mm $\times$ 50 mm skin–stiffener debonding tests were then used to assess the increase in computational cost with increasing cohesive interface length, considering only the coarsest mesh refinements that still ensure stable and accurate responses: 3D solid models with the 0.5 mesh and structural models with the 2.0 mm mesh. This enables a fair comparison of the CPU time required for failure analyses using 3D solid models with standard CEs and structural models with the proposed shell-to-shell CEs.

\subsubsection{Load–displacement curves}
The load-displacement curves obtained from the simulations of the 60 mm $\times$ 50 mm models are shown in Figures \ref{fig:FvsD_results_3D} and \ref{fig:FvsD_results_Structural}. The
designations “Solid CE” and “Structural CE” denote the results of the standard CE models and those of the proposed structural CE models, respectively. 

\begin{figure}[H]
\centering

\begin{minipage}[t]{0.48\columnwidth}
\centering
    	\begin{tikzpicture}[scale=0.65]
		\begin{axis}[%
			xlabel={Displacement, mm},
			ylabel={Applied load, N},
                xlabel style={font=\large},
                ylabel style={font=\large},
                xmin=0.0, xmax=0.50,
                ymin=0.0, ymax=60.0,
                xtick distance=0.1, 
                ytick distance=10.0, 
                legend style={at={(0.425,1.0)}, anchor=north west, column sep=1ex,
                legend cell align=left, font=\small},
			table/col sep=comma,height=9cm]

            \addplot[black,mark=none, line width=1.0pt]  table [x=-v, y=F] {4.Results_data/modeI_propagation_60x50x10/3Dsolids_alpha25/modeIprop_3Dsolids_0.5mm.txt};
		      \addlegendentry{Reference solution}

            \addplot[blue,mark=x, line width=1.0pt]  table [x=-v, y=F] {4.Results_data/modeI_propagation_60x50x10/3Dsolids_alpha50/modeIprop_3Dsolids_0.5mm.txt};
		      \addlegendentry{Solid CE model 0.5mm}

            \addplot[teal,mark=x, line width=1.0pt]  table [x=-v, y=F] {4.Results_data/modeI_propagation_60x50x10/3Dsolids_alpha50/modeIprop_3Dsolids_1.0mm.txt};
		      \addlegendentry{Solid CE model 1.0mm}

            \addplot[red,mark=x, line width=1.0pt]  table [x=-v, y=F] {4.Results_data/modeI_propagation_60x50x10/3Dsolids_alpha50/modeIprop_3Dsolids_2.0mm.txt};
		      \addlegendentry{Solid CE model 2.0mm}
		\end{axis}
	\end{tikzpicture}
\caption*{(a) Solid CE models, $\alpha_p=50$}
\end{minipage}
\hfill
\begin{minipage}[t]{0.48\columnwidth}
\centering
    	\begin{tikzpicture}[scale=0.65]
		\begin{axis}[%
			xlabel={Displacement, mm},
			ylabel={Applied load, N},
                xlabel style={font=\large},
                ylabel style={font=\large},
                xmin=0.0, xmax=0.50,
                ymin=0.0, ymax=60.0,
                xtick distance=0.1, 
                ytick distance=10.0, 
                legend style={at={(0.425,1.0)}, anchor=north west, column sep=1ex,
                legend cell align=left, font=\small},
			table/col sep=comma,height=9cm]

            \addplot[black,mark=none, line width=1.0pt]  table [x=-v, y=F] {4.Results_data/modeI_propagation_60x50x10/3Dsolids_alpha25/modeIprop_3Dsolids_0.5mm.txt};
		      \addlegendentry{Reference solution}

            \addplot[blue,mark=x, line width=1.0pt]  table [x=-v, y=F] {4.Results_data/modeI_propagation_60x50x10/3Dsolids_alpha5/modeIprop_3Dsolids_0.5mm.txt};
		      \addlegendentry{Solid CE model 0.5mm}

            \addplot[teal,mark=x, line width=1.0pt]  table [x=-v, y=F] {4.Results_data/modeI_propagation_60x50x10/3Dsolids_alpha5/modeIprop_3Dsolids_1.0mm.txt};
		      \addlegendentry{Solid CE model 1.0mm}

            \addplot[red,mark=x, line width=1.0pt]  table [x=-v, y=F] {4.Results_data/modeI_propagation_60x50x10/3Dsolids_alpha5/modeIprop_3Dsolids_2.0mm.txt};
		      \addlegendentry{Solid CE model 2.0mm}
		\end{axis}
	\end{tikzpicture}
\caption*{(b) Solid CE models, $\alpha_p=5$}
\end{minipage}
\caption{Load–displacement curves for the solid CE models.}
\label{fig:FvsD_results_3D}
\end{figure}

\begin{figure}[H]
\centering

\begin{minipage}[t]{0.48\columnwidth}
\centering
    	\begin{tikzpicture}[scale=0.65]
		\begin{axis}[%
			xlabel={Displacement, mm},
			ylabel={Applied load, N},
                xlabel style={font=\large},
                ylabel style={font=\large},
                xmin=0.0, xmax=0.50,
                ymin=0.0, ymax=50.0,
                xtick distance=0.1, 
                ytick distance=10.0, 
                legend style={at={(0.325,1.0)}, anchor=north west, column sep=1ex,
                legend cell align=left, font=\small},
			table/col sep=comma,height=9cm]

            \addplot[black,mark=none, line width=1.0pt]  table [x=-v, y=F] {4.Results_data/modeI_propagation_60x50x10/3Dsolids_alpha25/modeIprop_3Dsolids_0.5mm.txt};
		      \addlegendentry{Reference solution}

            \addplot[blue,mark=x, line width=1.0pt]  table [x=-v, y=F] {4.Results_data/modeI_propagation_60x50x10/Shell6DofsBF_alpha50/modeIprop_shell6DofsBF_0.5mm.txt};
		      \addlegendentry{Structural CE model 0.5mm}

            \addplot[teal,mark=x, line width=1.0pt]  table [x=-v, y=F] {4.Results_data/modeI_propagation_60x50x10/Shell6DofsBF_alpha50/modeIprop_shell6DofsBF_1.0mm.txt};
		      \addlegendentry{Structural CE model 1.0mm}

            \addplot[red,mark=x, line width=1.0pt]  table [x=-v, y=F] {4.Results_data/modeI_propagation_60x50x10/Shell6DofsBF_alpha50/modeIprop_shell6DofsBF_2.0mm.txt};
		      \addlegendentry{Structural CE model 2.0mm}
		\end{axis}
	\end{tikzpicture}
\caption*{(a) Structural CE models, $\alpha_p=50$}
\end{minipage}
\hfill
\begin{minipage}[t]{0.48\columnwidth}
\centering
    	\begin{tikzpicture}[scale=0.65]
		\begin{axis}[%
			xlabel={Displacement, mm},
			ylabel={Applied load, N},
                xlabel style={font=\large},
                ylabel style={font=\large},
                xmin=0.0, xmax=0.50,
                ymin=0.0, ymax=50.0,
                xtick distance=0.1, 
                ytick distance=10.0, 
                legend style={at={(0.325,1.0)}, anchor=north west, column sep=1ex,
                legend cell align=left, font=\small},
			table/col sep=comma,height=9cm]

            \addplot[black,mark=none, line width=1.0pt]  table [x=-v, y=F] {4.Results_data/modeI_propagation_60x50x10/3Dsolids_alpha25/modeIprop_3Dsolids_0.5mm.txt};
		      \addlegendentry{Reference solution}

            \addplot[blue,mark=x, line width=1.0pt]  table [x=-v, y=F] {4.Results_data/modeI_propagation_60x50x10/Shell6DofsBF_alpha5/modeIprop_shell6DofsBF_0.5mm.txt};
		      \addlegendentry{Structural CE model 0.5mm}

            \addplot[teal,mark=x, line width=1.0pt]  table [x=-v, y=F] {4.Results_data/modeI_propagation_60x50x10/Shell6DofsBF_alpha5/modeIprop_shell6DofsBF_1.0mm.txt};
		      \addlegendentry{Structural CE model 1.0mm}

            \addplot[red,mark=x, line width=1.0pt]  table [x=-v, y=F] {4.Results_data/modeI_propagation_60x50x10/Shell6DofsBF_alpha5/modeIprop_shell6DofsBF_2.0mm.txt};
		      \addlegendentry{Structural CE model 2.0mm}
		\end{axis}
	\end{tikzpicture}
\caption*{(b) Structural CE models, $\alpha_p=5$}
\end{minipage}
\caption{Load–displacement curves for the structural CE models.}
\label{fig:FvsD_results_Structural}
\end{figure}

\noindent The results for meshes of different element sizes are plotted together with a reference solution, obtained using a solid CE model with a fine mesh. Two values were considered for the penalty stiffness parameter $\alpha_{p}$ in Eq. \ref{Penalty_Stiffness}: $\alpha_{p}=50$, as suggested in \cite{Turonetal2007}, and $\alpha_{p}=5$. Although not shown herein, higher values of $\alpha_{p}$, such as $\alpha_{p}=100$, led to increasingly unstable cohesive responses and larger errors in the predicted peak load.

Fig. \ref{fig:FvsD_results_3D} shows that the standard CE model requires element sizes below 0.5 mm to achieve accurate results. For meshes of 1.0 mm and 2.0 mm, the peak-load error exceeds 9\% and 28\%, respectively, and the post-peak response becomes highly unstable. In contrast, Fig. \ref{fig:FvsD_results_Structural} shows that the proposed structural CE model remains in close agreement with the reference solution even with the 2.0 mm mesh, with peak-load errors below 8\% and a stable post-peak response.

Minor post-peak oscillations observed for coarser meshes are attributed to the larger integration-point spacing. The post-peak oscillations are significantly reduced {\color{blue}when} $\alpha_p=5$, indicating that lower penalty parameters than those recommended in \cite{Turonetal2007} for standard CEs can be effectively used. The load-displacement curves obtained from the simulations of the 120 mm $\times$ 50 mm models are shown in Figure \ref{fig:FvsD_results_3D_vs_Structural}. Despite minor peak-load differences, both models yield comparable responses, albeit at markedly different computational costs.

\begin{figure}[H]
\centering
    \begin{tikzpicture}[scale=0.60]
		\begin{axis}[%
            width=12cm,          
            height=10cm,          
			xlabel={Displacement, mm},
			ylabel={Applied load, N},
                xlabel style={font=\large},
                ylabel style={font=\large},
                xmin=0.0, xmax=1.0,
                ymin=0.0, ymax=40.0,
                xtick distance=0.2, 
                ytick distance=5.0, 
                legend style={at={(0.425,1.0)}, anchor=north west, column sep=1ex,
                legend cell align=left, font=\small},
			table/col sep=comma,height=9cm]

            \addplot[black,mark=x, line width=1.0pt]  table [x=-v, y=F] {4.Results_data/modeI_propagation_120x50x10/3Dsolids_alpha50/modeIprop_3Dsolids_0.5mm.txt};
		      \addlegendentry{Solid CE model 0.5mm}

            \addplot[red,mark=x, line width=1.0pt]  table [x=-v, y=F] {4.Results_data/modeI_propagation_120x50x10/Shell6DofsBF_alpha5/modeIprop_shell6DofsBF_2.0mm.txt};
		      \addlegendentry{Structural CE model 2.0mm}
		\end{axis}
	\end{tikzpicture}
\caption{Load–displacement curves - Solid CEs versus Structural CEs.}
\label{fig:FvsD_results_3D_vs_Structural}
\end{figure}

\subsubsection{\color{blue}Cohesive stresses}

{\color{blue}The cohesive stresses along the interface central line $\Gamma_c$ during crack propagation are presented in Figure \ref{benchmarks_CohesiveStresses_results} for the solid CE and structural CE models. The results show that the structural CE model with the coarse mesh, with an element size of 2.0 mm, delivered cohesive traction profiles similar to those provided by the most refined models. On the other hand, the solid CE model with the coarse mesh, with an element size of 2.0 mm, resulted in highly oscillatory traction profiles (see Figs. \ref{benchmarks_CohesiveStresses_results}a, \ref{benchmarks_CohesiveStresses_results}c and \ref{benchmarks_CohesiveStresses_results}e). 

It can be observed from Figure \ref{benchmarks_CohesiveStresses_results} that the numerically obtained cohesive zone length is smaller than 0.5 mm. The black vertical lines in Fig. \ref{benchmarks_CohesiveStresses_results} indicate the 2.0 mm spacing between the element boundaries adopted for the coarse mesh models. It is interesting to observe how the high-gradient traction profile in the cohesive region can be reproduced within a single high-order structural CE, which is the main reason for the superior performance of the proposed model on coarse meshes.

These results, together with the analytically obtained cohesive zone lengths for the benchmark tests and their corresponding structural responses in Figure \ref{benchmarks_LvsD_results2}, allow us to conclude that the proposed model is able to deliver very satisfactory responses, both in terms of load-displacement curves and cohesive traction profiles, for models with element sizes equal to or greater than the estimated cohesive zone length.}

\begin{figure}[H]
\centering

\begin{minipage}[t]{0.48\columnwidth}
\centering
	\begin{tikzpicture}[scale=0.60]
		\begin{axis}[%
        xlabel={Interface coordinate, mm},
        ylabel={Traction $\mathrm{t}_{\mathrm{I}}$, $\mathrm{MPa}$},
            xlabel style={font=\large},
            ylabel style={font=\large},
            xmin=0.0, xmax=60.0,
            ymin=-40.0, ymax=80.0,
            xtick distance=10.0, 
            ytick distance=20.0, 
                legend style={at={(0.0,1.00)}, anchor=north west, column sep=1ex,
                legend cell align=left, font=\small},
			table/col sep=comma,height=9cm]

            \addplot[blue, line width=1.0pt]  table [x=S, y=TI] {4.Results_data/modeI_propagation_60x50x10/3Dsolids_alpha50/Traction_3Dsolids_0.5mm_step1.txt};
            \addlegendentry{Solid CE model 0.5mm}

            \addplot[red, dashed, line width=3.0pt]  table [x=S, y=TI] {4.Results_data/modeI_propagation_60x50x10/3Dsolids_alpha50/Traction_3Dsolids_2.0mm_step1.txt};
            \addlegendentry{Solid CE model 2.0mm}

            \addplot[black, mark=|, mark size=3pt, line width=1.0pt]  table [x=S, y=o] {4.Results_data/modeI_propagation_60x50x10/Shell6DofsBF_alpha5/2.0mm_elements.txt};

		\end{axis}
	\end{tikzpicture}
\caption*{{\color{blue}(a) Solid CE - intermediary step 1}}
\end{minipage}
\hfill
\begin{minipage}[t]{0.48\columnwidth}
\centering
	\begin{tikzpicture}[scale=0.60]
		\begin{axis}[%
			xlabel={Interface coordinate, mm},
			ylabel={Traction $\mathrm{t}_{\mathrm{I}}$, $\mathrm{MPa}$},
                xlabel style={font=\large},
                ylabel style={font=\large},
                xmin=0.0, xmax=60.0,
                ymin=-40.0, ymax=80.0,
                xtick distance=10.0, 
                ytick distance=20.0, 
                legend style={at={(0.0,1.00)}, anchor=north west, column sep=1ex,
                legend cell align=left, font=\small},
			table/col sep=comma,height=9cm]

            \addplot[blue, line width=1.0pt]  table [x=S, y=TI] {4.Results_data/modeI_propagation_60x50x10/Shell6DofsBF_alpha5/Traction_shell6DofsBF_0.5mm_step1.txt};
            \addlegendentry{Structural CE model 0.5mm}

            \addplot[red, dashed, line width=3.0pt]  table [x=S, y=TI] {4.Results_data/modeI_propagation_60x50x10/Shell6DofsBF_alpha5/Traction_shell6DofsBF_2.0mm_step1.txt};
            \addlegendentry{Structural CE model 2.0mm}

            \addplot[black, mark=|, mark size=3pt, line width=1.0pt]  table [x=S, y=o] {4.Results_data/modeI_propagation_60x50x10/Shell6DofsBF_alpha5/2.0mm_elements.txt};
              
		\end{axis}
	\end{tikzpicture}
\caption*{{\color{blue}(b) Structural CE - intermediary step 1}}
\end{minipage}

\vspace{5pt}

\begin{minipage}[t]{0.48\columnwidth}
\centering
	\begin{tikzpicture}[scale=0.60]
		\begin{axis}[%
        xlabel={Interface coordinate, mm},
        ylabel={Traction $\mathrm{t}_{\mathrm{I}}$, $\mathrm{MPa}$},
            xlabel style={font=\large},
            ylabel style={font=\large},
            xmin=0.0, xmax=60.0,
            ymin=-40.0, ymax=80.0,
            xtick distance=10.0, 
            ytick distance=20.0, 
                legend style={at={(0.0,1.00)}, anchor=north west, column sep=1ex,
                legend cell align=left, font=\small},
			table/col sep=comma,height=9cm]

            \addplot[blue, line width=1.0pt]  table [x=S, y=TI] {4.Results_data/modeI_propagation_60x50x10/3Dsolids_alpha50/Traction_3Dsolids_0.5mm_step2.txt};
            \addlegendentry{Solid CE model 0.5mm}

            \addplot[red, dashed, line width=3.0pt]  table [x=S, y=TI] {4.Results_data/modeI_propagation_60x50x10/3Dsolids_alpha50/Traction_3Dsolids_2.0mm_step2.txt};
            \addlegendentry{Solid CE model 2.0mm}

            \addplot[black, mark=|, mark size=3pt, line width=1.0pt]  table [x=S, y=o] {4.Results_data/modeI_propagation_60x50x10/Shell6DofsBF_alpha5/2.0mm_elements.txt};
              
		\end{axis}
	\end{tikzpicture}
\caption*{\color{blue}(c) Solid CE - intermediary step 2}
\end{minipage}
\hfill
\begin{minipage}[t]{0.48\columnwidth}
\centering
	\begin{tikzpicture}[scale=0.60]
		\begin{axis}[%
        xlabel={Interface coordinate, mm},
        ylabel={Traction $\mathrm{t}_{\mathrm{I}}$, $\mathrm{MPa}$},
            xlabel style={font=\large},
            ylabel style={font=\large},
            xmin=0.0, xmax=60.0,
            ymin=-40.0, ymax=80.0,
            xtick distance=10.0, 
            ytick distance=20.0, 
                legend style={at={(0.0,1.00)}, anchor=north west, column sep=1ex,
                legend cell align=left, font=\small},
			table/col sep=comma,height=9cm]

            \addplot[blue, line width=1.0pt]  table [x=S, y=TI] {4.Results_data/modeI_propagation_60x50x10/Shell6DofsBF_alpha5/Traction_shell6DofsBF_0.5mm_step2.txt};
            \addlegendentry{Structural CE model 0.5mm}

            \addplot[red, dashed, line width=3.0pt]  table [x=S, y=TI] {4.Results_data/modeI_propagation_60x50x10/Shell6DofsBF_alpha5/Traction_shell6DofsBF_2.0mm_step2.txt};
            \addlegendentry{Structural CE model 2.0mm}

            \addplot[black, mark=|, mark size=3pt, line width=1.0pt]  table [x=S, y=o] {4.Results_data/modeI_propagation_60x50x10/Shell6DofsBF_alpha5/2.0mm_elements.txt};

		\end{axis}
	\end{tikzpicture}
\caption*{\color{blue}(d) Structural CE - intermediary step 2}
\end{minipage}

\vspace{5pt}

\begin{minipage}[t]{0.48\columnwidth}
\centering
	\begin{tikzpicture}[scale=0.60]
		\begin{axis}[%
        xlabel={Interface coordinate, mm},
        ylabel={Traction $\mathrm{t}_{\mathrm{I}}$, $\mathrm{MPa}$},
            xlabel style={font=\large},
            ylabel style={font=\large},
            xmin=0.0, xmax=60.0,
            ymin=-40.0, ymax=80.0,
            xtick distance=10.0, 
            ytick distance=20.0, 
                legend style={at={(0.0,1.00)}, anchor=north west, column sep=1ex,
                legend cell align=left, font=\small},
			table/col sep=comma,height=9cm]

            \addplot[blue, line width=1.0pt]  table [x=S, y=TI] {4.Results_data/modeI_propagation_60x50x10/3Dsolids_alpha50/Traction_3Dsolids_0.5mm_step2_other.txt};
            \addlegendentry{Solid CE model 0.5mm}

            \addplot[red, dashed, line width=3.0pt]  table [x=S, y=TI] {4.Results_data/modeI_propagation_60x50x10/3Dsolids_alpha50/Traction_3Dsolids_2.0mm_step2_other.txt};
            \addlegendentry{Solid CE model 2.0mm}

            \addplot[black, mark=|, mark size=3pt, line width=1.0pt]  table [x=S, y=o] {4.Results_data/modeI_propagation_60x50x10/Shell6DofsBF_alpha5/2.0mm_elements.txt};
              
		\end{axis}
	\end{tikzpicture}
\caption*{\color{blue}(e) Solid CE - intermediary step 3}
\end{minipage}
\hfill
\begin{minipage}[t]{0.48\columnwidth}
\centering
	\begin{tikzpicture}[scale=0.60]
		\begin{axis}[%
        xlabel={Interface coordinate, mm},
        ylabel={Traction $\mathrm{t}_{\mathrm{I}}$, $\mathrm{MPa}$},
            xlabel style={font=\large},
            ylabel style={font=\large},
            xmin=0.0, xmax=60.0,
            ymin=-40.0, ymax=80.0,
            xtick distance=10.0, 
            ytick distance=20.0, 
                legend style={at={(0.0,1.00)}, anchor=north west, column sep=1ex,
                legend cell align=left, font=\small},
			table/col sep=comma,height=9cm]

            \addplot[blue, line width=1.0pt]  table [x=S, y=TI] {4.Results_data/modeI_propagation_60x50x10/Shell6DofsBF_alpha5/Traction_shell6DofsBF_0.5mm_step2_other.txt};
            \addlegendentry{Structural CE model 0.5mm}

            \addplot[red, dashed, line width=3.0pt]  table [x=S, y=TI] {4.Results_data/modeI_propagation_60x50x10/Shell6DofsBF_alpha5/Traction_shell6DofsBF_2.0mm_step2_other.txt};
            \addlegendentry{Structural CE model 2.0mm}

            \addplot[black, mark=|, mark size=3pt, line width=1.0pt]  table [x=S, y=o] {4.Results_data/modeI_propagation_60x50x10/Shell6DofsBF_alpha5/2.0mm_elements.txt};
              
		\end{axis}
	\end{tikzpicture}
\caption*{\color{blue}(f) Structural CE - intermediary step 3}
\end{minipage}
\caption{\color{blue}Cohesive traction profiles: Solid CEs versus Structural CEs.}
\label{benchmarks_CohesiveStresses_results}
\end{figure}

\subsubsection{Computational performance}

The proposed structural model is able to reduce the computational time of the debonding simulations considerably by allowing larger elements to be used. The results of the structural model are compared against those of the solid model. The CPU times for the analysis of the 60 mm $\times$ 50 mm models are reported in Table \ref{CPU_time_60x50_ap_50}.

\begin{table}[h!]
\centering
\fontsize{9}{11}\selectfont
\caption{CPU time for the 60 mm $\times$ 50 mm models.}
\begin{tabular}{lccc}
    \hline
    \textbf{Models $\alpha_{p}=50$} & \textbf{3D CE 0.5\,mm} & \textbf{3D  CE 1.0\,mm} & \textbf{3D CE 2.0\,mm} \\
    \hline
    CPU time (s) & 4292. & 528.4 & 39.9 \\
    \hline 
    \textbf{Models $\alpha_{p}=50$} & \textbf{Struc. CE 0.5\,mm} & \textbf{Struc. CE 1.0\,mm} & \textbf{Struc. CE 2.0\,mm} \\
    \hline 
    CPU time (s) & 7418. & 1235. & 147.7 \\
    \hline 
    \textbf{Models $\alpha_{p}=5$} & \textbf{Struc. CE 0.5\,mm} & \textbf{Struc. CE 1.0\,mm} & \textbf{Struc. CE 2.0\,mm} \\
    \hline 
    CPU time (s) & 7563. & 791.1 & 90.9 \\
    \hline 
\end{tabular}
\label{CPU_time_60x50_ap_50}
\end{table}

A comparison between the 3D solid model with a 0.5 mm mesh and the structural model with a 2.0 mm mesh shows that the structural model reduces the computational time by more than 95\%, while still providing accurate predictions of the peak load and the overall response. The CPU time for the analysis of the 120 mm $\times$ 50 mm models are shown in Table \ref{CPU_time_3D_vs_StructuralCE_120x50}.
\begin{table}[H]
\centering
\fontsize{9}{11}\selectfont
\caption{CPU time for the 120 mm $\times$ 50 mm models.}
\begin{tabular}{lcc}
    \hline
    \textbf{Models}& \textbf{3D CE 0.5\,mm} $\alpha_{p} = 50$ & \textbf{Struc. CE 2.0\,mm} $\alpha_{p} = 5$\\
    \hline
    CPU time (s) & 9263. & 434.2\\
    \hline 
\end{tabular}
\label{CPU_time_3D_vs_StructuralCE_120x50}
\end{table}

\noindent Once again, the comparison between the 3D solid model and the structural model shows that the structural CE approach reduces the computational time of the debonding analysis by more than 95\%.

\subsubsection{\color{blue}Study on numerical integration}
This section examines the influence of the number of integration points on the simulation results, using the error in the load–displacement curves relative to the reference solution as the evaluation metric. The analysis is restricted to the structural CE model with a 2.0 mm mesh and $\alpha_{p}=5$. The corresponding load–displacement curves obtained with different numbers of integration points are shown in Figure \ref{fig:FvsD_results_integration_points}. The numerical integration of the cohesive element equations, described in Section \ref{subsec:Cohesive element equation}, may be performed using different numbers of integration points in both the through-thickness and crack-propagation directions. The influence of the number of integration points in each direction is shown in Figures \ref{fig:FvsD_results_integration_points}(a) and \ref{fig:FvsD_results_integration_points}(b), respectively.

\begin{figure}[H]
\centering

\begin{minipage}[t]{0.48\columnwidth}
\centering
    	\begin{tikzpicture}[scale=0.65]
		\begin{axis}[%
			xlabel={Displacement, mm},
			ylabel={Applied load, N},
                xlabel style={font=\large},
                ylabel style={font=\large},
                xmin=0.0, xmax=0.50,
                ymin=0.0, ymax=50.0,
                xtick distance=0.1, 
                ytick distance=10.0, 
                legend style={at={(0.40,1.0)}, anchor=north west, column sep=1ex,
                legend cell align=left, font=\small},
			table/col sep=comma,height=9cm]

            \addplot[black,mark=none, line width=1.0pt]  table [x=-v, y=F] {4.Results_data/modeI_propagation_60x50x10/3Dsolids_alpha25/modeIprop_3Dsolids_0.5mm.txt};
		      \addlegendentry{Reference solution}

            \addplot[red,mark=x, line width=1.0pt]  table [x=-v, y=F] {4.Results_data/modeI_propagation_60x50x10/Shell6DofsBF_alpha5/modeIprop_shell6DofsBF_2.0mm_13x1.txt};
		      \addlegendentry{13 x 1 integration points}

            \addplot[teal,mark=x, line width=1.0pt]  table [x=-v, y=F] {4.Results_data/modeI_propagation_60x50x10/Shell6DofsBF_alpha5/modeIprop_shell6DofsBF_2.0mm_13x3.txt};
		      \addlegendentry{13 x 3 integration points}

            \addplot[blue,mark=x, line width=1.0pt]  table [x=-v, y=F] {4.Results_data/modeI_propagation_60x50x10/Shell6DofsBF_alpha5/modeIprop_shell6DofsBF_2.0mm.txt};
		      \addlegendentry{13 x 5 integration points}
		\end{axis}
	\end{tikzpicture}
\caption*{(a) Through the thickness integration.}
\end{minipage}
\hfill
\begin{minipage}[t]{0.48\columnwidth}
\centering
    	\begin{tikzpicture}[scale=0.65]
		\begin{axis}[%
			xlabel={Displacement, mm},
			ylabel={Applied load, N},
                xlabel style={font=\large},
                ylabel style={font=\large},
                xmin=0.0, xmax=0.50,
                ymin=0.0, ymax=50.0,
                xtick distance=0.1, 
                ytick distance=10.0, 
                legend style={at={(0.40,1.0)}, anchor=north west, column sep=1ex,
                legend cell align=left, font=\small},
			table/col sep=comma,height=9cm]

            \addplot[black,mark=none, line width=1.0pt]  table [x=-v, y=F] {4.Results_data/modeI_propagation_60x50x10/3Dsolids_alpha25/modeIprop_3Dsolids_0.5mm.txt};
		      \addlegendentry{Reference solution}

            \addplot[red,mark=x, line width=1.0pt]  table [x=-v, y=F] {4.Results_data/modeI_propagation_60x50x10/Shell6DofsBF_alpha5/modeIprop_shell6DofsBF_2.0mm_9x1.txt};
		      \addlegendentry{9 x 1 integration points}

            \addplot[teal,mark=x, line width=1.0pt]  table [x=-v, y=F] {4.Results_data/modeI_propagation_60x50x10/Shell6DofsBF_alpha5/modeIprop_shell6DofsBF_2.0mm_11x1.txt};
		      \addlegendentry{11 x 1 integration points}

            \addplot[blue,mark=x, line width=1.0pt]  table [x=-v, y=F] {4.Results_data/modeI_propagation_60x50x10/Shell6DofsBF_alpha5/modeIprop_shell6DofsBF_2.0mm_13x1.txt};
		      \addlegendentry{13 x 1 integration points}

		\end{axis}
	\end{tikzpicture}
\caption*{(b) Integration in the propagation direction.}
\end{minipage}
\caption{Numerical responses for the 2.0 mm mesh structural CE models, $\alpha_{p}=5$.}
\label{fig:FvsD_results_integration_points}
\end{figure}

Figure \ref{fig:FvsD_results_integration_points}(a) shows that, for this Mode I–dominated debonding test, increasing the number of integration points through the stiffener thickness has no effect on the predicted response. In contrast, the number of integration points in the crack-propagation direction has a minor influence, as illustrated in Figure \ref{fig:FvsD_results_integration_points}(b). Analyses with seven or fewer integration points in the propagation direction did not converge beyond the peak load. As reported in \cite{Aietal2025}, increasing the number of integration points improves accuracy for coarse meshes but has negligible impact once the mesh is sufficiently refined, which appears to be the case for the 2.0 mm mesh considered here.

\subsection{End-notched flexure test in a complex stiffened panel}

\begin{figure}[H]
\centering
\scalebox{0.25}{\includegraphics{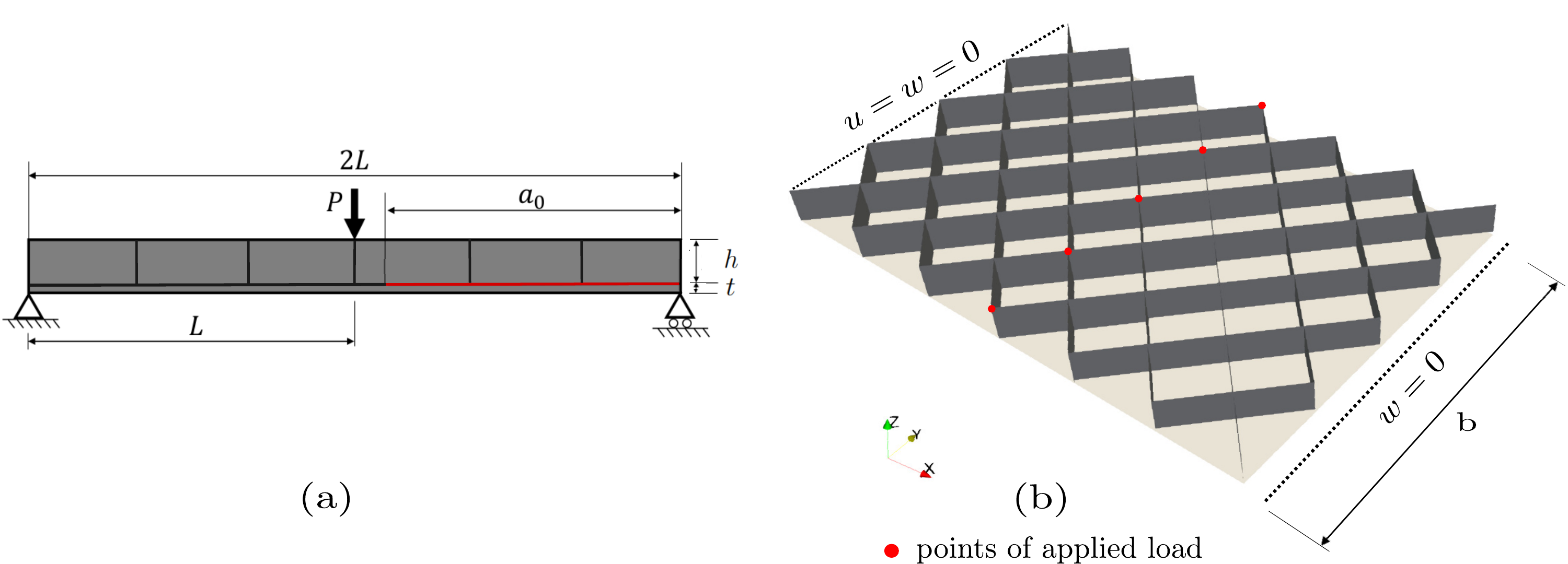}}
    \vspace{-5pt}
	\caption{Complex stiffened panel model and boundary conditions for the ENF test.}
	\label{fig:Complex stiffened panel}
\end{figure}

The end-notched flexure (ENF) complex stiffened panel test is illustrated in Figure \ref{fig:Complex stiffened panel}, with its geometrical parameters listed in Table \ref{ENF_complex_panel_parameters}. Figure \ref{fig:Complex stiffened panel}(a) shows a side view of the ENF-like test setup, while the complex stiffened configuration of the panel is shown in Figures \ref{fig:Complex stiffened panel}(b) in which the red dots indicate points of applied load. 
\vspace{-20pt}
\vspace{-20pt}
\begin{table}[H]
\centering
\footnotesize
\caption{Geometric parameters for the complex panel ENF panel test, parameters in $~\mathrm{mm}$.}
\begin{tabular}{@{}cccccc@{}}
    \toprule
    &   $2L$        & $a_0$  &  $h$ & $b$ & $t \text{(thickness)}$\\ \midrule
      & $150.0$ & $50.0$ & $10.0$  & $100.0$ & $2.5$  \\
    \bottomrule
\end{tabular}
\label{ENF_complex_panel_parameters}
\end{table}
\vspace{-5pt}
\noindent A pre-crack of length $a_{0}=50$ mm was introduced, cutting all skin–stiffener interfaces across the specimen width $b$. The thickness $t$ of the panel and the stiffeners was assumed to be 2.5 mm. {\color{blue} The problem is governed by the out-of-plane bending of the panel and the in-plane bending of the stiffeners. The panel has a thickness-to-span ratio of $t/2L=0.0167$, for which transverse shear effects are expected to be small. For thicker laminates, where transverse shear becomes significant, the multi-layer
modelling approach with compliant cohesive interfaces from \cite{Aietal2025} can still provide the transverse shear deformation needed in the laminate.} The elastic material properties adopted for the skin and the stiffener are those previously presented in Tables \ref{IM7/8552_parameters} and \ref{stiffener_parameters}, respectively. The interfacial fracture properties are also taken from Table \ref{IM7/8552_parameters}. These fracture properties are assumed to be the same as those adopted in \cite{Aietal2025} for the original ENF delamination tests.

\subsubsection{Description of the models}

\begin{figure}[H]
\centering
\scalebox{0.26}{\includegraphics{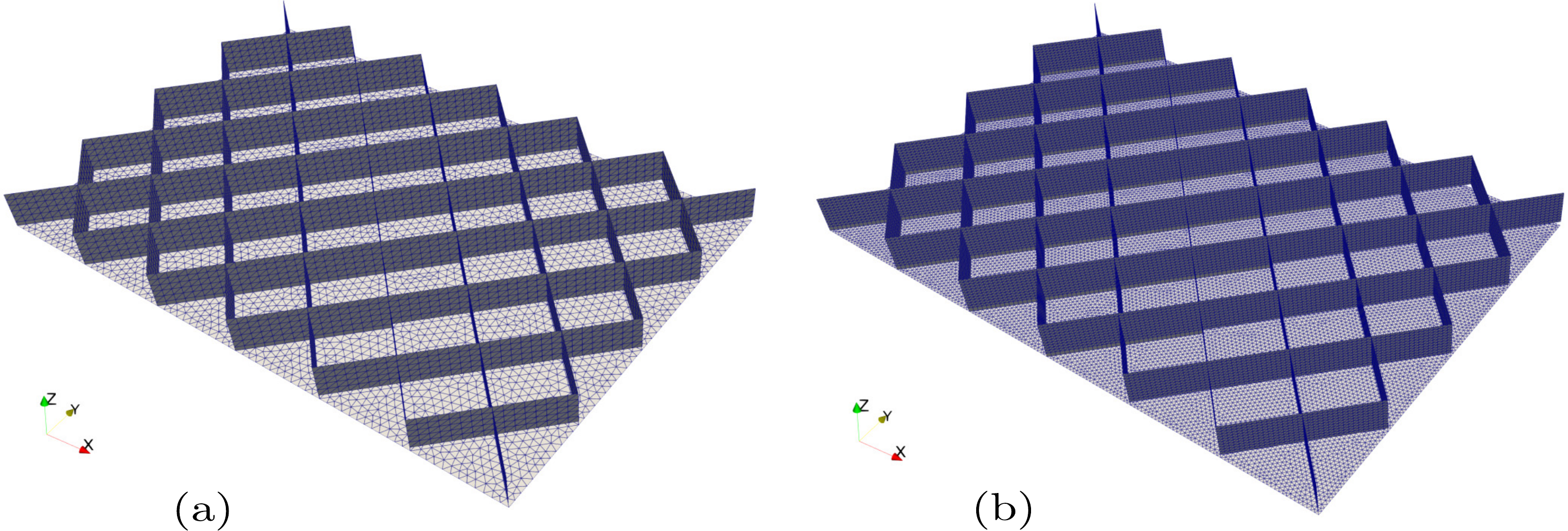}}
	\caption{Complex stiffened panel model and boundary conditions for the ENF test.}
	\label{fig:Complex stiffened panel_meshes}
\end{figure}

The structural finite element models employed shell elements for both the skin and stiffeners, with the proposed shell-to-shell {\color{blue}CEs} at the interfaces. Both components were discretized using a single layer of shell elements, and structural CEs were inserted in the uncracked region. Two mesh refinements were considered, with element sizes of 2.0 mm and 1.0 mm. Figure \ref{fig:Complex stiffened panel_meshes} illustrates the coarse and fine mesh models, which contain 8,485 and 36,110 nodes (corresponding to 50,910 and 216,660 DoFs), respectively.  

\subsubsection{Load–displacement curves}

{\color{blue}Figures \ref{fig:FvsD_results_ENF_complex_panel}(a) and \ref{fig:FvsD_results_ENF_complex_panel}(b) illustrate the load-displacement curves for the models with mesh elements of sizes 2.0 mm and 1.0 mm, respectively. The convergence of the structural response is observed by comparing the load-displacement responses of both models. Overall, the response resembles that of the original ENF test. However, several sharp snap-backs are observed, and the corresponding critical points are also illustrated in Fig. \ref{fig:FvsD_results_ENF_complex_panel}. The nature of these snapbacks is clarified by the damage evolution illustrated in Figures \ref{fig:Complex_panel_damage_map} and \ref{fig:Complex_panel_damage_mapB} for the 2.0 mm and 1.0 mm mesh models, respectively.}

Even before the first peak load, significant debonding occurs, and the abrupt load drop immediately afterward is associated with the formation of two new damaged regions (Fig. \ref{fig:Complex_panel_damage_map}a and {\color{blue}Fig. \ref{fig:Complex_panel_damage_mapB}a}).

\begin{figure}[H]
\centering

\begin{minipage}[t]{0.48\columnwidth}
\centering
    	\begin{tikzpicture}[scale=0.60]
		\begin{axis}[%
			xlabel={Displacement, mm},
			ylabel={Applied load, N},
                xlabel style={font=\large},
                ylabel style={font=\large},
                xmin=0.0, xmax=8.0,
                ymin=0.0, ymax=3500.0,
                xtick distance=1.0, 
                ytick distance=500.0,
                legend style={at={(0.0,1.0)}, anchor=north west, column sep=1ex,
                legend cell align=left, font=\small},
			table/col sep=comma,height=9cm]

            \addplot[red, dashed, line width=1.0pt]  table [x=-v, y=F] {4.Results_data/complex_panel/Grid45/Shell6DofsBF_alpha5/complexpanel_shell6DofsBF_2.0mm.txt};
            \addlegendentry{Structural CE model 2.0mm}

            \addplot[teal!60!black,mark=*, only marks, mark size=2pt]  table [x=-v, y=F] {4.Results_data/complex_panel/Grid45/Shell6DofsBF_alpha5/complexpanel_shell6DofsBF_2.0mm_critical_points.txt};

            \addplot[only marks, mark=none, nodes near coords, point meta=explicit symbolic, nodes near coords style={font=\large, anchor=west, xshift=-70pt, yshift=12pt}]
            table [x=-v, y=F, meta=label] {4.Results_data/complex_panel/Grid45/Shell6DofsBF_alpha5/complexpanel_shell6DofsBF_2.0mm_critical_point1.txt};

            \addplot[only marks, mark=none, nodes near coords, point meta=explicit symbolic, nodes near coords style={font=\large, anchor=west, xshift=0pt, yshift=4pt}]
            table [x=-v, y=F, meta=label] {4.Results_data/complex_panel/Grid45/Shell6DofsBF_alpha5/complexpanel_shell6DofsBF_2.0mm_critical_point2.txt};

            \addplot[only marks, mark=none, nodes near coords, point meta=explicit symbolic, nodes near coords style={font=\large, anchor=west, xshift=-5pt, yshift=12pt}]
            table [x=-v, y=F, meta=label] {4.Results_data/complex_panel/Grid45/Shell6DofsBF_alpha5/complexpanel_shell6DofsBF_2.0mm_critical_point3.txt};

            \addplot[only marks, mark=none, nodes near coords, point meta=explicit symbolic, nodes near coords style={font=\large, anchor=west, xshift=-6pt, yshift=12pt}]
            table [x=-v, y=F, meta=label] {4.Results_data/complex_panel/Grid45/Shell6DofsBF_alpha5/complexpanel_shell6DofsBF_2.0mm_critical_point4.txt};

            \addplot[only marks, mark=none, nodes near coords, point meta=explicit symbolic, nodes near coords style={font=\large, anchor=west, xshift=-5pt, yshift=12pt}]
            table [x=-v, y=F, meta=label] {4.Results_data/complex_panel/Grid45/Shell6DofsBF_alpha5/complexpanel_shell6DofsBF_2.0mm_critical_point5.txt};

            \addplot[only marks, mark=none, nodes near coords, point meta=explicit symbolic, nodes near coords style={font=\large, anchor=west, xshift=-21pt, yshift=10pt}]
            table [x=-v, y=F, meta=label] {4.Results_data/complex_panel/Grid45/Shell6DofsBF_alpha5/complexpanel_shell6DofsBF_2.0mm_critical_point6.txt};
		\end{axis}
	\end{tikzpicture}
\caption*{{\color{blue}(a) Load-displacement curve: 2.0 mm model}}
\end{minipage}
\hfill
\begin{minipage}[t]{0.48\columnwidth}
\centering
    	\begin{tikzpicture}[scale=0.60]
		\begin{axis}[%
			xlabel={Displacement, mm},
			ylabel={Applied load, N},
                xlabel style={font=\large},
                ylabel style={font=\large},
                xmin=0.0, xmax=8.0,
                ymin=0.0, ymax=3500.0,
                xtick distance=1.0, 
                ytick distance=500.0,
                legend style={at={(0.0,1.0)}, anchor=north west, column sep=1ex,
                legend cell align=left, font=\small},
			table/col sep=comma,height=9cm]



            \addplot[blue, dashed, line width=1.0pt]  table [x=-v, y=F] {4.Results_data/complex_panel/Grid45/Shell6DofsBF_alpha5/complexpanel_shell6DofsBF_1.0mm.txt};
		      \addlegendentry{Structural CE model 1.0mm}

            \addplot[teal!60!black,mark=*, only marks, mark size=2pt]  table [x=-v, y=F] {4.Results_data/complex_panel/Grid45/Shell6DofsBF_alpha5/complexpanel_shell6DofsBF_1.0mm_critical_points.txt};

            \addplot[only marks, mark=none, nodes near coords, point meta=explicit symbolic, nodes near coords style={font=\large, anchor=west, xshift=-70pt, yshift=12pt}]
            table [x=-v, y=F, meta=label] {4.Results_data/complex_panel/Grid45/Shell6DofsBF_alpha5/complexpanel_shell6DofsBF_1.0mm_critical_point1.txt};

            \addplot[only marks, mark=none, nodes near coords, point meta=explicit symbolic, nodes near coords style={font=\large, anchor=west, xshift=0pt, yshift=4pt}]
            table [x=-v, y=F, meta=label] {4.Results_data/complex_panel/Grid45/Shell6DofsBF_alpha5/complexpanel_shell6DofsBF_1.0mm_critical_point2.txt};

            \addplot[only marks, mark=none, nodes near coords, point meta=explicit symbolic, nodes near coords style={font=\large, anchor=west, xshift=-5pt, yshift=12pt}]
            table [x=-v, y=F, meta=label] {4.Results_data/complex_panel/Grid45/Shell6DofsBF_alpha5/complexpanel_shell6DofsBF_1.0mm_critical_point3.txt};

            \addplot[only marks, mark=none, nodes near coords, point meta=explicit symbolic, nodes near coords style={font=\large, anchor=west, xshift=7pt, yshift=25pt}]
            table [x=-v, y=F, meta=label] {4.Results_data/complex_panel/Grid45/Shell6DofsBF_alpha5/complexpanel_shell6DofsBF_1.0mm_critical_point4.txt};

            \addplot[only marks, mark=none, nodes near coords, point meta=explicit symbolic, nodes near coords style={font=\large, anchor=west, xshift=-10pt, yshift=-20pt}]
            table [x=-v, y=F, meta=label] {4.Results_data/complex_panel/Grid45/Shell6DofsBF_alpha5/complexpanel_shell6DofsBF_1.0mm_critical_point5.txt};

            \addplot[only marks, mark=none, nodes near coords, point meta=explicit symbolic, nodes near coords style={font=\large, anchor=west, xshift=-5pt, yshift=18pt}]
            table [x=-v, y=F, meta=label] {4.Results_data/complex_panel/Grid45/Shell6DofsBF_alpha5/complexpanel_shell6DofsBF_1.0mm_critical_point6.txt};
		\end{axis}
	\end{tikzpicture}
\caption*{{\color{blue}(b) Load-displacement curve: 1.0 mm model}}
\end{minipage}
\caption{{\color{blue}Load–displacement curves - ENF stiffened panel test.}}
\label{fig:FvsD_results_ENF_complex_panel}
\end{figure}

\noindent  Similar mechanisms govern the second and third load peaks, where the corresponding load drops are driven by the emergence of symmetric damaged regions (Figs. \ref{fig:Complex_panel_damage_map}b - \ref{fig:Complex_panel_damage_map}c, and {\color{blue}Figs. \ref{fig:Complex_panel_damage_mapB}b - \ref{fig:Complex_panel_damage_mapB}c}). The subsequent peaks are followed by the appearance of new asymmetric damaged regions (Figs. \ref{fig:Complex_panel_damage_map}d and \ref{fig:Complex_panel_damage_map}e) {\color{blue}for the 2.0 mm mesh model, and new symmetric damage regions (Figs. \ref{fig:Complex_panel_damage_mapB}d and \ref{fig:Complex_panel_damage_mapB}e) for the 1.0 mesh model.} Beyond a deflection of 7 mm, the damage pattern returns to an approximately symmetric configuration (Fig. \ref{fig:Complex_panel_damage_map}f and {\color{blue}Fig. \ref{fig:Complex_panel_damage_mapB}f}). 

\vspace{-10pt}

\begin{figure}[H]
\centering
\scalebox{0.27}{\includegraphics{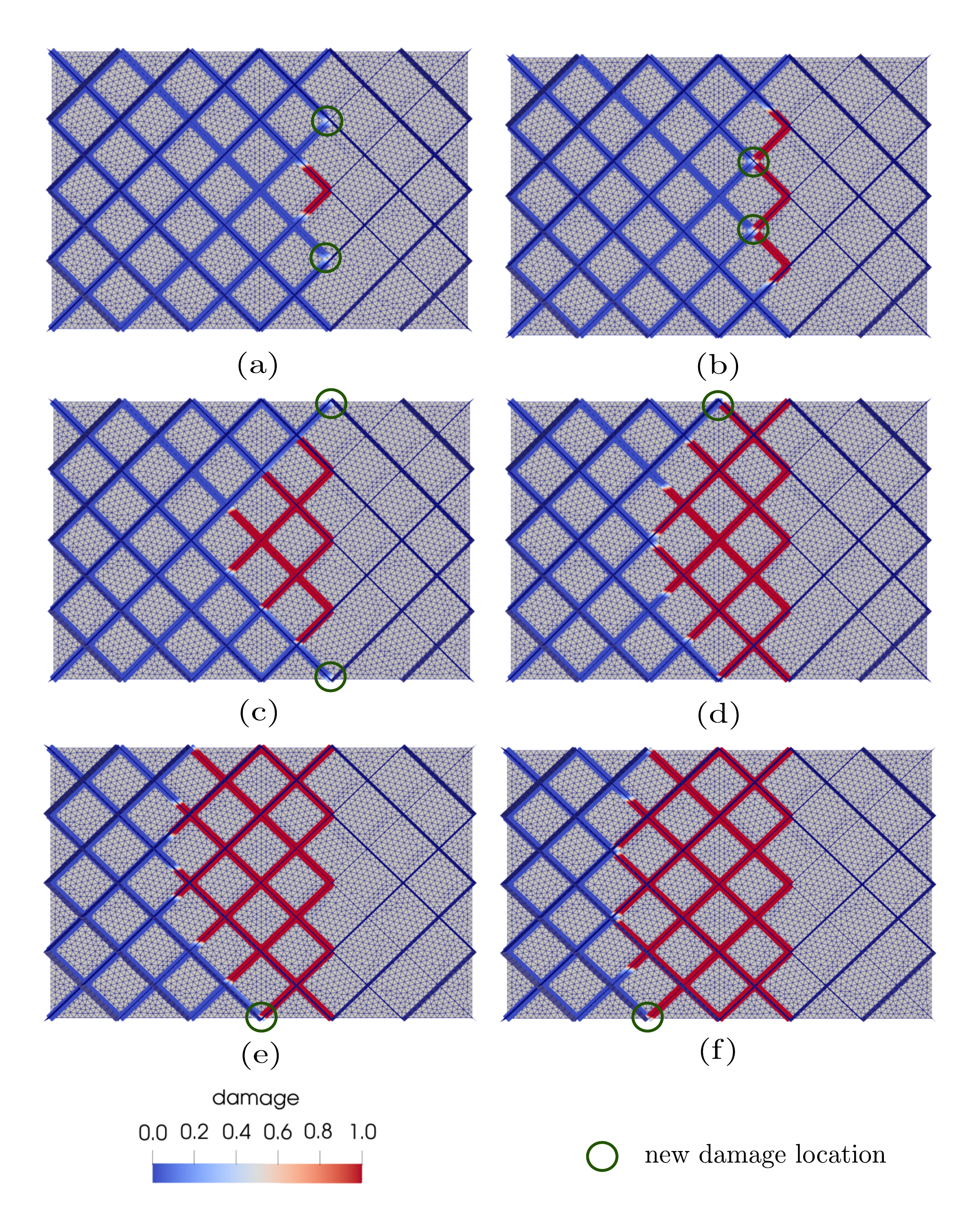}}
	\caption{Interface damage for different time steps - 2.0 mm model (see Fig. \ref{fig:FvsD_results_ENF_complex_panel}a).}
	\label{fig:Complex_panel_damage_map}
\end{figure}

\begin{figure}[H]
\centering
\scalebox{0.27}{\includegraphics{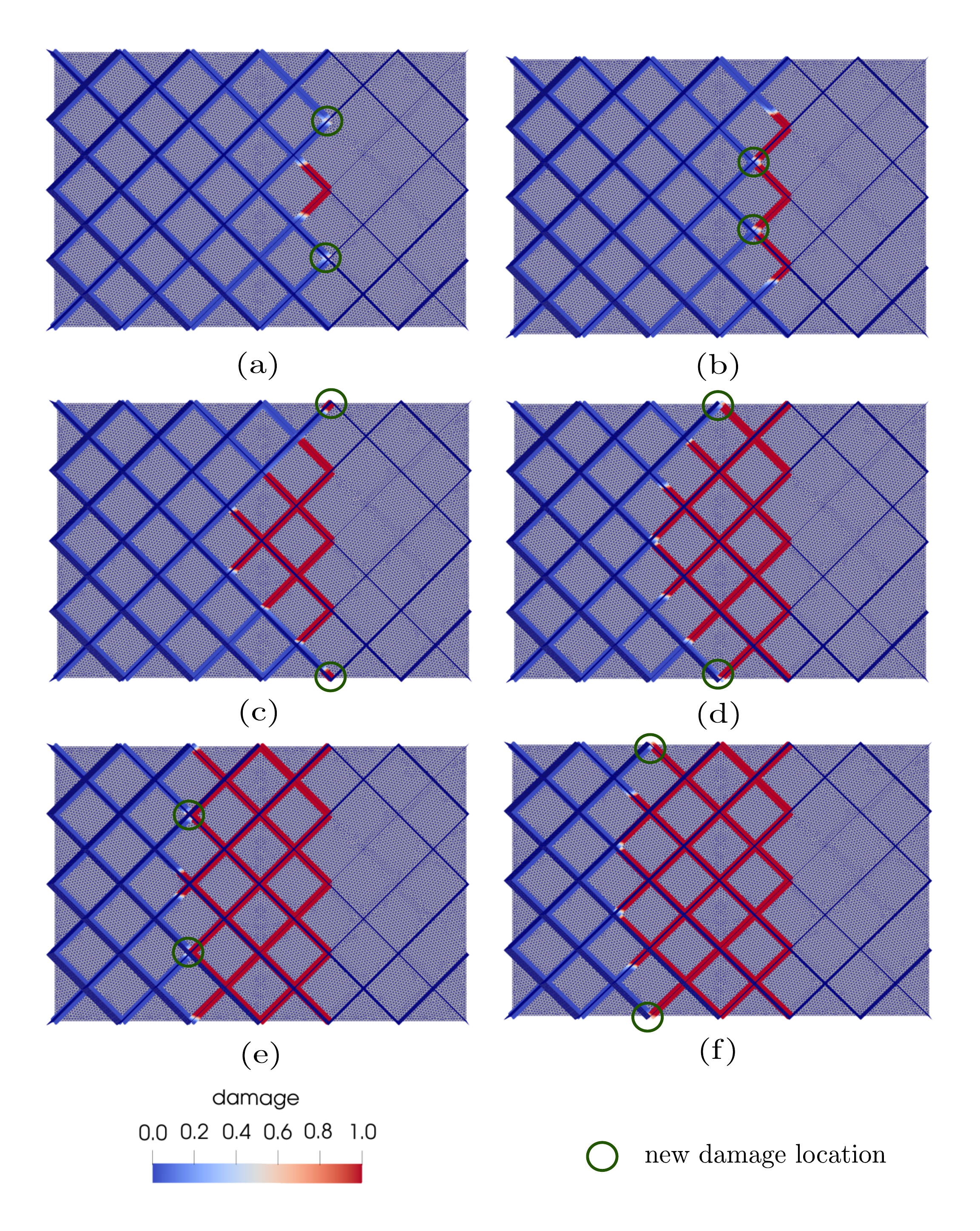}}
	\caption{{\color{blue}Interface damage for different time steps - 1.0 mm model (see Fig. \ref{fig:FvsD_results_ENF_complex_panel}b).}}
	\label{fig:Complex_panel_damage_mapB}
\end{figure}

The asymmetry observed in the numerical solution is initially triggered by asymmetry in the mesh. Minor asymmetry suffices for the extremely brittle failure events to occur sequentially on both sides of the panel instead of concurrently. For the model with the 1.0 mm mesh, the strong asymmetries previously observed in Figs. \ref{fig:Complex_panel_damage_map}(d) and \ref{fig:Complex_panel_damage_map}(e) did not occur, and the damage evolution was approximately symmetric {\color{blue}throughout} all load steps.

\begin{figure}[H]
\centering
\scalebox{0.27}{\includegraphics{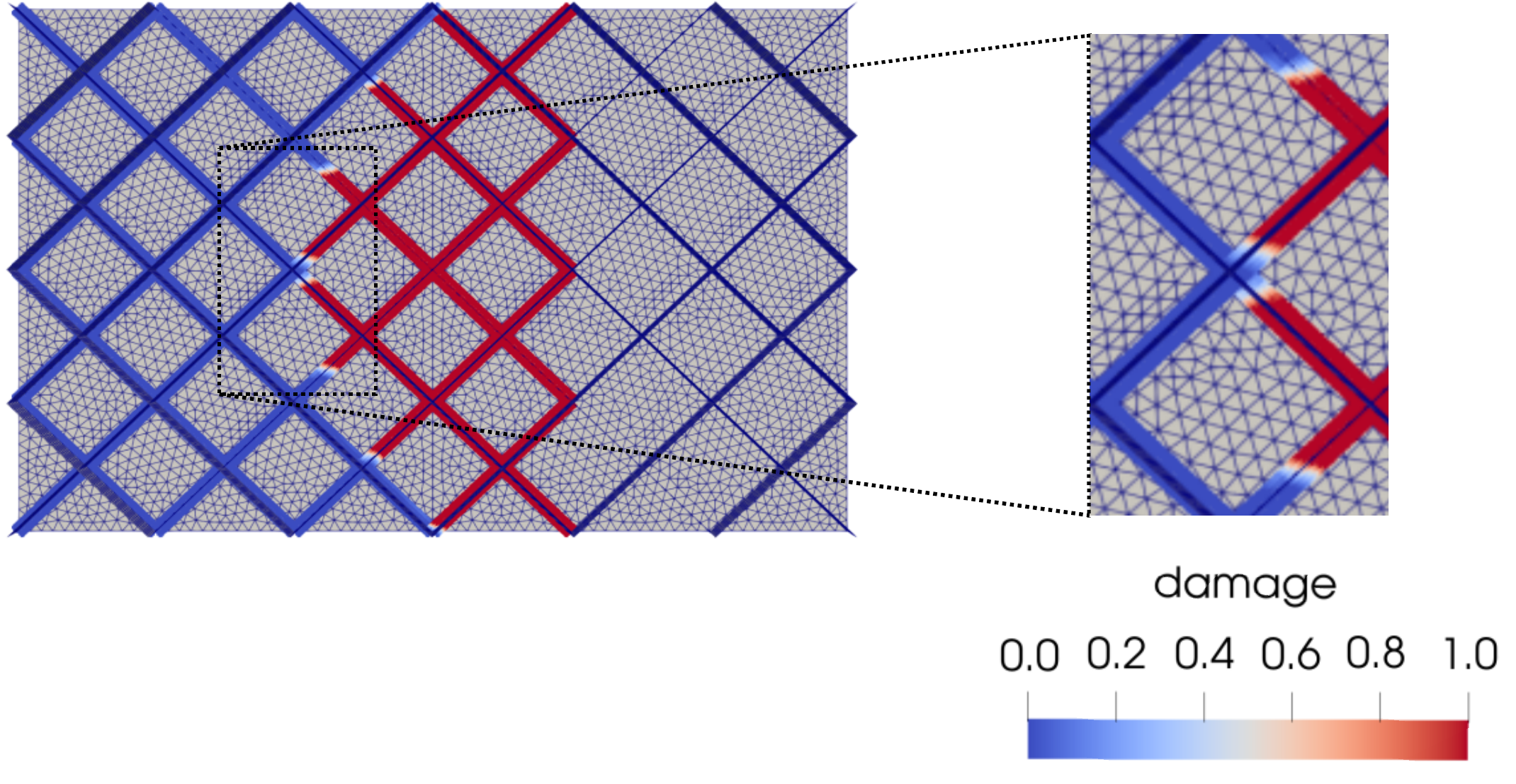}}
	\caption{{\color{blue}Interface damage - zoom view from Fig. \ref{fig:Complex_panel_damage_map}(d).}}
	\label{fig:Complex_panel_damage_mapC}
\end{figure}

{\color{blue}Additionally, Figure \ref{fig:Complex_panel_damage_mapC} shows a zoomed view of a region of interest, where it is possible to observe the cohesive zone - in which the damage field evolves from 0 to 1 - within the domain of a single element, as well as the non-constant damage values at opposite integration points through the stiffener thickness direction.}

{\color{blue}A more rigorous quantitative analysis of the stability and path-following sensitivity to improve the comprehension of the snap-back phenomena is carried out. To that end, the load-energy dissipation curves are presented in Figures \ref{benchmarks_LvsK_LvsD_results}(a) and \ref{benchmarks_LvsK_LvsD_results}(b) for the models with mesh element sizes of 2.0 mm and 1.0 mm, respectively.}

\begin{figure}[H]
\centering

\begin{minipage}[t]{0.48\columnwidth}
\centering
	\begin{tikzpicture}[scale=0.60]
		\begin{axis}[%
			xlabel={Total dissipation, N mm},
			ylabel={Applied load, N},
                xlabel style={font=\large},
                ylabel style={font=\large},
                xmin=0.0, xmax=1500.0,
                ymin=0.0, ymax=3500.0,
                xtick distance=250.0, 
                ytick distance=500.0,
                legend style={at={(0.0,1.0)}, anchor=north west, column sep=1ex,
                legend cell align=left, font=\small},
			table/col sep=comma,height=9cm]

            \addplot[red, dashed, line width=1.0pt]  table [x=Dis, y=F] {4.Results_data/complex_panel/Grid45/Shell6DofsBF_alpha5/lodi_dissipation_2.0mm.txt};

            \addplot[teal!60!black,mark=*, only marks, mark size=2pt]  table [x=Dis, y=F] {4.Results_data/complex_panel/Grid45/Shell6DofsBF_alpha5/lodi_dissipation_2.0mm_critical_points.txt};
		\end{axis}
	\end{tikzpicture}
\caption*{\color{blue}(a) load - energy dissipation: 2.0 mm model}
\end{minipage}
\hfill
\begin{minipage}[t]{0.48\columnwidth}
\centering
	\begin{tikzpicture}[scale=0.60]
		\begin{axis}[%
			xlabel={Total dissipation, N mm},
			ylabel={Applied load, N},
                xlabel style={font=\large},
                ylabel style={font=\large},
                xmin=0.0, xmax=1500.0,
                ymin=0.0, ymax=3500.0,
                xtick distance=250.0, 
                ytick distance=500.0,
                legend style={at={(0.0,1.0)}, anchor=north west, column sep=1ex,
                legend cell align=left, font=\small},
			table/col sep=comma,height=9cm]

            \addplot[blue, dashed, line width=1.0pt]  table [x=Dis, y=F] {4.Results_data/complex_panel/Grid45/Shell6DofsBF_alpha5/lodi_dissipation_1.0mm.txt};

            \addplot[teal!60!black,mark=*, only marks, mark size=2pt]  table [x=Dis, y=F] {4.Results_data/complex_panel/Grid45/Shell6DofsBF_alpha5/lodi_dissipation_1.0mm_critical_points.txt};
		\end{axis}
	\end{tikzpicture}
\caption*{\color{blue}(b) load - energy dissipation: 1.0 mm model}
\end{minipage}
\caption{\color{blue}load - energy dissipation curves.}
\label{benchmarks_LvsK_LvsD_results}
\end{figure}

{\color{blue}The critical load steps previously presented in Fig. \ref{fig:FvsD_results_ENF_complex_panel} are also indicated on those curves in order to clarify the physical nature of the snap-back phenomena. The energy dissipation is computed simply by summing up the work done by the cohesive tractions over the jump vector for all the integration points over all cohesive surfaces. The results presented in Figure \ref{benchmarks_LvsK_LvsD_results} show that the snap-backs occurring at the critical points are associated with significant energy dissipation, showing that the snapbacks are not numerical artifacts but related to physical crack growth events.}



\begin{figure}[H]
\centering
\scalebox{0.28}{\includegraphics{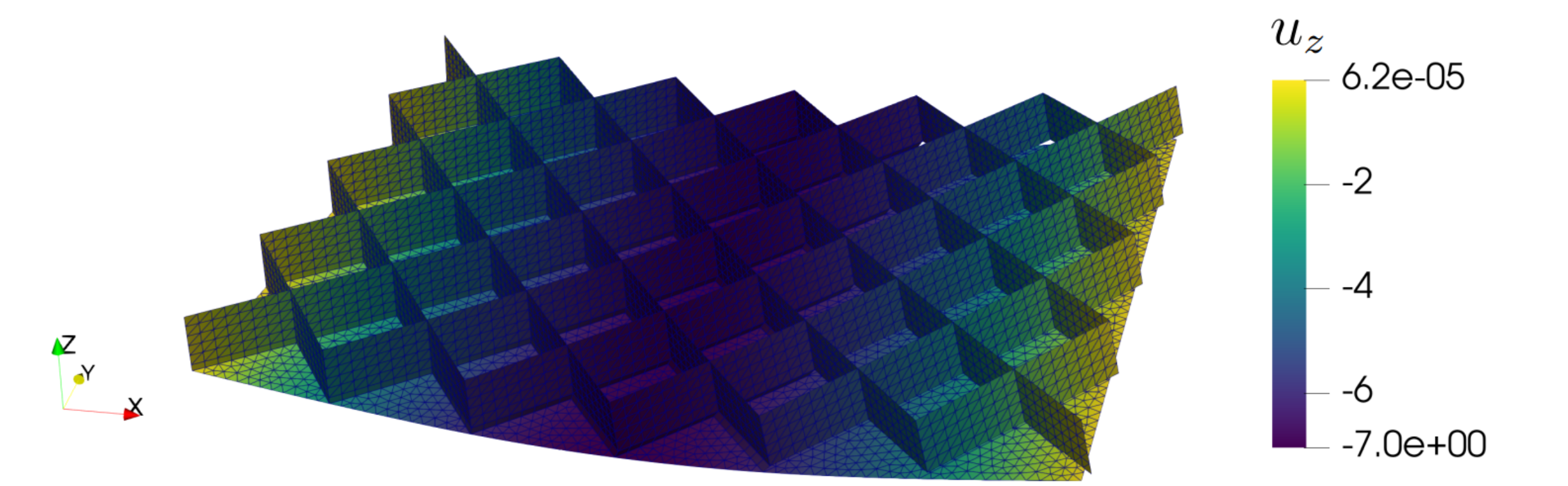}}
	\caption{Deformed configuration with unscaled displacements.}
	\label{fig:Complex_panel_deformed_configuration}
\end{figure}

Figure \ref{fig:Complex_panel_deformed_configuration} illustrates the deformed configuration for the model with a 2.0 mm mesh at a load step near the end of the analysis.

\subsubsection{Computational performance}

The CPU times for the analysis of the models are shown in Table \ref{CPU_time_Complex_panel}.
\begin{table}[H]
\centering
\fontsize{9}{11}\selectfont
\caption{CPU time for the ENF stiffened panel analyses.}
\begin{tabular}{lcc}
    \hline
    \textbf{Models} & \textbf{Struc. CE 1.0\,mm} & \textbf{Struc. CE 2.0\,mm}\\
    \hline
    CPU time (s) & 1.670e+05 \hspace{2pt} & 9757. \hspace{2pt}\\
    \hline 
\end{tabular}
\label{CPU_time_Complex_panel}
\end{table}
\vspace{-5pt}
\noindent The CPU time required for the 1.0 mm model {\color{blue}- 46.4 hours -} is prohibitive for structural engineering practice. In contrast, the 2.0 mm model provides essentially the same structural response at a feasible computational cost of {\color{blue}2.7 hours}.

\subsection{\color{blue}Pull-out test in a cross-notched complex stiffened}

\begin{figure}[H]
\centering
\scalebox{0.25}{\includegraphics{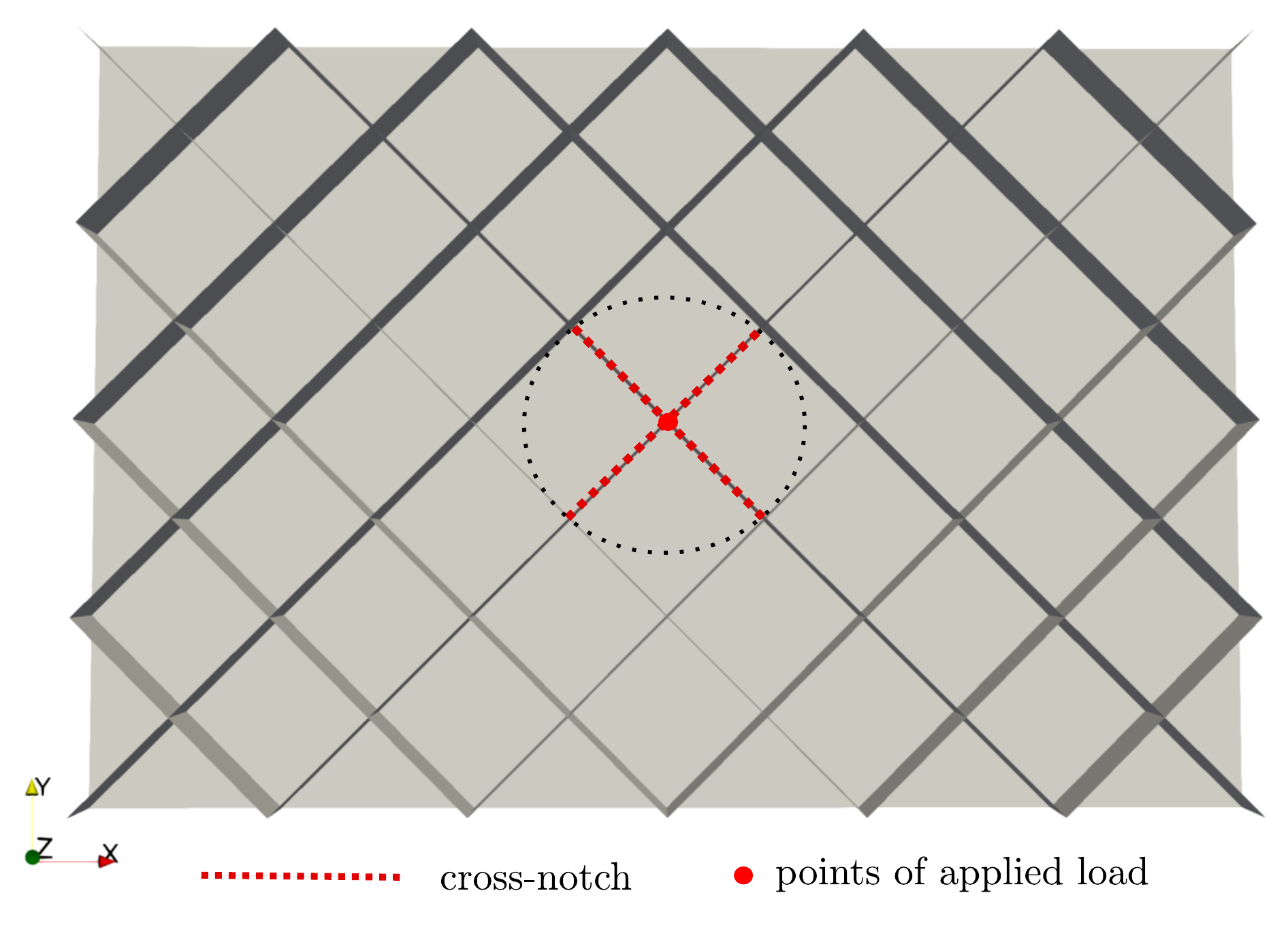}}
    \vspace{-5pt}
	\caption{\color{blue}Pull-out test in a cross-notched complex stiffened panel.}
	\label{fig:Complex stiffened panel_cross_notch}
\end{figure}

{\color{blue}In order to explore the robustness of the proposed element, the complex stiffened panel models from the previous section were studied under a different notch and load configuration. Instead of an initial edge rectangular notch and mid-span line load pattern, a central cross-notch within a circular region and a central point pull-out force were considered, with the panel fully constrained in the $w$ direction. Figure \ref{fig:Complex stiffened panel_cross_notch} presents the notch and load configuration considered for the analysis, in which the cross-notch is indicated by the red dashed line, and the central pull-out force is indicated by the red dot.} 
\vspace{-10pt}

{\color{blue}The propagation pattern for this type of notch/load configuration is known to produced radial propagation from the initial circular notch in delamination applications \cite{PradhanandTay1998}. A similar propagation pattern is also observed for the stiffener panel models studied herein, with cracks propagating in the radial direction from the initial circular region, as presented in the next subsection.}

\subsubsection{\color{blue}Load–displacement curves and damage propagation}

{\color{blue}Figures \ref{fig:FvsD_results_cross_notch_complex_panel}(a) and \ref{fig:FvsD_results_cross_notch_complex_panel}(b) illustrate the load-displacement curves for the models with mesh elements of sizes 2.0 mm and 1.0 mm, respectively. Very similar responses are obtained in terms of peak load for both models, with the advantage of the analysis with the coarse mesh model taking a fraction of the time to run compared to the analysis with the fine mesh model. The differences observed in the post-peak response may be attributed to mesh asymmetry effects, as discussed in the previous section. The structural response up to complete failure is highly complex, exhibiting several sharp snap-back events. The corresponding critical points associated with some of these snap-backs are also highlighted in Fig. \ref{fig:FvsD_results_cross_notch_complex_panel}.}

{\color{blue}Figure \ref{fig:Complex_panel_cross_damage_map} presents the radial propagation pattern of interface damage for the mesh model with 1.0 mm elements.}

\begin{figure}[H]
\centering

\begin{minipage}[t]{0.48\columnwidth}
\centering
    	\begin{tikzpicture}[scale=0.60]
		\begin{axis}[%
			xlabel={Displacement, mm},
			ylabel={Applied load, N},
                xlabel style={font=\large},
                ylabel style={font=\large},
                xmin=0.0, xmax=3.0,
                ymin=0.0, ymax=3500.0,
                xtick distance=0.5, 
                ytick distance=500.0,
                legend style={at={(0.0,1.0)}, anchor=north west, column sep=1ex,
                legend cell align=left, font=\small},
			table/col sep=comma,height=9cm]

            \addplot[red, dashed, line width=1.0pt]  table [x=v, y=F] {4.Results_data/complex_panel/Grid45_crossnotch/Shell6DofsBF_alpha5/complexpanel_crossnotch_shell6DofsBF_2.0mm.txt};
            \addlegendentry{Structural CE model 2.0mm}

            \addplot[teal!60!black,mark=*, only marks, mark size=2pt]  table [x=v, y=F] {4.Results_data/complex_panel/Grid45_crossnotch/Shell6DofsBF_alpha5/complexpanel_crossnotch_shell6DofsBF_2.0mm_critical_points.txt};

            \addplot[only marks, mark=none, nodes near coords, point meta=explicit symbolic, nodes near coords style={font=\large, anchor=west, xshift=-22pt, yshift=15pt}]
            table [x=v, y=F, meta=label] {4.Results_data/complex_panel/Grid45_crossnotch/Shell6DofsBF_alpha5/complexpanel_crossnotch_shell6DofsBF_2.0mm_critical_point1.txt};

            \addplot[only marks, mark=none, nodes near coords, point meta=explicit symbolic, nodes near coords style={font=\large, anchor=west, xshift=1pt, yshift=1pt}]
            table [x=v, y=F, meta=label] {4.Results_data/complex_panel/Grid45_crossnotch/Shell6DofsBF_alpha5/complexpanel_crossnotch_shell6DofsBF_2.0mm_critical_point2.txt};

            \addplot[only marks, mark=none, nodes near coords, point meta=explicit symbolic, nodes near coords style={font=\large, anchor=west, xshift=1pt, yshift=4pt}]
            table [x=v, y=F, meta=label] {4.Results_data/complex_panel/Grid45_crossnotch/Shell6DofsBF_alpha5/complexpanel_crossnotch_shell6DofsBF_2.0mm_critical_point3.txt};

            \addplot[only marks, mark=none, nodes near coords, point meta=explicit symbolic, nodes near coords style={font=\large, anchor=west, xshift=1pt, yshift=4pt}]
            table [x=v, y=F, meta=label] {4.Results_data/complex_panel/Grid45_crossnotch/Shell6DofsBF_alpha5/complexpanel_crossnotch_shell6DofsBF_2.0mm_critical_point4.txt};

            \addplot[only marks, mark=none, nodes near coords, point meta=explicit symbolic, nodes near coords style={font=\large, anchor=west, xshift=1pt, yshift=4pt}]
            table [x=v, y=F, meta=label] {4.Results_data/complex_panel/Grid45_crossnotch/Shell6DofsBF_alpha5/complexpanel_crossnotch_shell6DofsBF_2.0mm_critical_point5.txt};

            \addplot[only marks, mark=none, nodes near coords, point meta=explicit symbolic, nodes near coords style={font=\large, anchor=west, xshift=-12.5pt, yshift=12.5pt}]
            table [x=v, y=F, meta=label] {4.Results_data/complex_panel/Grid45_crossnotch/Shell6DofsBF_alpha5/complexpanel_crossnotch_shell6DofsBF_2.0mm_critical_point6.txt};

		\end{axis}
	\end{tikzpicture}
\caption*{{\color{blue}(a) Load-displacement curve: 2.0 mm model}}
\end{minipage}
\hfill
\begin{minipage}[t]{0.48\columnwidth}
\centering
    	\begin{tikzpicture}[scale=0.60]
		\begin{axis}[%
			xlabel={Displacement, mm},
			ylabel={Applied load, N},
                xlabel style={font=\large},
                ylabel style={font=\large},
                xmin=0.0, xmax=3.0,
                ymin=0.0, ymax=3500.0,
                xtick distance=0.5, 
                ytick distance=500.0,
                legend style={at={(0.0,1.0)}, anchor=north west, column sep=1ex,
                legend cell align=left, font=\small},
			table/col sep=comma,height=9cm]

            \addplot[blue, dashed, line width=1.0pt]  table [x=v, y=F] {4.Results_data/complex_panel/Grid45_crossnotch/Shell6DofsBF_alpha5/complexpanel_crossnotch_shell6DofsBF_1.0mm_DelftBlue.txt};
            \addlegendentry{Structural CE model 1.0mm}

            \addplot[teal!60!black,mark=*, only marks, mark size=2pt]  table [x=v, y=F] {4.Results_data/complex_panel/Grid45_crossnotch/Shell6DofsBF_alpha5/complexpanel_crossnotch_shell6DofsBF_1.0mm_critical_points.txt};

		\end{axis}
	\end{tikzpicture}
\caption*{{\color{blue}(b) Load-displacement curve: 1.0 mm model}}
\end{minipage}
\caption{{\color{blue}Load–displacement curves - pull-out test in a cross-notched stiffened panel.}}
\label{fig:FvsD_results_cross_notch_complex_panel}
\end{figure} 

\vspace{-10pt}

\begin{figure}[H]
\centering
\scalebox{0.21}{\includegraphics{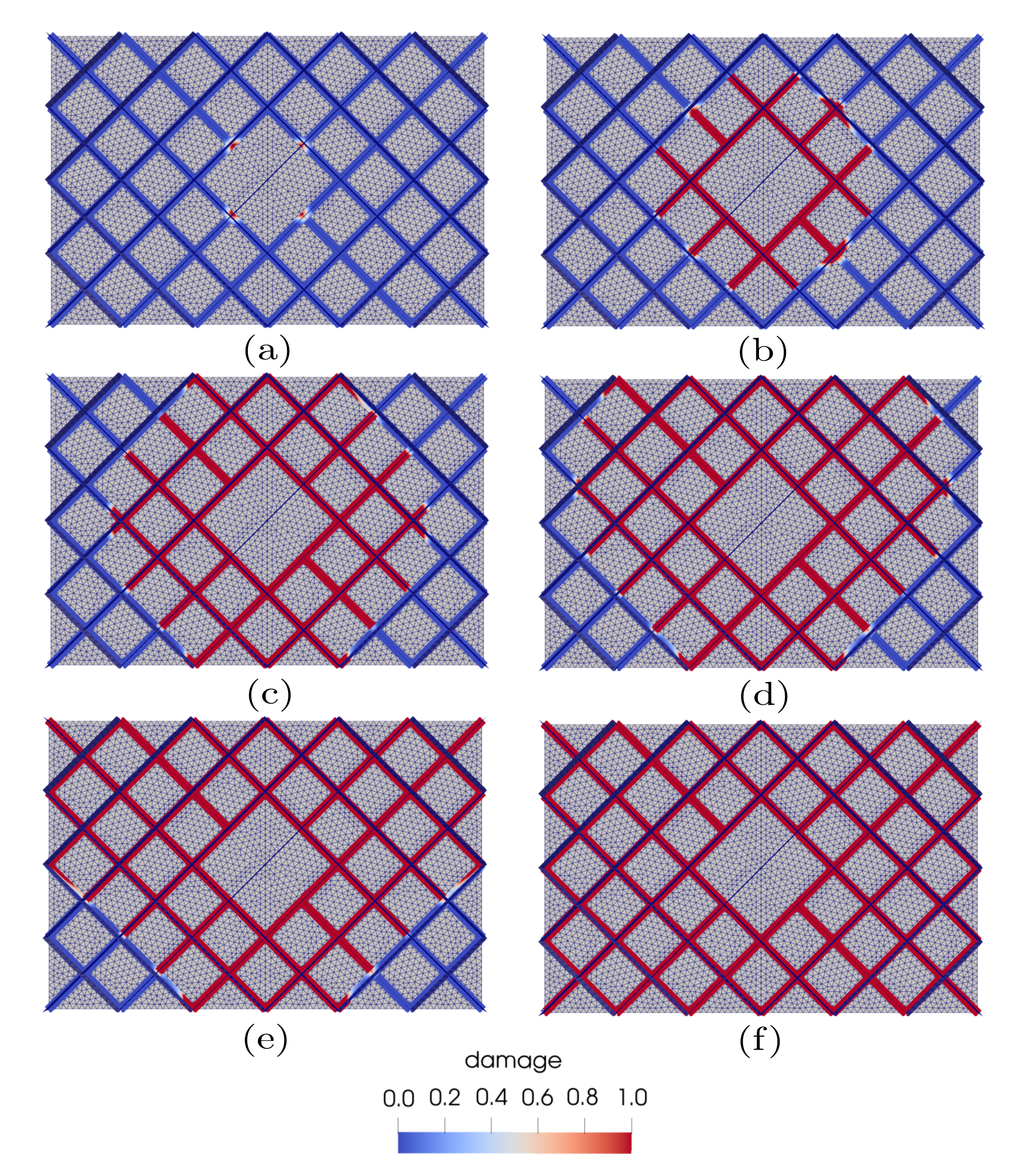}}
	\caption{{\color{blue}Radial interface damage evolution for different time steps (see Fig. \ref{fig:FvsD_results_cross_notch_complex_panel}a).}}
	\label{fig:Complex_panel_cross_damage_map}
\end{figure}

\begin{figure}[H]
\centering
\scalebox{0.28}{\includegraphics{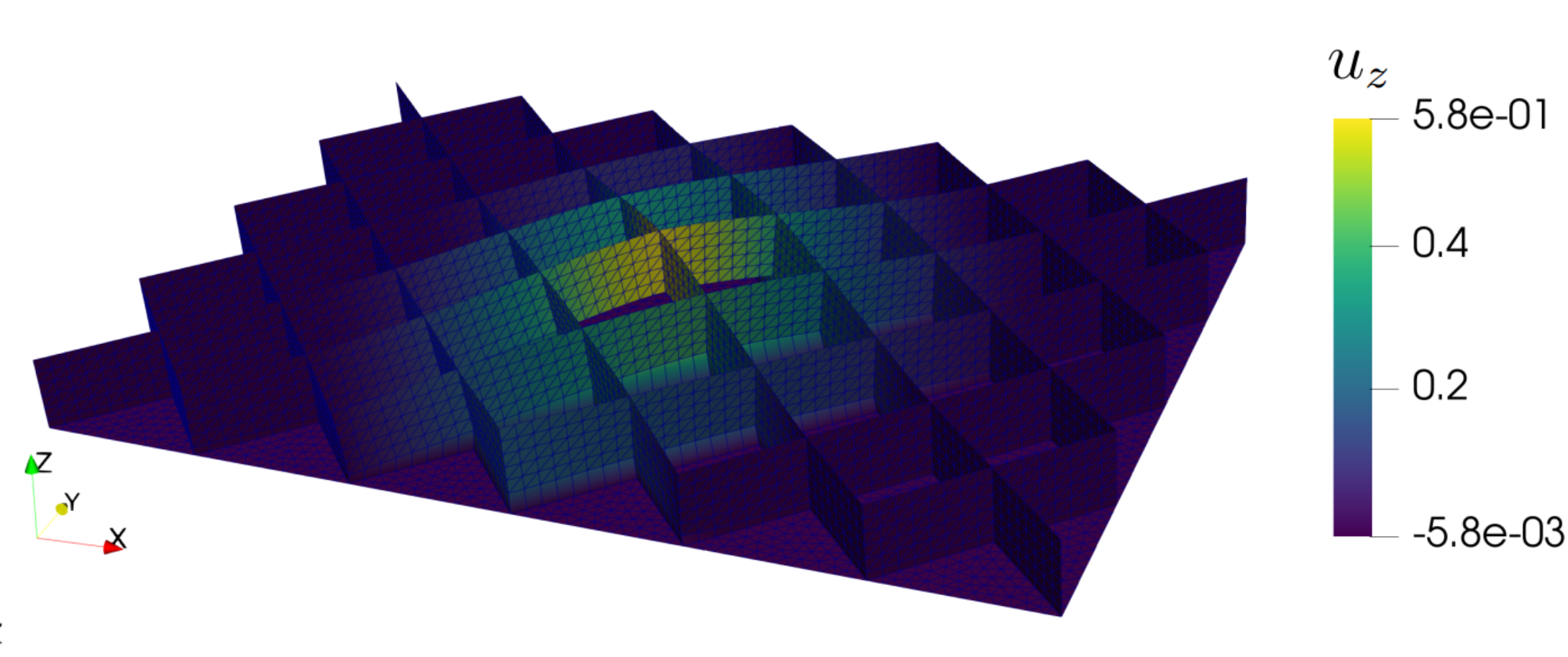}}
	\caption{Deformed configuration with displacements magnified by a factor of 10.}
	\label{fig:Complex_panel_cross_notch_deformed_configuration}
\end{figure}

{\color{blue}The damage maps in Fig. \ref{fig:Complex_panel_cross_damage_map} correspond to the damage states at the critical load steps highlighted in Fig. \ref{fig:FvsD_results_cross_notch_complex_panel}(a). Additionally, Figure \ref{fig:Complex_panel_damage_mapD} shows a zoomed view of a region of interest, where it is possible to observe one more time the cohesive zone within the domain of a single element.

\begin{figure}[H]
\centering
\scalebox{0.27}{\includegraphics{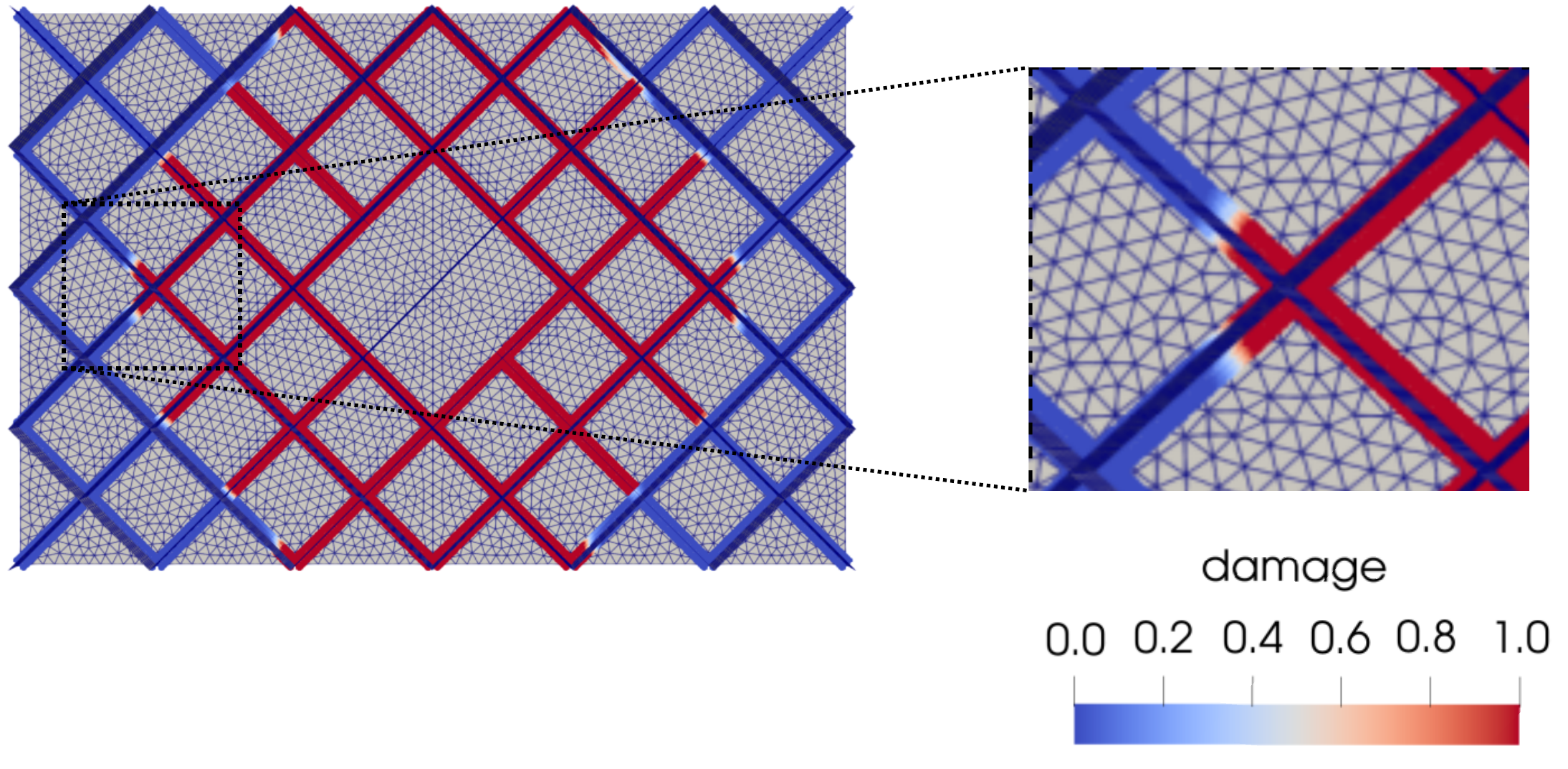}}
	\caption{{\color{blue}Interface damage - zoom view from Fig. \ref{fig:Complex_panel_cross_damage_map}(c).}}
	\label{fig:Complex_panel_damage_mapD}
\end{figure}

The damage evolution and load–displacement response of the models demonstrate the performance of the proposed formulation under a more challenging mixed-mode propagation scenario, characterized by nonlinear crack front evolution. Figure \ref{fig:Complex_panel_cross_notch_deformed_configuration} illustrates the deformed configuration for the model with a 2.0 mm mesh at an intermediary load step.}


\vspace{-5pt}

\section{Conclusions}
\label{sec:conclusions}

A novel structural cohesive element is proposed for the efficient modeling of debonding in composite panels with overmolded stiffeners. {\color{blue}Three-node higher-order hybrid/mixed shell elements based on the Kirchhoff hypothesis are used to model the panels and the stiffeners. The in-plane bending response of the element is enhanced by the Free Formulation (FF). The proposed cohesive element connects to T-jointed orthogonal shells. An appropriate definition of the displacement jump, consistent with shell kinematics, enables to compute the jump vector at any point over the cohesive surface from the shell displacement approximations evaluated at the element edges. The higher-order weakly continuous fields adopted for the evaluation of the jump vector enable reliable debonding analyses using relatively coarse meshes. The framework is suitable for analyzing debonding in skin–stiffener structures where non-constant damage through the stiffener thickness is expected.}

The model is verified for mode I, mode II, and mixed-mode benchmark problems. The results show that the structural cohesive models based on the FF membrane delivered superior performance compared to those employing the Constant Strain Triangle (CST) membrane in most cases. A panel–stiffener debonding problem designed to promote stable crack propagation is also analyzed using both standard 3D cohesive elements and the proposed structural cohesive element with the FF membrane. The results demonstrate that the proposed models can employ coarser meshes than the standard cohesive models, achieving more than a 95\%  reduction in CPU time, while maintaining comparable accuracy. Consequently, the proposed element enables efficient analysis of stiffener debonding in laminated panels with complex overmolded stiffener grids. {\color{blue}The model can be coupled in future studies with the cohesive element in \cite{Aietal2025} to simulate both delamination and panel–stiffener debonding in progressive failure analyses. It is also applicable to skin–stiffener debonding in overmolded panels with thickness-dependent interface properties, a feature of significant interest in thermoplastic applications.}

\section*{Declaration of competing interest}
\label{sec:DeclarationofCompetingInterest}

The authors declare that they have no known competing financial interests or personal relationships that could have appeared to influence the work reported in this paper.

\section*{Data availability}
\label{sec:DataAvailability}

Data presented in this article will be available at the 4TU.ResearchData repository through \url{https://doi.org/10.4121/9eaeb151-d8c1-42a7-b060-7fea8085e2d2}.

\section*{Acknowledgements}
\label{sec:Acknowledgements}

This research was carried out as part of the project ENLIGHTEN (project number N21010 g) in the framework of the Partnership Program of the Materials innovation institute M2i (\url{www.m2i.nl}) and the Netherlands Organization for Scientific Research (\url{www.nwo.nl}).

\appendix
\renewcommand{\thefigure}{A.\arabic{figure}}
\setcounter{figure}{0}
\setcounter{table}{0} 
\renewcommand{\thetable}{A.\arabic{table}}

\section{Shape function approximations and derivatives}
\label{app:shapefunctions}

The approximations of the in-plane displacement approximations $\tilde{u}$, $\tilde{v}$ and their derivatives are expressed in terms of the membrane DoFs $\overline{\mathbf{u}}$. The out-of-plane approximation $\tilde{w}$ and its derivatives are expressed in terms of the plate DoFs $\overline{\mathbf{w}}$.

\subsection{\texorpdfstring{Shape functions of $\tilde{u}$, $\tilde{v}$ and their derivatives}{Shape functions of u, v and their derivatives}}

Referring back to Eq. (\ref{Displacement_decomposition}) and the transformation (\ref{qm_to_au_transformation}), the membrane displacement approximations can be written in terms of the membrane DoFs $\overline{\mathbf{u}}$ as 
\begin{align}
    \tilde{\mathbf{u}}=
    \left\{\begin{array}{l}
        \tilde{u} \\
        \tilde{v}
    \end{array}\right\}=
    \left[\begin{array}{ccc}
        \boldsymbol{\phi}_{r}\mathbf{H}_{\mathrm{r}} +  \boldsymbol{\phi}_{c}\mathbf{H}_{\mathrm{c}} + \boldsymbol{\phi}_{h}\mathbf{H}_{\mathrm{h}} 
    \end{array}\right]\bar{\mathbf{u}}=
    \left[\begin{array}{c}
        \mathbf{N}_{\mathrm{u}} \\
        \mathbf{N}_{\mathrm{v}}
    \end{array}\right]\bar{\mathbf{u}},
    \label{Membrane_displacements_Dofs_II}
\end{align}

\noindent where the ($1\times9$) matrices $\mathbf{N}_{\mathrm{u}}$ and $\mathbf{N}_{\mathrm{v}}$ are the shape functions of $ \tilde{u}$ and $ \tilde{v}$, respectively. The ($2\times3$) matrices $\boldsymbol{\phi}_{r}$, $\boldsymbol{\phi}_{c}$ and $\boldsymbol{\phi}_{h}$ are the rigid body, constant strain and higher-order modes, as given in \cite{BerganandFelippa1985}, and the ($3\times9$) matrices $\mathbf{H}_{\mathrm{c}}$, $\mathbf{H}_{\mathrm{r}}$ and $\mathbf{H}_{\mathrm{h}}$ are defined in Eq. (\ref{qm_to_au_transformation}).

The derivatives of the approximation (\ref{Membrane_displacements_Dofs_II}) are 
\begin{align}
    \tilde{\mathbf{u}}_{\color{blue},x}=
    \left\{\begin{array}{l}
        \tilde{u}_{\color{blue},x} \\
        \tilde{v}_{\color{blue},x}
    \end{array}\right\}=
    \left[\begin{array}{ccc}
        \boldsymbol{\phi}_{r\color{blue},x}\mathbf{H}_{\mathrm{r}} + \boldsymbol{\phi}_{c\color{blue},x}\mathbf{H}_{\mathrm{c}} + \boldsymbol{\phi}_{h\color{blue},x}\mathbf{H}_{\mathrm{h}} 
    \end{array}\right]\bar{\mathbf{u}}=
    \left[\begin{array}{c}
        \mathbf{N}_{\mathrm{u}_{\color{blue},x}} \\
        \mathbf{N}_{\mathrm{v}_{\color{blue},x}}
    \end{array}\right]\bar{\mathbf{u}},
    \label{Membrane_displacements_derivatives_x}
\end{align}
\begin{align}
    \tilde{\mathbf{u}}_{\color{blue},y}=
    \left\{\begin{array}{l}
        \tilde{u}_{\color{blue},y} \\
        \tilde{v}_{\color{blue},y}
    \end{array}\right\}=
    \left[\begin{array}{ccc}
        \boldsymbol{\phi}_{r\color{blue},y}\mathbf{H}_{\mathrm{r}} + \boldsymbol{\phi}_{c\color{blue},y}\mathbf{H}_{\mathrm{c}} + \boldsymbol{\phi}_{h\color{blue},y}\mathbf{H}_{\mathrm{h}} 
    \end{array}\right]\bar{\mathbf{u}}=
    \left[\begin{array}{c}
        \mathbf{N}_{\mathrm{u}_{\color{blue},y}} \\
        \mathbf{N}_{\mathrm{v}_{\color{blue},y}}
    \end{array}\right]\bar{\mathbf{u}},
    \label{Membrane_displacements_derivatives_y}
\end{align}

 \noindent where the derivatives $\boldsymbol{\phi}_{k\color{blue},x}=\partial\boldsymbol{\phi}_{k}/\partial x$ and $\boldsymbol{\phi}_{k\color{blue},y}=\partial\boldsymbol{\phi}_{k}/\partial y$, $k=r,c,h$, are computed from the modes $\boldsymbol{\phi}_{r}$, $\boldsymbol{\phi}_{c}$ and $\boldsymbol{\phi}_{h}$ defined in \cite{BerganandFelippa1985}.

\subsection{\texorpdfstring
  {Shape functions of $\tilde{w}$ and its derivatives}
  {Shape functions of w and derivatives}
}

Starting from the cubic approximation (\ref{Plate_Approximations}), the displacement $\tilde{w}$ and its derivatives $\tilde{w}_{\color{blue},x} = \partial \tilde{w} / \partial x$ and $\tilde{w}_{\color{blue},y} = \partial \tilde{w} / \partial y$ were expressed in terms of the plate DoFs $\overline{\mathbf{w}}$ in \cite{Aietal2025}. The resulting approximations are given by
\begin{align}  \label{Plate_fields_Dofs_I}
    \tilde{w} &= \left[\mathbf{S}^{\mathrm{T}} \mathbf{M}_A^{-1}\left(\mathbf{B}_A-\mathbf{M}_B \mathbf{C}\right) +\mathbf{R}^{\mathrm{T}} \mathbf{C}\right] \overline{\mathbf{w}}=\mathbf{N}_w \overline{\mathbf{w}} \\[4pt] 
    \label{Plate_fields_Dofs_II}
    \tilde{w}_{\color{blue},x} &=\left[\mathbf{S}_{\color{blue},x} \mathbf{M}_A^{-1}\left(\mathbf{B}_A-\mathbf{M}_B \mathbf{C}\right) +\mathbf{R}_{\color{blue},x}^{\mathrm{T}} \mathbf{C}\right] \overline{\mathbf{w}}=\mathbf{N}_{w_{\color{blue},x}} \overline{\mathbf{w}} \\[4pt]
    \label{Plate_fields_Dofs_III}
    \tilde{w}_{\color{blue},y} &=\left[\mathbf{S}_{\color{blue},y} \mathbf{M}_A^{-1}\left(\mathbf{B}_A-\mathbf{M}_B \mathbf{C}\right) +\mathbf{R}_{\color{blue},y}^{\mathrm{T}} \mathbf{C}\right] \overline{\mathbf{w}}=\mathbf{N}_{w_{\color{blue},y}} \overline{\mathbf{w}} 
\end{align}

\noindent where the $(1 \times 9)$ matrices $\mathbf{N}_w$, $\mathbf{N}_{w_{\color{blue},x}}$ and $\mathbf{N}_{w_{\color{blue},y}}$ are the shape functions of $ \tilde{w}$, $\tilde{w}_{\color{blue},x}$ and $\tilde{w}_{\color{blue},y}$, respectively. The matrices $\mathbf{S}$, $\mathbf{R}$, $\mathbf{R}_{\color{blue},x}$ and $\mathbf{R}_{\color{blue},y}$ are the ones that are not constants. They are defined as
\begin{align} \nonumber
    \mathbf{S} &=\left[\begin{array}{lll}
    1 & x & y
    \end{array}\right]^{\mathrm{T}}, \hspace{10pt} \mathbf{R}=\left[\begin{array}{lllllll}
    x^2 & xy & y^2 & x^3 & x^2y & xy^2 & y^3
    \end{array}\right]^{\mathrm{T}},
\end{align}

\vspace{-20pt}

\begin{align} \nonumber
    \mathbf{R}_{\color{blue},x} &=\left[\begin{array}{lllllll}
    2x & y & 0 & 3x^2 & 2xy & y^2 & 0
    \end{array}\right]^{\mathrm{T}},  \\[8pt] \nonumber
    \mathbf{R}_{\color{blue},y} &=\left[\begin{array}{lllllll}
    0 & x & 2y & 0 & x^2 & 2xy & 3y^2
    \end{array}\right]^{\mathrm{T}}.
\end{align}

The constant matrix $\mathbf{C}$, which depends on material and geometrical properties of plate element, is defined as
\begin{align} \nonumber
    \mathbf{C}=(\mathbf{H})^{-1}(\mathbf{B}\mathbf{T})
\end{align}
\noindent where the matrices $\mathbf{H}$, $\mathbf{B}$ and $\mathbf{T}$ are the same as those from Eq. (\ref{alpha_Dofs_Relation}). The matrices $\mathbf{M}_A$, $\mathbf{M}_B$, $\mathbf{B}_A$, $\mathbf{S}_{\color{blue},x}$ and $\mathbf{S}_{\color{blue},y}$ are also constant and given by
\begin{align} \nonumber
   \mathbf{M}^{\alpha}_A=\left[\begin{array}{lll}
    1 & x_1 & y_1 \\
    1 & x_2 & y_2 \\
    1 & x_3 & y_3
    \end{array}\right], \hspace{10pt} \mathbf{M}^{\alpha}_B=\left[\begin{array}{lllllll}
    x_1^2 & x_1 y_1 & y_1^2 & x_1^3 & x_1^2 y_1 & x_1 y_1^2 & y_1^3 \\
    x_2^2 & x_2 y_2 & y_2^2 & x_2^3 & x_2^2 y_2 & x_2 y_2^2 & y_2^3 \\
    x_3^2 & x_3 y_3 & y_3^2 & x_3^3 & x_3^2 y_3 & x_3 y_3^2 & y_3^3
    \end{array}\right],
\end{align}
\begin{align} \nonumber
   \mathbf{B}_A=\left[\begin{array}{lllllllll}
    1 & 0 & 0 & 0 & 0 & 0 & 0 & 0 & 0 \\
    0 & 0 & 0 & 1 & 0 & 0 & 0 & 0 & 0 \\
    0 & 0 & 0 & 0 & 0 & 0 & 1 & 0 & 0
    \end{array}\right], 
\end{align}
\begin{align} \nonumber
    \mathbf{S}_{\color{blue},x}=\left[\begin{array}{lll}
    0 & 1 & 0
    \end{array}\right]^{\mathrm{T}}, \hspace{10pt} \mathbf{S}_{\color{blue},y}=\left[\begin{array}{lll}
    0 & 0 & 1
    \end{array}\right]^{\mathrm{T}},
    \label{Sx_Sy_matrices}
\end{align}
\noindent in which $x_i,y_i$, $i=1,2,3$, are the nodal coordinates of the element. Notice that $\mathbf{R}_{\color{blue},x}$, $\mathbf{R}_{\color{blue},y}$, $\mathbf{S}_{\color{blue},x}$ and $\mathbf{S}_{\color{blue},y}$ are just the derivatives of $\mathbf{R}$ and $\mathbf{S}$ with respect to $x$ and $y$.

 \bibliographystyle{elsarticle-num} 
 \bibliography{cas-refs}

@article{Dugdale1960,
    Author = {Dugdale, D S},
    title = {Yielding of steel sheets containing slits},
    journal = {Journal of the Mechanics and Physics of Solids},
    volume = {8},
    year = {1960},
    pages = {100-104},
}

@article{Barenblatt1962,
    Author = {Barenblatt, G I},
    title = {The mathematical theory of equilibrium cracks in brittle fracture},
    journal = {Advances in Applied Mechanics},
    volume = {7},
    year = {1962},
    pages = {55-129},
}

@article{Bell1969,
    Author = {Allman, D J},
    title = {A refined triangular plate bending finite element},
    journal = {International Journal for Numerical Methods in Engineering},
    volume = {1},
    year = {1969},
    pages = {101-22},
}

@article{Allman1976,
    Author = {Allman, D J},
    title = {A simple cubic displacement element for plate bending},
    journal = {International Journal for Numerical Methods in Engineering},
    volume = {10},
    year = {1976},
    pages = {263-281},
}

@article{BerganandNygard1984,
    Author = {Bergan, P G and Nygård, M K},
    title = {Finite elements with increased freedom in choosing shape functions},
    journal = {International Journal for Numerical Methods in Engineering},
    volume = {50},
    issue = {1},
    year = {1985},
    pages = {25-69},
}

@article{BerganandFelippa1985,
    Author = {Bergan, P G and Felippa, C A},
    title = {A triangular membrane element with rotational degrees of freedom},
    journal = {Computer Methods in Applied Mechanics and Engineering},
    volume = {50},
    issue = {1},
    year = {1985},
    pages = {25-69},
}

@article{Felippa1989,
    Author = {Felippa, C A},
    title = {Parametrized multifield variational principles in elasticity: II. hybrid functionals and the free formulation},
    journal = {Communications in Applied Numerical Methods},
    volume = {5},
    year = {1989},
    pages = {89-98},
}

@article{Williams1989,
    Author = {Williams, J G},
    title = {The fracture mechanics of delamination tests},
    journal = {Journal of Strain Analysis},
    volume = {24},
    issue = {4},
    year = {1989},
    pages = {207-14},
}

@article{BenzeggaghandKenane1996,
    Author = {Benzeggagh, M L and Kenane, M},
    title = { Measurement of mixed-mode delamination fracture toughness of unidirectional glass/epoxy composites with mixed mode bending apparatus},
    journal = {Composites Science and Technology},
    volume = {56},
    year = {1996},
    pages = {439-449},
}

@article{PradhanandTay1998,
  author = { S. C. Pradhan and T. E. Tay},
  title = {Three-dimensional finite element modelling of delamination growth in notched composite laminates under compression loading},
  journal = {Engineering Fracture Mechanics},
  volume = {98},
  year = {1998},
  pages = {157-171},
}

@article{Qiuetal2001,
    Author = {Qiu, Y and Crisfield, M A and Alfano, G},
    title = {An interface element formulation for the simulation of delamination with buckling},
    journal = {Engineering Fracture Mechanics},
    volume = {68},
    year = {2001},
    pages = {1755-76},
}

@techreport{CamanhoandDavila2002,
  author      = {Camanho, P P and Dávila, C G},
  title       = {Mixed-Mode Deco\-hesion Finite Elements for the Simulation of Delamination in Composite Materials},
  institution = {NASA},
  number      = {NASA/TM-2002-211737},
  year        = {2002},
  type        = {Technical Report},
}

@article{Felippa2003,
    Author = {Felippa, C A},
    title = {A study of optimal membrane triangles with drilling freedoms},
    journal = {Computer Methods in Applied Mechanics and Engineering},
    volume = {192},
    year = {2003},
    pages = {2125-68},
}

@article{YangandCox2005,
    Author = {Yang, Q and Cox, B},
    title = {Cohesive models for damage evolution in laminated composites},
    journal = {International Journal of Fracture},
    volume = {133},
    year = {2005},
    pages = {107-37},
}

@article{Turonetal2006,
    Author = {Turon, A and Camanho, P P and Costa, J and Dávila, C G},
    title = {A damage model for the simulation of delamination in advanced composites under variable-mode loading},
    journal = {Mechanics of Materials},
    volume = {38},
    issue = {11},
    year = {2006},
    pages = {1072-89},
}

@article{Turonetal2007,
    Author = {Turon, A and Dávila, C G and Camanho, P P and Costa, J},
    title = {An engineering solution for mesh size effects in the simulation of delamination using cohesive zone models},
    journal = {Engineering Fracture Mechanics},
    volume = {74},
    issue = {10},
    year = {2007},
    pages = {1665–82},
}

@misc{ASTM_D5528_2007,
  title = {{ASTM D5528 / D5528M-21}, Standard test method for mode I interlaminar fracture toughness of unidirectional fiber-reinforced polymer matrix composites},
  howpublished = {West Conshohocken, PA: ASTM International},
  year         = {2007},
}

@article{HarperandHallett2008,
    Author = {Harper, P W and Hallett, S R},
    title = {Cohesive zone length in numerical simulations of composite delamination},
    journal = {Engineering Fracture Mechanics},
    volume = {75},
    year = {2008},
    pages = {4774-4792},
}

@article{Verhooseletal2009,
    Author = {Verhoosel, C V and Remmers, J J C and Gutierrez, M A},
    title = {A dissipation-based arc-length method for robust simulation of brittle and ductile failure},
    journal = {Int. J. Numer. Meth. Engng},
    volume = {77},
    year = {2009},
    pages = {1290–1321},
}

@article{BalzaniandWagner2010,
    Author = {Balzani, C and Wagner, W},
    title = {Numerical treatment of damage propagation in axially compressed composite airframe panels},
    journal = {International Journal of Structural Stability and Dynamics},
    volume = {10},
    issue = {4},
    year = {2010},
    pages = {683–703},
}

@article{Turonetal2010,
    Author = {Turon, A and Camanho, P and Costa, J and Renart, J},
    title = {Accurate simulation of delamination growth under mixed-mode loading using cohesive elements: Definition of interlaminar strengths and elastic stiffness},
    journal = {Composite Structures},
    volume = {92},
    issue = {8},
    year = {2010},
    pages = {1857–64},
}

@article{vanderMeerandSluys2010,
    Author = {van der Meer, F and Sluys, L J},
    title = {Mesh-independent modeling of both distributed and discrete matrix cracking in interaction with delamination in composites},
    journal = {Engineering Fracture Mechanics},
    volume = {77},
    issue = {4},
    year = {2010},
    pages = {719-35},
}

@article{Yangetal2010,
    Author = {Yang, Q D and Fang, X J and Shi, J X and Lua, J},
    title = {An improved cohesive element for shell delamination analyses},
    journal = {International journal for numerical methods in engineering},
    volume = {83},
    year = {2010},
    pages = {611-41},
}

@article{vanderMeeretal2012,
    Author = {van der Meer, F and Moës, N and Sluys, L J},
    title = {A level set model for delamination–modeling crack growth without cohesive zone or stress singularity},
    journal = {Engineering Fracture Mechanics},
    volume = {79},
    year = {2012},
    pages = {191-212},
}

@article{vanderMeer2012,
    Author = {van der Meer, F P},
    title = {Mesolevel Modeling of Failure in Composite Laminates: Constitutive, Kinematic and Algorithmic Aspects},
    journal = {Archives of Computational Methods in Engineering},
    volume = {19},
    year = {2012},
    pages = {381-425},
}

@article{Doetal2013,
    Author = {Do, B and Liu, W and Yang, Q and Su, X},
    title = {Improved cohesive stress integration schemes for cohesive zone elements},
    journal = {Engineering Fracture Mechanics},
    volume = {107},
    year = {2013},
    pages = {14-28},
}

@article{Álvarezetal2014,
    Author = {Álvarez, D and Blackman, B and Guild, F and Kinloch, A},
    title = {Mode I fracture in adhesively bonded joints: A mesh-size independent modelling approach using cohesive elements},
    journal = {Engineering Fracture Mechanics},
    volume = {115},
    year = {2014},
    pages = {73-95},
}

@article{Krueger2015,
    Author = {Krueger, R},
    title = {A summary of benchmark examples to assess the performance of quasi-static delamination propagation prediction capabilities in finite element codes},
    journal = {Journal of Composite Materials},
    volume = {49},
    issue = {26},
    year = {2015},
    pages = {3297-316},
}

@article{Luetal2018,
    Author = {Lu, X and Chen, B-Y and Tan, V B and Tay, T-E},
    title = {Adaptive floating node method for modelling cohesive fracture of composite materials},
    journal = {Engineering Fracture Mechanics},
    volume = {194},
    year = {2018},
    pages = {240-61},
}

@article{Bazilevsetal2018,
    Author = {Bazilevs, Y and Pigazzini, M S and Ellison, A and Kim, H},
    title = {A new multi-layer approach for progressive damage simulation in composite laminates based on isogeometric analysis and Kirchhoff–Love shells. Part I: basic theory and modeling of delamination and transverse shear},
    journal = {Computational Mechanics},
    volume = {62},
    year = {2018},
    pages = {563-85},
}

@article{Akterskaiaetal2018,
    Author = {Akterskaia, M and Jansen, E and Hallett, S R and Weaver, P and Rolfes, R},
    title = {Analysis of skin-stringer debonding in composite panels through a two-way global-local method},
    journal = {Composite Structures},
    volume = {202},
    year = {2018},
    pages = {1280-94},
}

@inproceedings{Valverdeetal2018,
  author    = {Valverde, M A and Kupfer, R and Kawashita, L F and Gude, M and Hallett, R},
  title     = {Effect of processing
parameters on quality and strength in thermoplastic composite injection overmoulded components},
  booktitle = {18th European Conference on Composite Materials},
  year      = {2018},
  pages = {1-8},
  publisher = {Applied Mechanics Laboratory},
}

@article{Luetal2019,
    Author = {Lu, X and Ridha, M and Chen, B and Tan, V and Tay, T},
    title = {On cohesive element parameters and delamination modelling},
    journal = {Engineering Fracture Mechanics},
    volume = {206},
    year = {2019},
    pages = {278–96},
}

@article{Akkermanetal2020,
    Author = {Akkerman, R and Bouwman, M and Wijskamp, S},
    title = {Analysis of the Thermoplastic Composite Overmolding Process: Interface Strength},
    journal = {Frontiers in Materials},
    volume = {7},
    year = {2020},
    pages = {102925},
}

@article{GiustiandLucchetta2020,
    Author = {Giusti, R and Lucchetta, G},
    title = {Modeling the Adhesion Bonding Strength in Injection Overmolding of Polypropylene Parts},
    journal = {Polymers},
    volume = {12},
    year = {2020},
    pages = {2063},
}

@article{Valverdeetal2020,
    Author = {Valverde, M and Kupfer, R and Wollmann, T and Kawashita, L and Gude, M and Hallett, S},
    title = {Influence
of component design on features and properties in thermoplastic overmoulded composites},
    journal = {Composites Part A: Applied Science and Manufacturing},
    volume = {132},
    year = {2020},
    pages = {105823},
}

@article{Nguyen-Thanhetal2020,
    Author = {Nguyen-Thanh, C and Nguyen, V P and de Vaucorbeil, A and Mandal, T K and Wu, J-Y},
    title = {Jive: An open source, research-oriented C++ library for solving partial differential equations},
    journal = {Advances in Engineering Software},
    volume = {150},
    year = {2020},
    pages = {102925},
}

@article{RussoandChen2020,
    Author = {Russo, R and Chen, B},
    title = {Overcoming the cohesive zone limit in composites delamination: modeling with slender structural elements and higher-order adaptive integration},
    journal = {International Journal for Numerical Methods in Engineering},
    volume = {121},
    issue = {24},
    year = {2020},
    pages = {5511-45},
}

@article{Boutagouga2021,
    Author = {Boutagouga, D},
    title = {A Review on Membrane Finite Elements with Drilling Degree of Freedom},
    journal = {Archives of Computational Methods in Engineering},
    volume = {28},
    year = {2021},
    pages = {3049-65},
}

@misc{ASTM_D6671_2022,
  title        = {{ASTM D6671 / D6671M-22}: Standard test method for mixed mode I–mode II interlaminar fracture toughness of unidirectional fiber-reinforced polymer matrix composites},
  howpublished = {West Conshohocken, PA: ASTM International},
  year         = {2022},
}

@article{Neveuetal2022,
    Author = {Neveu, F and Cornu, C and Olivier, P and Castanié, B},
    title = {Manufacturing and impact behaviour of aeronautic overmolded grid-stiffened thermoplastic carbon plates},
    journal = {Composite Structures},
    volume = {284},
    year = {2022},
    pages = {115228},
}

@article{GiustiandLucchetta2023,
    Author = {Giusti, R and Lucchetta, G},
    title = {Cohesive Zone Modeling of the Interface Fracture in Full-Thermoplastic Hybrid Composites for Lightweight Application},
    journal = {Polymers},
    volume = {15},
    year = {2023},
    pages = {4459},
}

@article{BalducciandChen2024,
    Author = {Balducci, G T and Chen, B},
    title = {Overcoming the cohesive zone limit in the modelling of composites delamination with TUBA cohesive elements},
    journal = {Composites Part A: Applied Science and Manufacturing},
    volume = {185},
    year = {2024},
    pages = {108356},
}

@article{Aietal2025,
    Author = {Ai, X and Chen, B and Kassapoglou, C},
    title = {Structural cohesive element for the modelling of delamination in composite laminates without the cohesive zone limit},
    journal = {Engineering Fracture Mechanics},
    volume = {329},
    year = {2025},
    pages = {111586},
}

@article{Hofmanetal2026,
    Author = {Hofman, P and van der Meer, F P and Sluys, L J},
    title = {Computational analysis of fracture and fatigue in overmolded thermoplastic composites: time-homogenized viscoplasticity, cohesive fracture and processing effects},
    journal = {International Journal of Solids and Structures},
    volume = {338},
    year = {2026},
    pages = {114092},
}
\end{document}